\documentclass[twocolumn]{aastex63}
\usepackage{natbib}
\usepackage{rotating}
\usepackage{multirow}
\usepackage{hyperref}
\usepackage{amsmath}
\usepackage{color}
\shorttitle{Czernik 41 and NGC 1342}
\shortauthors{\"Ozt\"urk et al.}
\graphicspath{{./}{Figures/}}
\begin{document}
\title {A Comprehensive Study of Czernik 41 and NGC 1342 Using CCD UBV and Gaia DR3 Data}

\correspondingauthor{Bur\c cin Tan\i k \"Ozt\"urk}
\email{burcin.tanik@istanbul.edu.tr}

\author[0000-0002-5657-6194]{Bur\c cin Tan\i k \"Ozt\"urk}
\affiliation{Istanbul University, Institute of Graduate Studies in Science, Programme of Astronomy and Space Sciences, 34116, Istanbul, Turkey}

\author[0000-0003-3510-1509]{Sel\c{c}uk Bilir}
\affiliation{Istanbul University, Faculty of Science, Department of Astronomy and Space Sciences, 34119, Beyaz\i t, Istanbul, Turkey}

\author[0000-0002-5657-6194]{Talar Yontan}
\affiliation{Istanbul University, Faculty of Science, Department of Astronomy and Space Sciences, 34119, Beyaz\i t, Istanbul, Turkey}

\author[0000-0002-0435-4493]{Olcay Plevne}
\affiliation{Istanbul University, Faculty of Science, Department of Astronomy and Space Sciences, 34119, Beyaz\i t, Istanbul, Turkey}

\author[0000-0002-0688-1983]{Tansel Ak}
\affiliation{Istanbul University, Faculty of Science, Department of Astronomy and Space Sciences, 34119, Beyaz\i t, Istanbul, Turkey}
\affiliation{Istanbul University, Observatory Research and Application Center, Istanbul University, 34119, Istanbul, Turkey}

\author[0000-0002-0912-6019]{Serap Ak}
\affiliation{Istanbul University, Faculty of Science, Department of Astronomy and Space Sciences, 34119, Beyaz\i t, Istanbul, Turkey}

\author[0000-0003-2575-9892]{Remziye Canbay}
\affiliation{Istanbul University, Institute of Graduate Studies in Science, Programme of Astronomy and Space Sciences, 34116, Istanbul, Turkey}

\author[0000-0001-9445-4588]{Timothy Banks}
\affiliation{Nielsen, 675 6th Ave, New York, NY 10011, USA}
\affiliation{Departments of Physical Science and Engineering, Harper College, 1200 W Algonquin Rd, IL 60067, USA}

\begin{abstract}
In this study, the structural, astrophysical, kinematic, and Galactic orbital parameters of the open clusters Czernik 41 and NGC 1342, as well as their dynamical evolution, are investigated using CCD {\it UBV} photometry and {\it Gaia} data. By applying the {\sc UPMASK} algorithm to {\it Gaia} astrometric data for the estimation of cluster membership probabilities, we have determined that 382 stars in Czernik 41 and 111 stars in NGC 1342 exhibit the highest statistical likelihood of being cluster members. Fundamental parameters (including reddening, metallicity, distance, and age) were derived using both classical methods, where parameters are determined separately, and Markov Chain Monte Carlo (MCMC) methods, where parameters are estimated simultaneously. The results obtained from both approaches are in agreement, confirming the reliability of the derived parameters and demonstrating their robustness against potential degeneracies. The distances to Czernik 41 and NGC 1342 were determined as $2485 \pm 151$ pc and $645 \pm 42$ pc, respectively, while their ages were estimated to be $69 \pm 10$ Myr and $1000 \pm 50$ Myr. The metallicity values ([Fe/H]) were found to be $0.07 \pm 0.09$ dex for Czernik 41 and $-0.14 \pm 0.07$ dex for NGC 1342. The stellar mass functions for both clusters were derived, yielding slopes of $\Gamma = 1.67 \pm 0.23$ for Czernik 41 and $\Gamma = 1.56 \pm 0.41$ for NGC 1342. Kinematic orbit analysis indicates that Czernik 41 originated within the Solar circle, whereas NGC 1342 formed outside it.
\end{abstract}


%-------------------------------------------------------------------------------------

\section{Introduction}
\label{sec:introduction}

Open clusters (OCs) are essential stellar systems for exploring the processes of star formation, stellar evolution, and the structural dynamics of the Galactic disc. These relatively young and metal-rich systems are typically located near the Galactic plane and consist of stars that share similar physical properties, such as age and chemical composition, despite significant differences in mass \citep{Lada03, Maurya20}. Formed through the collapse of molecular clouds, the stars within OCs initially remain gravitationally bound, exhibiting coherent positional and kinematic traits \citep{McKee07}. These unique characteristics make OCs indispensable for understanding the broader mechanisms of stellar evolution and dynamics in the Milky Way.

Photometric and astrometric studies have been proven crucial for determining fundamental properties of OCs, including age, distance, and metallicity. Such parameters, often inaccessible for single stars, are derived through the collective analysis of cluster members \citep{Dias21}. Recent advancements in astrometric precision, particularly through the {\it Gaia} mission \citep{Gaia16} data releases, have enhanced our ability to trace the motions of OC members, enabling detailed reconstruction of their orbits and Galactic dynamics \citep{Tarricq21, Rangwal24, Cinar24, Cakmak24, Tasdemir25}. These datasets have revealed systematic variations in the properties of OCs by location, providing clearer pictures of star formation and the chemical evolution of the Galactic disc \citep{Sestito08, Frinchaboy13, Reddy20}.

The dynamic evolution of OCs is another key area of research, with studies indicating that most clusters dissolve within a few hundred million years due to tidal interactions, stellar feedback, and encounters with giant molecular clouds \citep{Portegies-Zwart10}. Despite their ephemeral nature, the kinematic imprints of dissolved clusters contribute significantly to the stellar population of the Galactic disk. Investigating these dynamics offers insights into the processes governing stellar migration and the distribution of stellar populations across the Milky Way \citep{Piskunov06, Bastian18}. Furthermore, the internal dynamics of OCs or globular clusters, such as mass segregation and binary star interactions, play critical roles in determining their evolutionary trajectories \citep[cf.][]{Spitzer87}. The role of OCs in tracing the Galactic structure is increasingly being recognized. By combining their astrophysical properties with their kinematic data, researchers have successfully mapped the spiral arms \citep{Castro-Ginard21, Hao21}, warp \citep{Vazquez10, Hidayat19}, and radial metallicity gradients of the Milky Way \citep{Netopil22, Spina22, Joshi24}. Such studies provide valuable constraints on the formation and evolution of the Galactic disk, revealing the interplay between star formation, chemical enrichment, and dynamical processes. 

Within the scope of the systematic OC surveys that we initiated approximately 15 years ago \citep[c.f.][]{Bilir10, Yontan15, Bostanci15, Ak16}, we combine {\it UBV} photometric data with the high-precision astrometric and radial velocity measurements provided by the {\it Gaia} satellite. This approach enables us to analyze the structural, astrophysical, kinematical, and dynamical orbital properties of OCs, thereby allowing us to investigate their birthplaces. In this study, we focus on Czernik 41 and NGC 1342 to expand our understanding of relatively unexamined OCs in the Galaxy. Due to their proximity to the Galactic plane, these OCs are significantly influenced by the presence of field stars, necessitating the application of statistical methods to mitigate field contamination. This investigation represents a continuation of a systematic research initiative on under-explored OCs, contributing to a broader understanding of Galactic structure and evolution \citep{Yontan19, Yontan21, Yontan22, Yontan23a, Banks20, Akbulut21, Koc22, Gokmen23}. By employing CCD {\it UBV} photometric data in conjunction with the most up-to-date datasets provided by the {\it Gaia} mission, we aim to derive the fundamental parameters of Czernik 41 and NGC 1342 with improved precision.

%---------------------------------------------------------------

Czernik~41 ($\alpha=19^{\rm h} 51^{\rm m} 01^{\rm s}\!\!.\,0$, $\delta= +25^{\rm o} 17^{\rm '} 24.0^{\rm''}$, $l=62^{\rm o}\!\!.\,0238$, $b=-00^{\rm o}\!\!.\,6899$) was identified by \cite{Czernik66} following examination of the Palomar Sky Atlas. It was classified by \citet{Ruprecht66} as Trumpler type IV3m, denoting a very loose, low-density cluster of a wide range in stellar brightness and having a moderate number of member stars, although with the additional comment of `doubtful'. \cite{Maciejewski07} included it in their {\it BV} CCD survey of 42 clusters, which derived estimates for the limiting and core radii through \cite{King62} radial density profile fits as well as extinction, ages, and distances via isochrone fitting. Values for the parameter estimates relevant to the current study are given in Table~\ref{tab:literature}, which also contains key estimates from later papers discussed in this paragraph. \cite{Kharchenko12} and \cite{Kharchenko13} analyzed 2MASS infrared photometry \citep{Skrutzie16}, combined with PPMXL \citep{Roeser10} positions and proper motions, of 642 OCs in the second Galactic quadrant. They derived estimates for spatial parameters (such as cluster radii), distances, ages, and reddening through profile and isochrone fitting. \cite{Dias14} presented a catalog of 724 clusters, providing the mean proper motions and membership probabilities of stars located in the direction of the clusters based on the second U.S. Naval Observatory CCD Astrograph Catalog (UCAC2, \citeauthor{Zacharias04} \citeyear{Zacharias04}). {\cite{Loktin17} included the ages, distances, and color excesses of 496 OCs in their catalog, which was based on their redeterminations from published photometry (including 2MASS). \cite{Cantat-Gaudin18} compiled a catalog of 1,229 open clusters, including Czernik~41, deriving fundamental parameters based on {\em Gaia} DR2 \citep{Gaia18} data combined with an application of the {\sc UPMASK} \citep{Krone-Martins14} technique to assign membership probabilities for stars in the sight lines of the clusters (identified from a literature review). Subsequent studies \citep{Liu19, Cantat-Gaudin20, Cantat-Gaudin_Anders20, Tarricq21} made use of the proper motions and supplemented them with new estimates of the cluster age, metallicity, distance, and radial velocity. For example, \cite{Dias21} presented an analysis of the {\it Gaia} DR2 data for 1,743 OCs (again deriving distance, age, and extinction), while \cite{Tarricq21} presented weighted mean velocities and galactic orbital parameters for 1,382 OCs based on {\it Gaia} DR2 radial velocities and literature data. Other studies, such as \cite{Medina21, Zhou21, Hao22} were interested in a cepheid belonging to Czernik~41.

NGC 1342 ($\alpha=03^{\rm h} 31^{\rm m} 34^{\rm s}.6$, $\delta= +37^{\rm o} 22^{\rm '} 48^{\rm''}$, $l=154^{\rm o}\!\!.9402$, $b=-15^{\rm o}\!\!.3463$) was discovered by William Herschel on 28 December 1799 \citep{Herchel02}, including it into his sixth list of objects (``coarsely scattered clusters of stars''), noting an about 15 arcminute diameter. The cluster is included in the ``Herschel 400'' list popular with amateur astronomers. Given the size of the cluster and its early discovery, it has been included in many studies and catalogs (although there are only a handful of comprehensive studies of the cluster). Table~\ref{tab:literature} lists parameter estimates, relevant to the current paper, from the literature. A few key studies are described further in this paragraph. The cluster was classified by \citet{Ruprecht66} as a Trumpler type III3p, which indicates a loosely structured, low-density cluster with a wide range of stellar brightness and a relatively small number of member stars. \cite{Pena94} estimated the age ($4 \times 10^{8}$ yr), distance ($530 \pm 118$ pc), and reddening ($E(b-y)=0.297\pm 0.112$) of the cluster based on their $uvby$-$\beta$ photometry, noting `patchy' extinction across NGC~1342. \cite{Reddy15} collected high-dispersion echelle spectra of three red giant members of NGC~1342, estimating their radial velocities and composition, along with age (of 0.5 Gyr) via isochrone fits using photometry from \cite{Hoag61}. \cite{Sarajedini95} mentions in a BAAS summary their {\em BVI} observations and subsequent color-magnitude diagrams of the cluster, but, unfortunately, do not appear to have further published their results. \cite{Cantat-Gaudin18} applied the data and methodology described above for Czernik~41 to NGC~1342, leading to estimates used in many later catalogs. Similarly, \cite{Dias21} included the cluster in their study.

In this study, we analyzed the OCs Czernik 41 and NGC 1342 by determining the membership probabilities of their stars, mean proper motions, and distances. Our analysis combines ground-based {\it UBV} photometric observations with high-precision astrometry and photometry data from the {\it Gaia} third data release ({\it Gaia} DR3) \citep{Gaia23}. We report the fundamental parameters, luminosity and mass functions, the dynamical properties of mass segregation, and the kinematic and Galactic orbital properties of both OCs.

% Table 1
\begin{table*}
\setlength{\tabcolsep}{3pt}
\renewcommand{\arraystretch}{0.8}
\small
  \caption{Fundamental parameters for Czernik 41 and NGC 1342 obtained in this study and compiled from the literature: Color excess ($E(B-V$)), distance ($d$), iron abundance ([Fe/H]), age ($t$), proper-motion components ($\langle\mu_{\alpha}\cos\delta\rangle$, $\langle\mu_{\delta}\rangle$), radial velocity ($V_{\rm R}$), and reference (Ref).}
  \begin{tabular}{ccccccccc}
    \hline
    \hline
    \multicolumn{9}{c}{Czernik 41}\\
        \hline
        \hline
$E(B-V)$                    & $d$                   & [Fe/H]            & $t$               &  $\langle\mu_{\alpha}\cos\delta\rangle$  &  $\langle\mu_{\delta}\rangle$ & $V_{\rm R}$       & Ref  & \\
(mag)                       & (pc)                  & (dex)             & (Myr)             & (mas yr$^{-1}$)       & (mas yr$^{-1}$)   & (km s$^{-1})$     &      \\
    \hline
$1.28_{-0.17}^{+0.14}$      & $1360_{-650}^{+400}$  & ---               & 500               & ---                   & ---               & ---               & (1) & \\
1.280                       & 1540                  & ---               & 300               & $-1.34$               & $-3.30$           & ---               & (2) & \\
 ---                        & ---                   & ---               & ---               & $-2.49\pm5.08$        & $-2.60\pm1.22$    & ---               & (3)& \\
1.336                       & 3032                  & ---               & 30                & $-1.250\pm0.659$      & $-2.863\pm0.676$  & ---               & (4) & \\
\hline \hline
---                         & $2511_{-17}^{+16}$    & ---               & ---               & $-2.932\pm0.138$      & $-6.164\pm0.128$  & ---               & (5) & \multirow{7}{*}{\rotatebox{90}{Gaia Era}} \\
---                         & $786\pm450$           & ---               & 100               & $-2.900\pm0.230$      & $-6.072\pm0.297$  & ---               & (6) & \\
1.326                       & 2543                  & ---               & 13                & $-2.932\pm0.138$      & $-6.164\pm0.128$  & ---               & (7) & \\
---                         & $2511_{-17}^{+16}$    & ---               & ---               & $-2.932\pm0.138$      & $-6.164\pm0.128$  & ---               & (8) & \\
$1.237\pm0.008$             & $2620\pm212$          & $0.002\pm0.108$   & $6\pm1$           &$-2.937\pm0.137$       & $-6.153\pm0.128$  &$-13.40\pm0.53$    & (9) & \\
---                         & 2451                  &  ---              & 14                & $-2.932\pm0.138$      & $-6.164\pm0.128$  &$-2.41\pm11.67$    & (10) & \\
$1.500\pm0.035$             & $2485\pm151$          & $0.07\pm0.09$     & $69\pm10$         &$-2.963\pm0.068$       & $-6.163\pm0.101$  & $2.41\pm1.92$     & (11) &\\

  \hline
  \hline
    \multicolumn{8}{c}{NGC 1342}\\
        \hline
        \hline
$E(B-V)$ & $d$ & [Fe/H] & $t$ &  $\langle\mu_{\alpha}\cos\delta\rangle$ &  $\langle\mu_{\delta}\rangle$ & $V_{\rm R}$ & Ref & \\
    (mag) &  (pc)  & (dex) & (Myr) & (mas yr$^{-1}$) & (mas yr$^{-1}$) & (km s$^{-1}$) & & \\
    \hline
0.42            & 610                   & $-0.163$          &  ---              & ---               &  ---                  & ---               & (12) & \\
---             & 550                   & $-0.44$           &  ---              & ---               &---                    & ---               & (13) & \\
---             & 550                   & $-0.38$           & 300               & ---               & ---                   & ---               & (14) & \\  
$0.401\pm0.151$ & $530\pm118$           &  ---              & 400               & ---               & ---                   & ---               & (15) & \\
---             & 665                   & ---               & ---               & $-0.12\pm0.27$    & $-3.10\pm0.37$        & ---               & (16) & \\
0.32            & 665                   & ---               & 160               & $0.13\pm0.80$     & $2.72\pm0.77$         & $-10.90\pm0.30$   & (17) & \\
---             & ---                   & ---               & ---               & ---               & ---                   & $-10.67\pm0.11$   & (18) & \\
0.331           & 665                   &  $-0.160$         & 400               & $-3.12$           & $-5.50$               & $-11.50\pm0.40$   & (02) & \\
---             & ---                   & ---               & ---               & $-2.99\pm2.45$    & $-1.82\pm1.76$        & ---               & (03) & \\
---             & $665\pm133$           &  $-0.14\pm0.05$   & 452               & $-1.15\pm0.087$   & $-2.80\pm0.87$        & $-10.67\pm0.11$   & (19) & \\
0.345           & 667                   & ---               & 452               & $-2.704\pm0.413$  & $-8.046\pm0.386$      & $-11.50\pm0.40$   & (04) & \\
    ---         & 665                   &  $-0.160$         & 160               & $0.13\pm0.80$     & $-2.72\pm0.077$       & $-10.90\pm0.10$   & (20) & \\
\hline \hline
---             & $653_{-0.7}^{+1.0}$   & ---               & ---               & $0.520 \pm0.233$  & $-1.604\pm0.195$      & ---               & (05) & \multirow{9}{*}{\rotatebox{90}{Gaia Era}}\\\\
9.95            & $653_{-0.7}^{+1.0}$   & ---               & ---               & $0.520\pm0.273$   & $-1.595\pm0.254$      & ---               & (21) & \\
---             & $668\pm25$            & ---               & 871               & $0.516\pm0.004$   & $-1.172\pm0.004$      & ---               & (06) & \\
0.331           & 665                   & $-0.205$          & 400               & $0.520\pm0.233$   & $-1.604\pm0.195$      & $-18.85$          & (22) & \\
0.229           & 686                   & ---               & 813               & $0.520\pm0.233$   & $-1.604\pm0.195$      & ---               & (07) & \\
---             & $653_{-40}^{+46}$     & ---               & ---               & $0.520\pm0.233$   & $-1.604\pm0.195$      & ---               & (08) & \\
$0.396\pm0.053$ & $626\pm19$            & $-0.115\pm0.030$  & $580\pm190$       & $0.523\pm0.264$   & $-1.598\pm0.242$      & $-9.41\pm1.06$    & (09) & \\
---             & 675                   & ---               & 850               & $0.520\pm0.233$   & $-1.604\pm0.195$      & $-10.35\pm0.19$   & (10) & \\
$0.270\pm0.043$ & $645\pm42$            & $-0.14\pm0.07$    & $1000\pm50$       & $0.419\pm0.041$   & $-1.662\pm 0.031$     & $-10.48\pm0.20$   & (11) & \\
  \hline
  \hline
    \end{tabular}
    \\
(1) \citet{Maciejewski07}, (2) \citet{Kharchenko13}, (3) \citet{Dias14}, (4) \citet{Loktin17}, (5) \citet{Cantat-Gaudin18}, (6) \citet{Liu19}, (7) \citet{Cantat-Gaudin20}, (8) \citet{Cantat-Gaudin_Anders20}, (9) \citet{Dias21}, (10) \citet{Tarricq21}, (11) This study, (12) \citet{Cameron85}, (13) \citet{Strobel-Poland91}, (14) \citet{Strobel91}, (15) \citet{Pena94}, (16) \citet{Loktin03}, (17) \citet{Kharchenko05}, (18) \citet{Mermilliod08}, (19) \citet{Reddy15}, (20) \citet{Conrad17}, (21) \citet{Soubiran18}, (22) \citet{Zhong20}
  \label{tab:literature}%
\end{table*}%

% FIGURE 1
\begin{figure*}
\centering
\includegraphics[scale=0.7, angle=0]{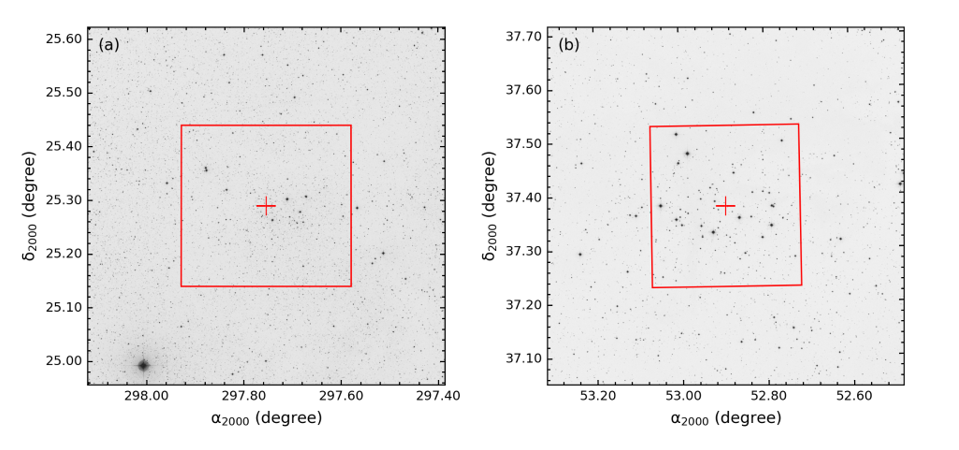}
\caption{Star fields for Czernik 41 (a) and NGC 1342 (b) in the equatorial coordinate system. The red boundaries indicate the fields observed with the T100 telescope. North and east correspond to the up and left directions, respectively.} 
\label{fig:charts}
\end {figure*}

%---------------------------------------------------------------

% Table 2
\begin{table}[t]
\setlength{\tabcolsep}{10pt}
\renewcommand{\arraystretch}{1}
  \centering
  \caption{Observational data for both OCs. The table includes the cluster names, observation dates (day-month-year), and the filters of the photometric {\it UBV} system used. For each filter, the rows specify the exposure times (in seconds) along with the corresponding number of exposures ($N$).}
  \medskip
    \begin{tabular}{ccc}
\hline
\hline  
  \multicolumn{3}{c}{Filter/Exp, Time (s) $\times N$}   \\
    \hline
$U$           & $B$          & $V$          \\
    \hline
    \hline
  \multicolumn{3}{c}{Czernik 41}   \\
  \multicolumn{3}{c}{Observation Date: 21-07-2020}\\
   \hline
   \hline
~~~80$\times$4 & ~~~8$\times$6 &~~2$\times$10 \\
1800$\times$2 & 100$\times$5 & ~20$\times$5 \\
1500$\times$3 & 600$\times$3 & 300$\times$4 \\
\hline
\hline
  \multicolumn{3}{c}{NGC 1342}   \\
  \multicolumn{3}{c}{Observation Date: 29-09-2019}\\
\hline
\hline
~~60$\times$3 & ~~~6$\times$5 &  ~~~4$\times$5 \\
 ~600$\times$3 & ~~60$\times$5 &~40$\times$5 \\
1800$\times$3 & 600$\times$3 & 400$\times$3 \\
    \hline
    \end{tabular}%
  \label{tab:exposures}%

\end{table}%
 
%-------------------------------------------------------------------------------------------
\section{Observations}
The CCD {\it UBV} photometric observations for the OCs Czernik 41 and NGC 1342 were conducted on 2020 July 21 and 2019 September 29, respectively. These observations utilized the 100-cm $f/10$ Ritchey-Chr\'etien telescope (T100) at the T\"UB\.ITAK National Observatory (TUG)\footnote{www.tug.tubitak.gov.tr} in Turkey. Imaging was performed with a back-illuminated 4k$\times$4k pixel CCD, providing an image scale of $0''\!\!.31$ pixel$^{-1}$ and a total field of view of $21'\!\!.5 \times 21'\!\!.5$. The CCD specifications included a gain of 0.55 e$^{-}$~ADU$^{-1}$ and a readout noise of 4.19 e$^{-}$. Identification charts for the OCs are shown in Figure~\ref{fig:charts}, and a detailed observation log is provided in Table~\ref{tab:exposures}. Shorter exposure times were used to obtain unsaturated images of brighter stars, while longer exposures facilitated accurate photometry for fainter stars. For photometric calibration, standard stars from \citet{Landolt09} were observed on the same nights as the clusters. A total of 14 Landolt fields were observed across an airmass range of 1.233 to 2.061, comprising 97 standard stars (Table~\ref{tab:standard_stars}). Technical information on the observing conditions of the TUG site can be found in \citet{Ak24}. The raw science images underwent standard bias subtraction and flat fielding using IRAF\footnote{IRAF was distributed by the National Optical Astronomy Observatories}. The instrumental magnitudes of the \citet{Landolt09} standard stars were derived using IRAF's aperture photometry tools. Multiple linear regression analysis of these magnitudes yielded photometric extinction and transformation coefficients for the two observing nights, which are presented in Table~\ref{tab:coefficients}. For cluster images, astrometric corrections were performed using PyRAF\footnote{PyRAF is a product of the Space Telescope Science Institute, operated by AURA for NASA} and astrometry.net\footnote{http://astrometry.net}. Instrumental magnitudes of objects within the cluster regions were measured using Source Extractor and PSF Extractor routines \citep{Bertin96}, with aperture corrections subsequently applied. Finally, instrumental magnitudes were transformed to standard {\em UBV} magnitudes based on the transformation equations provided by \citet{Janes13}

% Table 3
\begin{table}[t]
\setlength{\tabcolsep}{5pt}
\renewcommand{\arraystretch}{0.8}
  \centering
  \caption{Selected \citet{Landolt09} standard star fields. The columns denote the observation date (day-month-year), star field name from Landolt, the number of standard stars ($N_{\rm st}$) observed in a given field, the number of observations for each field ($N_{\rm obs}$), and the airmass range the fields were observed over.}
\medskip
    \begin{tabular}{lcc}
    \hline 
    \multicolumn{3}{c}{Observation Date: 29.09.2019}\\
     \multicolumn{3}{c}{Air Mass: 1.244--1.864}\\
    \hline
    \hline
Star Field	& $N_{\rm st}$ & $N_{\rm obs}$\\
\hline
SA93      &  4 	  & 1	\\
SA106     &  2	  & 1	\\
SA107	  &  7	  & 1	\\
SA108     &  2	  & 2	\\
SA110SF2  & 10	  & 1   \\
SA111	  &  5	  & 1	\\
SA112	  &  6	  & 1	\\
SA113	  & 15	  & 1	\\
SA114	  &  5	  & 1	\\
\hline	 
\multicolumn{3}{c}{Observation Date: 21.07.2020}\\
     \multicolumn{3}{c}{Air Mass: 1.233--2.061}\\
     \hline
\hline
SA92SF2   &  15   & 1	\\
SA93      &  4	  & 1	\\
SA96	  &  2 	  & 1	\\
SA98      &  19   & 1	\\
SA108	  &  2	  & 1   \\
SA109     &  2	  & 1	\\
SA109SF2  &  3	  & 1	\\
SA110SF2  &  10	  & 1	\\
SA111	  &  5	  & 1	\\
SA112	  &  6	  & 1	\\	
SA113	  &  15	  & 1	\\
SA114	  &  5	  & 1	\\	       
    \hline \end{tabular}%
  \label{tab:standard_stars}%
\end{table}%

% Table 4
\begin{table*}
\renewcommand{\arraystretch}{0.8}
  \centering
  \caption{Transformation and extinction coefficients obtained for the two observation nights: $k$ and $k'$ are the primary and secondary extinction coefficients, while $\alpha$ and $C$ are the transformation coefficients and zero points, respectively. Dates are day-month-year.}   
  \medskip
    \begin{tabular}{lcccc}
    \hline
    & \multicolumn{4}{c}{Observation Date: 21.07.2020}\\
    \hline
Filter/color & $k$   & $k'$             & $\alpha$          & $C$             \\
    \hline
$U$     & $0.533 \pm 0.124$   & $-0.322 \pm 0.098$   & ---                   & ---                 \\
$B$     & $0.307 \pm 0.052$   & $-0.106 \pm 0.040$   & $1.057 \pm 0.064$   & $2.089 \pm 0.079$ \\
$V$     & $0.119 \pm 0.018$   & ---                  & ---                   & ---                 \\
$U-B$   &  ---                & ---                  & $1.304 \pm 0.157$   & $4.421 \pm 0.182$ \\
$B-V$   &  ---                & ---                  & $0.071 \pm 0.009$   & $2.202 \pm 0.025$ \\
\hline
    & \multicolumn{4}{c}{Observation Date: 29.09.2019}\\
\hline
$U$     &  $0.471 \pm 0.074$ & $0.101 \pm  0.114$        & ---                   & ---                 \\ 
$B$     &  $0.298 \pm 0.061$ & $-0.004 \pm 0.072$      & $0.900 \pm 0.107$   & $1.783 \pm 0.091$ \\
$V$     &  $0.175 \pm 0.020$ & ---                     & ---                   & ---                 \\
$U-B$   &  ---                 & ---                   & $0.684 \pm 0.166$   & $4.176 \pm 0.110$ \\
$B-V$   &  ---                 & ---                   & $0.062 \pm 0.008$   & $1.834 \pm 0.032$ \\
\hline
    \end{tabular}%
  \label{tab:coefficients}%
\end{table*}%

\begin{align}\label{eq:01}
v =& V + \alpha_{\rm bv}(B-V) + k_{\rm v}X + C_{\rm bv}, \\ \nonumber
b =& V + \alpha_{\rm b}(B-V) + k_{\rm b}X + k'_{\rm b}X(B-V) + C_{\rm b}, \\ \nonumber
u =& V + (B-V) + \alpha_{\rm ub}(U-B) + k_{\rm u}X + k'_{\rm u}X(U-B) \\ \nonumber
&+ C_{\rm ub}
\end{align}

In the above equations, $U$, $B$, and $V$ are the magnitudes in the standard photometric system. $u$, $b$, and $v$ indicate the instrumental magnitudes. $X$ is the airmass. $k$ and $k'$ represent primary and secondary extinction coefficients. $\alpha$ and  $C$ are the transformation coefficients to the standard system and zero points, respectively.

%---------------------------------------------------------------

\section{Data analysis}
\subsection{Photometric Data}
Photometric catalogs were constructed for stars in the directions of Czernik 41 and NGC 1342. These include the stars' equatorial coordinates ($\alpha, \delta$), $V$ magnitudes, and $U-B$ and $B-V$ color indices, along with associated uncertainties. The relevant catalog for Czernik 41 contains data for 2,527 stars with apparent magnitudes in the range $9<V~{\rm (mag)}<22.5$, while the catalog for NGC 1342 includes 906 stars with magnitudes spanning $8<V~{\rm (mag)}<22.5$. To further characterize these clusters, catalogs based on {\it Gaia} DR3 data \citep{Gaia23} were compiled to investigate the clusters' positions in the {\it Gaia} photometric color spaces, compute membership probabilities, and compare cluster distances derived from photometric and trigonometric parallaxes. The {\it Gaia} catalogs include stars within a $40' \times 40'$ radius of the clusters' central coordinates, as adopted from \citet{Cantat-Gaudin20}. These catalogs provide photometric data ($G$, $G_{\rm BP}$, $G_{\rm RP}$), proper-motion (PM) components ($\mu_{\alpha}\cos\delta$, $\mu_\delta$), trigonometric parallaxes ($\varpi$), radial velocities ($V_{\rm R}$), and their corresponding uncertainties, which are listed in Table~\ref{tab:all_cat}. The {\it Gaia} DR3 catalog for Czernik 41 includes 217,879 stars within the magnitude range $5<G~{\rm (mag)}<22.5$, while the catalog for NGC 1342 contains 20,820 stars with magnitudes between $7<G~{\rm (mag)}<22$. Figure~\ref{fig:charts} (on page~\pageref{fig:charts}) displays the $40' \times 40'$ cluster fields in the {\it Gaia} system, with the coverage of the {\it UBV} data highlighted by red boundaries. Membership probabilities ($P$) for the cluster stars were derived from the astrometric data in the {\it Gaia} DR3 catalogs and incorporated into the final {\it Gaia} cluster catalogs. To ensure consistent analysis, stars in the {\it Gaia} catalogs were cross-matched with those in the {\it UBV} catalogs using a maximum separation limit of $8''$. This allowed the determination of membership probabilities for the stars to be included in the cluster analysis based on the {\it UBV} system.

To determine the precise fundamental parameters of the clusters, the photometric measurements were examined prior to the analysis to establish the faint magnitude limits for the stars in the cluster regions. These limits, known as photometric completeness limits, indicate the sensitivity of photometric measurements based on the exposure times used in the observations. The same method was applied to define the photometric completeness limits in both the {\it UBV} and {\it Gaia} systems. Histograms of the star count as a function of apparent magnitudes ($V$ and $G$) were created for the stars present in the two catalogs. The magnitude at which the number of stars reaches its maximum in these distributions was adopted as the completeness limit for $V$ and $G$ magnitudes. The histograms for the $V$ and $G$ apparent magnitudes of stars in the cluster fields are presented in Figure~\ref{fig:histograms}. Panels (a) and (c) show the distributions of star counts as a function of $V$ magnitudes for Czernik 41 and NGC 1342, respectively. In the case of Czernik 41 (Figure~\ref{fig:histograms}a), the number of stars increases toward fainter magnitudes until $V = 20$ mag, beyond which there is a sharp decline. A similar trend is observed for NGC 1342 (Figure~\ref{fig:histograms}c), where the decrease in star counts begins at $V = 19$ mag. However, the star counts at $V = 19$ and $20$ mag are found to be comparable. Consequently, the photometric completeness limit in the {\it UBV} system was adopted as $V = 20$ mag for both clusters. Similarly, panels (b) and (d) of Figure~\ref{fig:histograms} depict the distributions of star counts in $G$ magnitudes with 0.5 mag bins for Czernik 41 and NGC 1342, respectively. The magnitudes at which the star counts reach their highest values are identified as $G = 21$ mag for Czernik 41 (Figure~\ref{fig:histograms}b) and $G = 21$ mag for NGC 1342 (Figure~\ref{fig:histograms}d).

To check the photometric completeness limits, we generated synthetic star populations for the regions of both OCs using the Besan\c con Galaxy Model\footnote{\url{https://model.obs-besancon.fr/}} \citep{Robin03, Robin12}. This model incorporates the three-dimensional interstellar medium extinction map developed by \citet{Marshall06}, enabling a realistic representation of extinction along different lines of sight. The simulations were performed within the magnitude intervals of $10 < V~{\rm (mag)} \leq 23$ and $9 < G~{\rm (mag)} \leq 23$, consistent with the observational star counts and taking into account the three-dimensional extinction values from \citet{Marshall06}. Since the {\it UBV} and {\it Gaia} photometric data sets span different areas of the sky, solid angles of 0.1225 and 0.4450 square degrees, respectively, were adopted in the model.

The resulting synthetic star counts are presented alongside the observational data in Figure~\ref{fig:histograms}. The magnitude intervals in which the number of observed stars exceeds that of the model, particularly at the bright end of the $V$ and $G$ bands, represent the dominant magnitude ranges of the OCs (i.e., completeness is $\sim$ 100\%). Conversely, the apparent magnitudes at which the model star counts begin to surpass the observed counts correspond to the photometric completeness limits.  Therefore, we conclude that there is no loss of stars within the previously defined photometric completeness limits in this study. 
 
The apparent magnitude and color index uncertainties of the stars detected in the cluster fields, derived from the {\it UBV} and {\it Gaia} photometric systems, are listed in Table~\ref{tab:phiotometric_errors} as a function of apparent magnitudes in $V$ and $G$. The magnitude and color uncertainties were determined based on the statistical uncertainties in instrumental magnitudes and the propagation of errors in the corresponding bands. For Czernik 41, considering the photometric completeness limit in the $V$ band, the mean uncertainties in apparent magnitudes within the range $9<V~{\rm (mag)}\leq20$ were found to be no greater than 0.05 mag for $V$, and 0.16 mag for the $U-B$ and $B-V$ color indices. For NGC 1342, in the range $8<V~{\rm (mag)}\leq20$, the mean uncertainties were even smaller, not exceeding 0.025 mag for $V$, and 0.05 mag for the $U-B$ and $B-V$ color indices. An examination of the mean uncertainties in {\it Gaia} photometry showed that for $5<G~{\rm (mag)}\leq21$ in Czernik 41 and $7<G~{\rm (mag)}\leq21$ in NGC 1342, the mean uncertainties in $G$ magnitudes were no greater than 0.007 mag, while the uncertainties in the $ G_{\rm BP}-G_{\rm RP}$ color indices did not exceed 0.16 mag.

% FIGURE 2
\begin{figure*}
\centering
\includegraphics[width=\textwidth]{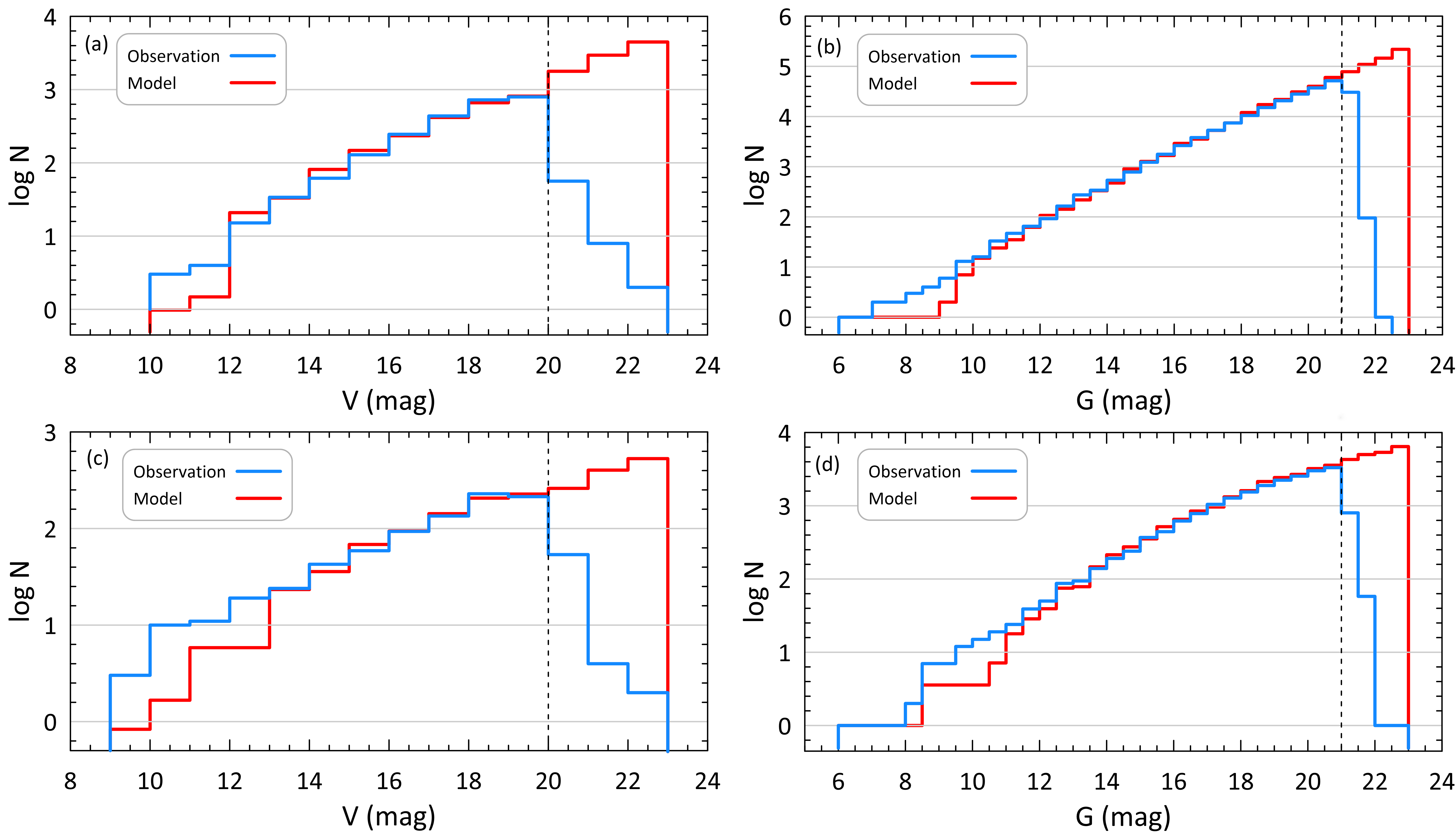}
\caption{Histograms of stars in Czernik 41 (panels a and b) and NGC 1342 (panels c and d) as a function of $V$ and $G$ magnitudes. The black dashed lines show the photometric completeness limits for each band. The blue and red color bins represent the observed and the Besan\c con Galaxy model values, respectively.} 
\label{fig:histograms}
\end {figure*} 

% TABLE 5
% otherwise.  
\begin{sidewaystable}
\setlength{\tabcolsep}{1.3pt}
\renewcommand{\arraystretch}{0.7}
\footnotesize
  \centering
 \caption{The catalogs for Czernik 41 and NGC 1342. The complete table can be found electronically.}
    \begin{tabular}{rrrrcrcrrrrr}
\hline
\multicolumn{11}{c}{Czernik 41}\\
\hline
ID	 & RA           &	DEC	        &      $V$	    &	$U-B$      & $B-V$	      &	$G$	          & 	$G_{\rm BP}-G_{\rm RP}$	 & 	$\mu_{\alpha}\cos\delta$ & 	$\mu_{\delta}$ & 	$\varpi$	& $P$ \\

	 & (hh:mm:ss.ss)           &	(dd:mm:ss.ss)	&      (mag)	    &	(mag)      & (mag)	      &	(mag)	          & 	(mag)	 & 	(mas yr$^{-1}$) & 	(mas yr$^{-1}$) & 	(mas)	&  \\
\hline
0001 & 19:50:14.48 & +25:08:58.48 & 15.114(0.014) & 0.695(0.041) & 1.098(0.020)  & 14.931(0.003) & 1.241(0.005) & $-5.555$(0.022)  & 2.968(0.014)    & 1.662(0.023)   & 0.000\\
0002 & 19:50:14.54 & +25:10:44.54 & 18.013(0.011) & 0.406(0.054) & 1.378(0.019) & 17.504(0.003) & 1.896(0.011) & $-0.043$(0.078)  & 1.373(0.052)    & 0.460(0.079)    & 0.090 \\
0003 & 19:50:14.66 & +25:12:55.64 & 16.593(0.042) & ---          & 1.489(0.067) & 16.063(0.003) & 1.863(0.006) & 6.217(0.037)     & 4.404(0.024)    & 0.835(0.037)   & 0.448 \\
0004 & 19:50:15.03 & +25:12:03.83 & 19.067(0.027) & ---          & 1.620(0.053)  & 18.382(0.003) & 2.087(0.027) & $-11.576$(0.134) & $-0.643$(0.088) & 0.504(0.134)   & 0.000 \\
...  & ...         & ...	      & ...           & ...          & ...          & ...           & ...          & ...              & ...             & ...            & ... \\
...  & ...         & ...	      & ...           & ...          & ...          & ...           & ...          & ...              & ...             & ...            & ... \\
...  & ...         & ...	      & ...           & ...          & ...          & ...           & ...          & ...              & ...             & ...            & ... \\
2524 & 19:51:44.80 &+25:23:42.50  &18.562(0.016)  & ---          &1.650(0.031)   &18.219(0.003)  &2.019(0.027)  &1.117(0.208)      & $-9.733$(0.116) &$-4.356$(0.078) & 0.251\\
2525 & 19:51:44.88 &+25:23:13.43  &16.053(0.027)  & ---          &1.296(0.041)  &15.712(0.003)  &1.531(0.005)  &0.734(0.313)      &7.504(0.029)     &8.315(0.021)    & 0.000\\
2526 & 19:51:45.13 &+25:25:01.08  &18.223(0.013)  & ---          &1.720(0.026)   &17.399(0.003)  &2.268(0.010)   &1.602(0.553)      &$-4.235$(0.071)  &$-1.583$(0.052) & 0.298 \\
2527 & 19:51:45.16 &+25:26:53.32  &17.462(0.007)  & ---          &1.745(0.015)  &16.611(0.003)  &2.291(0.007)  &0.701(0.212)      &$-5.339$(0.043)  &$-2.873$(0.033) & 0.235 \\
\hline
\multicolumn{11}{c}{NGC 1342}\\
\hline
ID	 & RA           &	DEC	        &      $V$	    &	$U-B$      & $B-V$	      &	$G$	          & 	$G_{\rm BP}-G_{\rm RP}$	 & 	$\mu_{\alpha}\cos\delta$ & 	$\mu_{\delta}$ & 	$\varpi$	& $P$ \\
	 & (hh:mm:ss.ss)           &	(dd:mm:ss.ss)	&      (mag)	    &	(mag)      & (mag)	      &	(mag)	          & 	(mag)	 & 	(mas yr$^{-1}$) & 	(mas yr$^{-1}$) & 	(mas)	&  \\
\hline
001 & 03:30:46.83  & +37:27:21.01 & 19.013(0.023)  & ---          & 1.879(0.058)  & 18.543(0.007) & 2.413(0.046)  &$-6.62$(0.323)  & 4.979(0.420)    &  0.032(0.364) & 0.162 \\
002 & 03:30:46.93  & +37:28:24.89 & 18.882(0.013)  & ---          & 1.060(0.021)   & 18.538(0.003) & 1.371(0.036)  &$-2.166$(0.188) & 2.457(0.235)   &  0.077(0.220)  & 0.209 \\
003 & 03:30:47.03  & +37:23:37.49 & 15.557(0.002)  & 0.618(0.006) & 1.114(0.003)  & 15.096(0.003) & 1.538(0.005)  &$-5.353$(0.036) & 6.682(0.041)   &  0.642(0.032) & 0.000 \\
004 & 03:30:47.15  & +37:27:57.87 & 17.085(0.004)  & 0.497(0.014) & 1.013(0.006)  & 16.774(0.003) & 1.318(0.008)  & 0.218(0.064)   & 2.006(0.079)   &  0.285(0.073) & 0.000 \\
... & ...          & ...	      & ...            & ...          & ...           & ...           & ...          & ...             & ...            & ...           & ...  \\
... & ...          & ...	      & ...            & ...          & ...           & ...           & ...          & ...             & ...            & ...           & ...  \\
... & ...          & ...	      & ...            & ...          & ...           & ...           & ...          & ...             & ...            & ...           & ...  \\
903 & 03:32:29.39  & +37:24:00.47 & 15.988(0.002)  & 1.039(0.011) & 1.367(0.004)  & 15.43(0.003)  & 1.719(0.006) & 1.099(0.035)    & 16.948(0.046)  & 0.727(0.041)  & 0.000 \\
904 & 03:32:29.82  & +37:30:03.34 & 18.117(0.008)  & ---          & 1.778(0.017)  & 17.185(0.003) & 2.208(0.016) &$-2.082$(0.088)  &$-0.059$(0.105) & 1.621(0.095)  & 0.310 \\
905 & 03:32:30.05  & +37:24:07.67 & 19.024(0.014)  & ---          & 1.026(0.023)  & 18.600(0.003)   & 1.443(0.049) &$-0.945$(0.204)   &$-0.892$(0.261)& 0.430(0.239)   & 0.000 \\
906 & 03:32:30.30  & +37:32:04.81 & 19.448(0.023)  & ---          & 2.101(0.063)  & 19.449(0.004) & ---          & 1.191(0.973)    &  11.215(0.856) & 3.434(0.522)  & 0.000 \\
\hline
    \end{tabular}
      \label{tab:all_cat}%
\end{sidewaystable}

% Table 6
\begin{table}
  \centering
\setlength{\tabcolsep}{2.85pt}
\renewcommand{\arraystretch}{1}
  \caption{Mean internal photometric uncertainties for each magnitude bin in $V$ and $G$ magnitudes.}
    \begin{tabular}{ccccccccc}
      \hline
    \multicolumn{5}{c}{Czernik 41} & \multicolumn{4}{c}{NGC 1342} \\
    \hline
  $V$ & $N$ & $\sigma_{\rm V}$ & $\sigma_{\rm U-B}$ & $\sigma_{\rm B-V}$ & $N$ & $\sigma_{\rm V}$ & $\sigma_{\rm U-B}$ & $\sigma_{\rm B-V}$\\
  \hline
[08, 12] &  10 & 0.002 & 0.003 & 0.003 &  26 & 0.001 & 0.003 & 0.002 \\
(12, 14] &  49 & 0.007 & 0.014 & 0.011 &  43 & 0.004 & 0.009 & 0.006 \\
(14, 15] &  62 & 0.011 & 0.036 & 0.017 &  43 & 0.004 & 0.011 & 0.006 \\
(15, 16] & 129 & 0.016 & 0.070 & 0.031 &  59 & 0.002 & 0.013 & 0.005 \\
(16, 17] & 249 & 0.019 & 0.106 & 0.023 &  94 & 0.003 & 0.014 & 0.013 \\
(17, 18] & 436 & 0.023 & 0.137 & 0.046 & 135 & 0.006 & 0.025 & 0.026 \\
(18, 19] & 725 & 0.032 & 0.155 & 0.058 & 231 & 0.012 & 0.037 & 0.045 \\
(19, 20] & 801 & 0.044 & 0.162 & 0.125 & 215 & 0.024 & 0.052 & 0.057 \\
(20, 22] &  66 & 0.057 & 0.193 & 0.168 &  60 & 0.034 & 0.069 & 0.065 \\
  \hline
  $G$ & $N$ & $\sigma_{\rm G}$ &   \multicolumn{2}{c}{$\sigma_{G_{\rm BP}-G_{\rm RP}}$} &$N$ & $\sigma_{\rm G}$ & \multicolumn{2}{c}{$\sigma_{G_{\rm BP}-G_{\rm RP}}$}\\
  \hline
  [06, 12] &    6 & 0.003 & \multicolumn{2}{c}{0.006} & ~34 & 0.003 & \multicolumn{2}{c}{0.005} \\
  (12, 14] &   46 & 0.003 & \multicolumn{2}{c}{0.006} & 104 & 0.003 & \multicolumn{2}{c}{0.005} \\
  (14, 15] &   84 & 0.003 & \multicolumn{2}{c}{0.007} & ~82 & 0.003 & \multicolumn{2}{c}{0.005} \\
  (15, 16] &  228 & 0.003 & \multicolumn{2}{c}{0.012} & 147 & 0.003 & \multicolumn{2}{c}{0.006} \\
  (16, 17] &  566 & 0.003 & \multicolumn{2}{c}{0.014} & 212 & 0.003 & \multicolumn{2}{c}{0.009} \\
  (17, 18] & 1334 & 0.003 & \multicolumn{2}{c}{0.025} & 354 & 0.003 & \multicolumn{2}{c}{0.019} \\
  (18, 19] & 2703 & 0.003 & \multicolumn{2}{c}{0.052} & 581 & 0.003 & \multicolumn{2}{c}{0.042} \\
  (19, 20] & 5121 & 0.004 & \multicolumn{2}{c}{0.094} & 770 & 0.004 & \multicolumn{2}{c}{0.082} \\
  (20, 23] & 5881 & 0.006 & \multicolumn{2}{c}{0.154} & 475 & 0.007 & \multicolumn{2}{c}{0.157} \\
      \hline
    \end{tabular}%
  \label{tab:phiotometric_errors}%
\end{table}%

%-------------------------------------------------------------------------------------

\subsection{The Radial Density Profile}
{\it Gaia} photometric data were utilized in the radial density profile (RDP) analysis of both OCs. The photometric $G$ magnitudes of stars were considered to calculate the structural parameters from the RDP. Stellar number density profiles were subsequently generated. The central coordinates of the OCs in the equatorial coordinate system were taken from the catalog of \citet{Cantat-Gaudin20}. In the analysis, stars within concentric annuli of varying radii, defined from the cluster center, were initially counted. The stellar number density for each annulus was then computed by dividing the number of stars by the area of the annulus (i.e., $\rho(r_i) = N_i / A_i$; $N$ and $A$ represent the number of stars in the defined annulus and the area of the annulus, respectively, with the index $i$ denoting the corresponding annulus). RDPs were constructed as a function of distance from the cluster center and compared with the \citet{King62} model which is described as $\rho(r)=f_{\rm bg}+ [f_{\rm 0}/(1+(r/r_{\rm c})^2)]$ (where $r$, $f_{\rm 0}$,  $r_{\rm c}$, and  $f_{\rm bg}$ represent the radius from the cluster centre, the central density, core radius, and background density, respectively). To derive the best-fitting model, a $\chi^2$ minimization method was applied to the RDPs. The final parameter estimates were determined as the values that minimized the computed $\chi^2$. The resulting RDPs for Czernik 41 and NGC 1342, along with the best-fit  \citet{King62} models, are presented in Figure~\ref{fig:king}. The structural parameters derived from these models follow as the central stellar density ($f_{\rm 0}$), core radius ($r_{\rm c}$), and background stellar density ($f_{\rm bg}$) of both OCs were determined as follows: for Czernik 41, $f_{\rm 0} = 7.622 \pm 0.750$ stars arcmin$^{-2}$, $r_{\rm c} = 1.728 \pm 0.266$ arcmin, and $f_{\rm bg} = 4.481 \pm 0.171$ stars arcmin$^{-2}$, while for NGC 1342, these values were found to be $f_{\rm 0} = 1.733 \pm 0.240$ stars arcmin$^{-2}$, $r_{\rm c} = 2.155 \pm 0.568$ arcmin, and $f_{\rm bg} = 1.766 \pm 0.090$ stars arcmin$^{-2}$. As shown in Figure~\ref{fig:king}, the background density, represented by the horizontal grey band, intersects the RDP model at the limiting radius for two clusters. Consequently, the limiting radius for both OCs was assumed as $r_{\rm lim} = 11'$. In the subsequent analyses, only stars within these limiting radii were considered, while stars located beyond the limiting radius, where field stars dominate, were excluded from the following statistical analysis of the clusters' fundamental parameters.

The concentration parameter of a cluster is an indicator for understanding its internal structure and evolutionary process, reflecting the combined effects of formation mechanisms, dynamical interactions, and environmental influences within the Galaxy. In general, the concentration parameter is defined using the limiting radius ($r_{\rm lim}$) instead of the tidal radius ($r_{\rm tidal}$) as in the classical King model definition, $C=\log(r_{\rm tidal}/r_{\rm core})$. We acknowledge that these two radii are conceptually distinct; however, in practice, they are often found to be closely related, particularly in systems where the tidal cutoff is not sharply defined or cannot be robustly constrained observationally. Several previous studies have adopted a similar approach, using $r_{\rm lim}$ as a proxy for $r_{\rm tidal}$ due to observational limitations and the practical difficulties in distinguishing between the two radii in low-surface-brightness systems. For example, \citet{Hill06} and \citet{Munoz10} employed limiting radii derived from surface brightness or star count profiles as effective substitutes for the formal tidal radius in their structural analyses of dwarf galaxies. Likewise, \citet{Miocchi13} discussed that for globular clusters, particularly those with extended halos, the limiting radius obtained from observational profiles can serve as a reasonable approximation of the tidal radius.

In this study, $C$ values were determined as $0.804 \pm 0.062$ and $0.708 \pm 0.102$ for Czernik 41 and NGC 1342, respectively. The derived concentration parameters for both OCs are relatively low, indicating a mild stellar density gradient between the core and the outer regions. This reflects a weak central concentration, suggesting that the clusters exhibit structural characteristics typical of dynamically unevolved or loosely bound systems. The effect of mass segregation in these clusters is expected to be weaker as the migration of more massive stars toward the center has not yet fully progressed. Consequently, clusters with low concentration parameters may be dynamically young or may have lost their central density over time due to external influences \citep{Elson87}.

% FIGURE 3
\begin{figure}[t]
\centering
\includegraphics[width=\columnwidth]{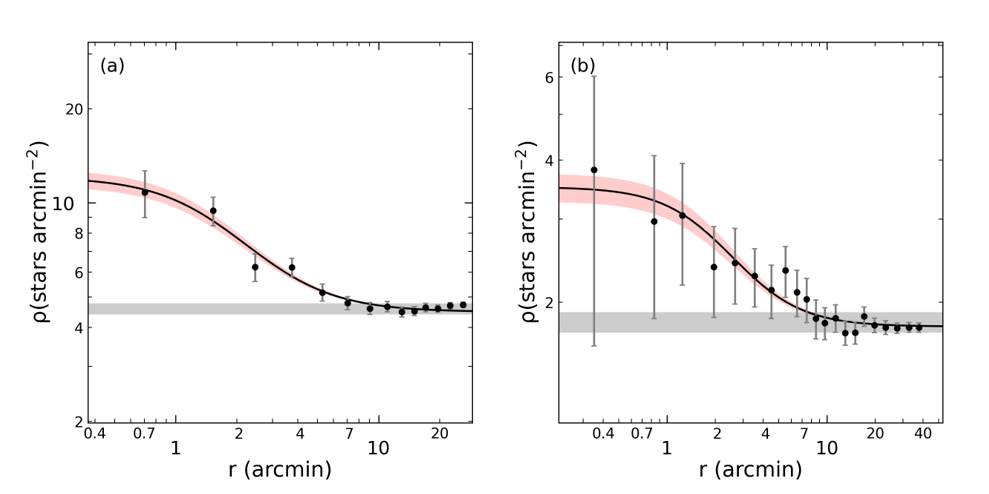}
\caption{King profiles of Czernik 41 (a) and NGC 1342 (b). The black solid curves represent the best-fit models, while the shaded red and grey regions indicate the confidence intervals of the model and the background stellar density, respectively. The dashed vertical lines denote the limit radii of the OCs.} 
\label{fig:king}
\end {figure} 

%----------------------------------------------------------------------------------

\subsection{CMDs and Cluster Membership}
\label{section:cmds}
The membership probabilities of stars within the fields of the OCs studied in this study were calculated using the {\sc UPMASK} method \citep{Krone-Martins14}. The {\it Gaia} DR3 \citep{Gaia23} catalogs created for Czernik 41 and NGC 1342 were utilized, employing a five-dimensional parameter space comprising positional and astrometric data to compute the membership probabilities of stars. Specifically, the input parameters for the program included the equatorial coordinates ($\alpha$, $\delta$), trigonometric parallaxes ($\varpi$), PM components ($\mu_{\alpha}\cos\delta$, $\mu_\delta$), and their associated uncertainties for each star in the catalogs. {\sc UPMASK} was run with $k$-means values ranging from 6 to 30 and 25 iterations. The optimal $k$-means values were selected based on the configurations that best represented the cluster structure through the computed membership probabilities. Consequently, the $k$-means values for Czernik 41 and NGC 1342 were set to 8 and 25, respectively. 

Color-magnitude diagrams (CMDs) serve as a fundamental tool for revealing the relationship between magnitude and color, thereby enabling the determination of the age, mass, and evolutionary status of member stars in OCs. Moreover, CMDs play a crucial role in testing stellar evolution models and deriving cluster parameter estimates with high precision. As discussed above, in this study stars located in the direction of the clusters were analyzed using {\it Gaia} DR3 astrometric data, leading to high-probability members being selected via the {\sc UPMASK} algorithm. The morphological features observed in the CMDs are essential for deriving key astrophysical parameters such as color excess, distance, and age of the OCs. Therefore, isochrone fitting based on a sample composed exclusively of stars with reliable membership probabilities allows for a more accurate determination of these parameters. We note that unresolved binary systems may cause a noticeable broadening in the CMD, complicating the interpretation of the stellar distribution. To better assess this effect and to improve the accuracy of the derived cluster parameters, high-probability member stars were plotted on CMDs constructed using both {\it UBV} and {\it Gaia} photometric magnitude and colors. Based on the {\it Gaia} data, stars within the cluster direction satisfying the conditions of limiting magnitude ($G\leq21$), membership probability ($P\geq 0.5$), and limiting radius ($r_{\rm lim}=11'$) were identified and shown in the $G\times (G_{\rm BP}-G_{\rm RP}$) CMDs. Under these criteria, 382 stars were identified as members of Czernik 41, and 111 stars as members of NGC 1342. Stars selected based on the specified criteria are presented in Figure~\ref{fig:cmds} on the $G\times (G_{\rm BP}-G_{\rm RP}$) CMDs as a function of their membership probabilities.

In the {\em UBV} photometric system, a consistent method was employed to ensure uniformity in the determination of cluster structure. Following the \textit{Gaia}-based approach, the lower boundary of the cluster main sequence was again adjusted to account for photometric dispersion and unresolved binaries, but using a different reference. Instead of theoretical isochrones, the observational ZAMS from \citet{Sung13} was applied to the $V \times (B-V)$ CMDs. The ZAMS curve was shifted empirically in both $V$ and $B-V$ directions to encapsulate the combined effects of interstellar extinction, unresolved binary presence, and photometric uncertainties. This refinement allowed for the selection of probable main-sequence members and bright stars near the cluster turn-off point and giant regions with $P \geq 0.5$. By applying these photometric criteria, the member selection process was harmonized across the two photometric systems, improving the reliability of the cluster analysis. In conclusion, for the {\em UBV} system, the most probable members of the clusters were defined as stars within $r_{\rm lim} \leq 11'$, with $V \leq 20$ mag, $P \geq 0.5$, and located within the ZAMS-constrained regions. The number of these stars is 172 for Czernik 41 and 89 for NGC 1342. These stars were subsequently utilized in determining the clusters' {\em UBV}-based parameter estimates. The distributions of stars identified in the $UBV$ system in the $V \times (B-V)$ diagrams, along with the ZAMS-constrained bands, are shown in Figure~\ref{fig:cmds}a (Czernik 41) and Figure~\ref{fig:cmds}c (NGC 1342), where the continuous and dashed blue curves represent the ZAMS boundaries. Additionally, the $G \times (G_{\rm BP} - G_{\rm RP})$ diagrams for stars detected in the {\it Gaia} data are displayed in Figure~\ref{fig:cmds}b (Czernik 41) and Figure~\ref{fig:cmds}d (NGC 1342). In both the {\em UBV}- and {\it Gaia}-based diagrams, stars marked according to their color scale represent the most probable cluster members, which satisfy $r_{\rm lim} \leq 11'$ and $P \geq 0.5$.

% FIGURE 4
\begin{figure*}[t]
\centering
\includegraphics[scale=0.50, angle=0]{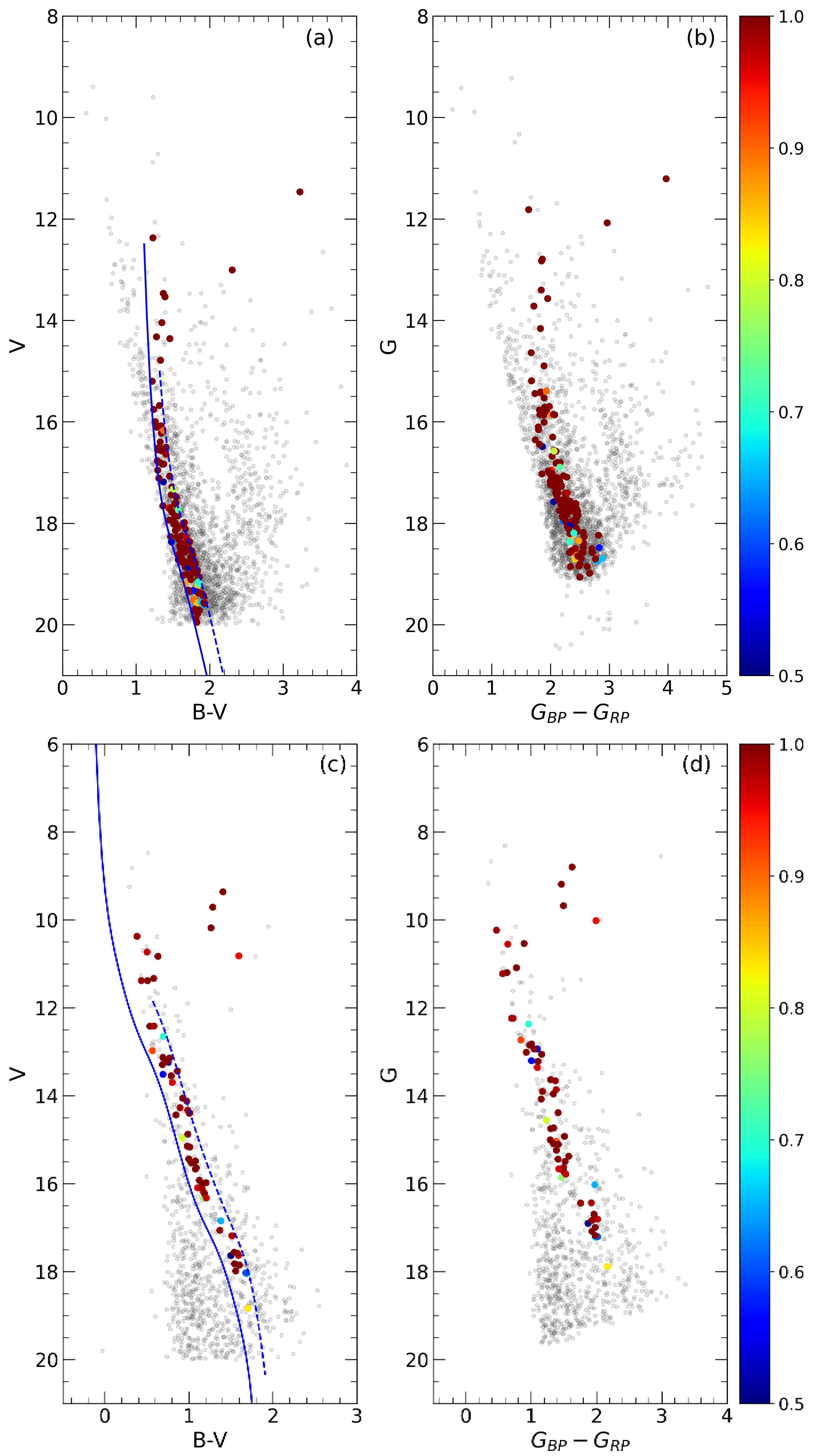}
\caption{CMDs of Czernik 41 (a-b) and NGC 1342 (c-d) in the {\it UBV} (a and c) and {\it Gaia} (b and d) photometric systems. Gray circles represent stars with membership probabilities $P<0.5$, while colored circles indicate stars with $P \geq 0.5$. The continuous and dashed blue curves denote the ZAMS lines used to define the cluster main sequences \citep{Sung13}.}
\label{fig:cmds}
\end {figure*}

Figure~\ref{fig:prob_hists} presents the probability histograms of stars identified in the cluster {\em Gaia}-based catalogs. These histograms include stars located within the limiting radii ($r_{\rm lim} \leq 11'$) and that are brighter than the photometric $G$ completeness limits. The analysis of stars with membership probabilities $P \geq 0.5$ within the limiting radii reveals that their proportions relative to all stars in the $40'\times40'$ cluster field, spanning probabilities $0 \leq P \leq 1$, are $2.34\%$ for Czernik 41 and $3.77\%$ for NGC 1342. This shows that field contamination is very high, confirming that these are not strongly populated clusters.

%Analyzing stars with membership probabilities $P \geq 0.5$ within the limiting radii reveals that their proportions relative to all stars in the $40'\times40'$ cluster field, spanning probabilities $0 \leq P \leq 1$, are $2.34\%$ for Czernik 41 and $3.77\%$ for NGC 1342, confirming that these are not strongly populous clusters.

The proper motion (PM) distributions of stars in the {\it Gaia} catalogs for the fields of Czernik 41 and NGC 1342, along with their vectorial motion directions in equatorial coordinates, are presented in Figure~\ref{fig:VPD_all}. An examination of the PM diagrams for the most probable cluster members (colored circles) reveals that the open clusters Czernik 41 (Figure~\ref{fig:VPD_all}a) and NGC 1342 (Figure~\ref{fig:VPD_all}c) are embedded within the field stars (gray circles). However, it is evident that stars with membership probabilities $P\geq 0.5$ are clustered within specific regions, with Czernik 41 containing a denser stellar population compared to NGC 1342. In panels of Figures~\ref{fig:VPD_all}b and d, the most probable member stars (represented by colored arrows) exhibit a consistent vectorial alignment across the sky. The mean PM components $\langle \mu_{\alpha}\cos\delta, \mu_{\delta} \rangle$ derived from the most probable member stars of Czernik 41 and NGC 1342 are calculated as $(-2.963 \pm 0.068, -6.163 \pm 0.101)$~mas~yr$^{-1}$ and $(0.419 \pm 0.041, -1.662 \pm 0.031)$~mas~yr$^{-1}$, respectively. In Figures~\ref{fig:VPD_all}a (Czernik 41) and c (NGC 1342), the intersections of the dashed blue lines indicate the mean PM values of the OCs.

The mean trigonometric parallaxes  $\langle \varpi \rangle$ of the clusters were determined by fitting Gaussian functions to the parallax histograms of the most probable cluster members. To ensure precise measurements, a relative parallax error constraint of $\sigma_\varpi / \varpi < 0.2$ was applied. Based on this, the $\langle \varpi \rangle$ of Czernik 41 and NGC 1342 were calculated as $0.381 \pm 0.048$~mas and $1.523 \pm 0.051$~mas, respectively. The parallax histograms and Gaussian fit for both clusters are shown in Figure~\ref{fig:plx_hist}. Using the linear relation $d_{\rm \varpi}(\mathrm{pc}) = 1000 / \varpi(\mathrm{mas})$, the distances were derived as $2625 \pm 331$~pc for Czernik 41 and $657 \pm 22$~pc for NGC 1342.

% FIGURE 5
\begin{figure}[!t]
\centering
\includegraphics[width=\columnwidth]{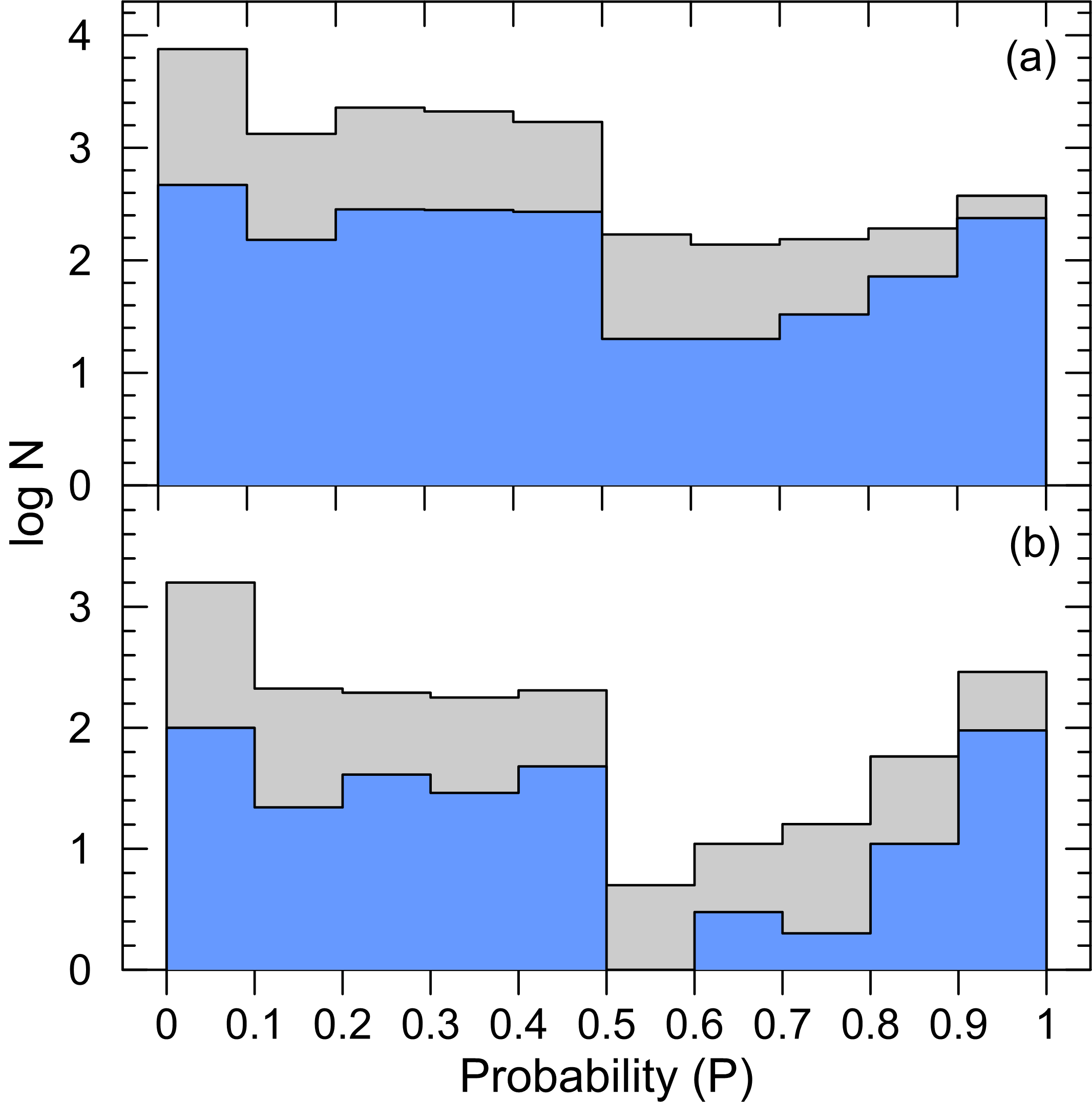}
\caption{Membership probability distributions for all stars detected in the cluster fields of Czernik 41 (a) and NGC 1342 (b), constructed for {\it Gaia} data. The filled grey histograms represent the membership probabilities of all stars identified within the cluster fields, while the filled blue histograms correspond to stars within the limit radius ($r_{\rm lim} \leq 11'$).}
\label{fig:prob_hists} 
\end {figure}

% FIGURE 6
\begin{figure}[t]
\centering
\includegraphics[width=\columnwidth]{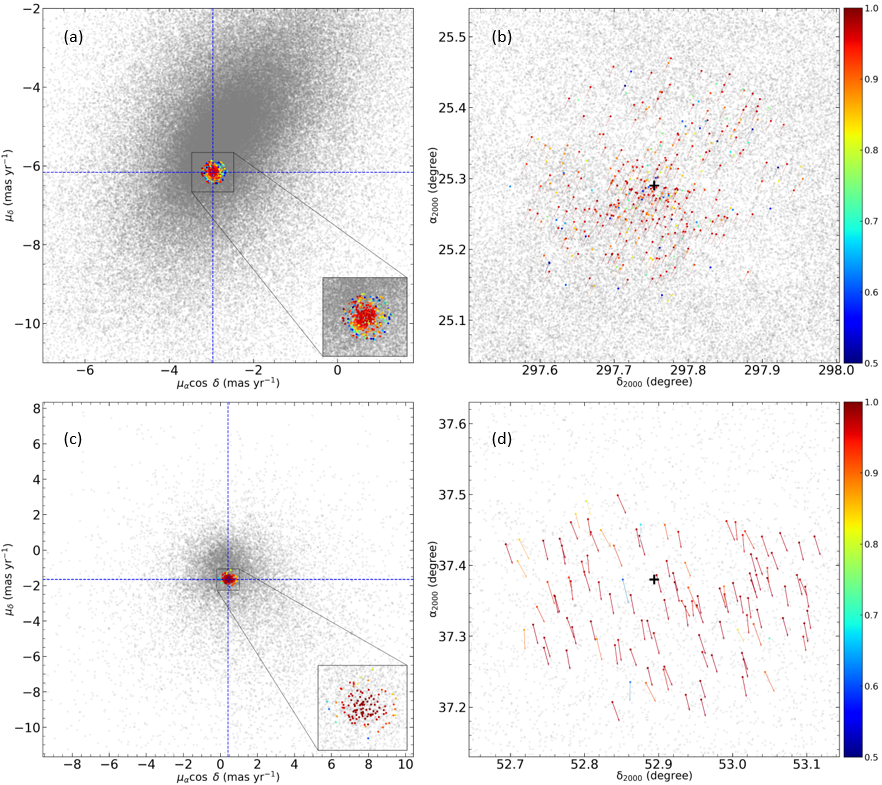}
\caption{Czernik 41 (a-b) and NGC 1342 (c-d) OC fields: {\it Gaia} DR3 PM components (left panels) and sky orientation vectors in equatorial coordinates (right panels). The color scale and circles are identical to Figure~\ref{fig:cmds}. Black plus signs in panels (b) and (d) denote the central equatorial coordinates of the OCs.}
\label{fig:VPD_all} 
 \end {figure}

% FIGURE 07
\begin{figure}[t]
\centering
\includegraphics[width=\columnwidth]{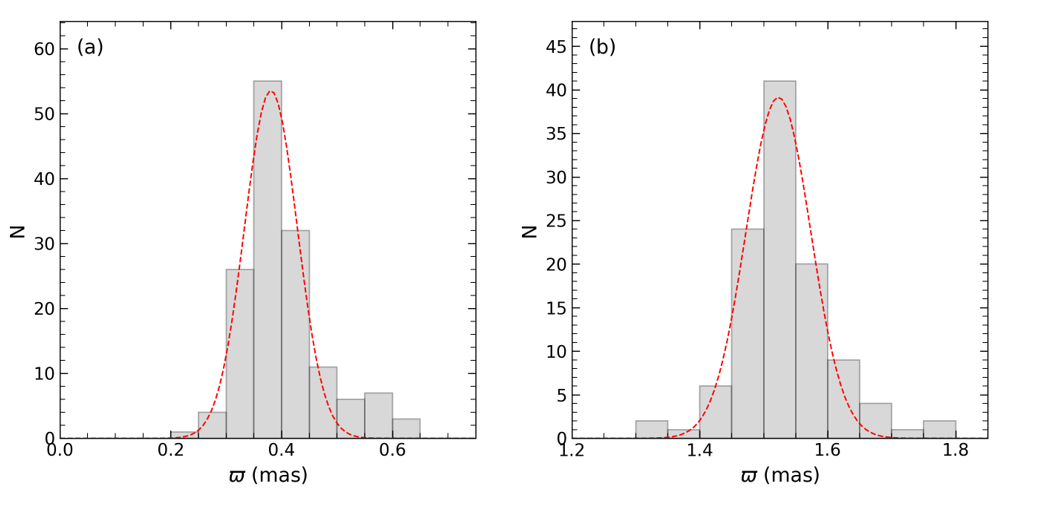}
\caption{Trigonometric parallax histograms for the members of Czernik 41 (a) and NGC 1342 (b) within the limiting radius ($r_{\rm lim} \leq 11'$) and with a relative parallax error of $\sigma_{\varpi}/\varpi < 0.2$. The Gaussian functions applied to the distributions are shown as dashed red lines.}
\label{fig:plx_hist}
\end {figure}

%---------------------------------------------------------------

\section{Astrophysical Parameters of the Clusters}

This section provides a summary of the methodologies employed in the astrophysical analysis of the star clusters Czernik 41 and NGC 1342. The reddening and photometric metallicities were independently estimated using two-color diagrams (TCDs). With these parameters held constant, the distance moduli and ages of the clusters were simultaneously derived from the CMDs. For a comprehensive explanation of the methodology, see the studies as follows \citet{Yontan23b, Bilir16, Bostanci18}.

%---------------------------------------------------------------

% FIGURE 08
\begin{figure}
\centering
\includegraphics[scale=0.25, angle=0]{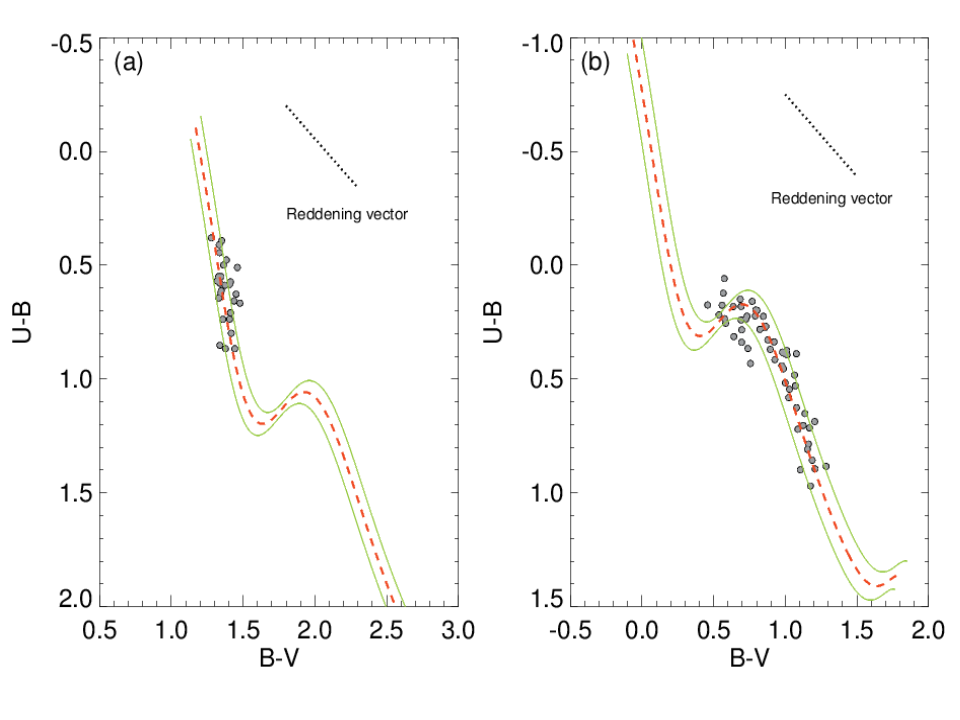}
\caption{The TCDs of the most probable main-sequence stars in the regions of the OCs Czernik 41 (a) and NGC 1342 (b) are presented. The reddened ZAMS, as defined by \citet{Sung13}, is depicted by the red dashed curves, while the green solid curves indicate the $\pm1\sigma$ standard deviations.
\label{fig:tcds}} 
\end{figure}

%FIGURE 09
\begin{figure}
\centering
\includegraphics[width=\columnwidth]{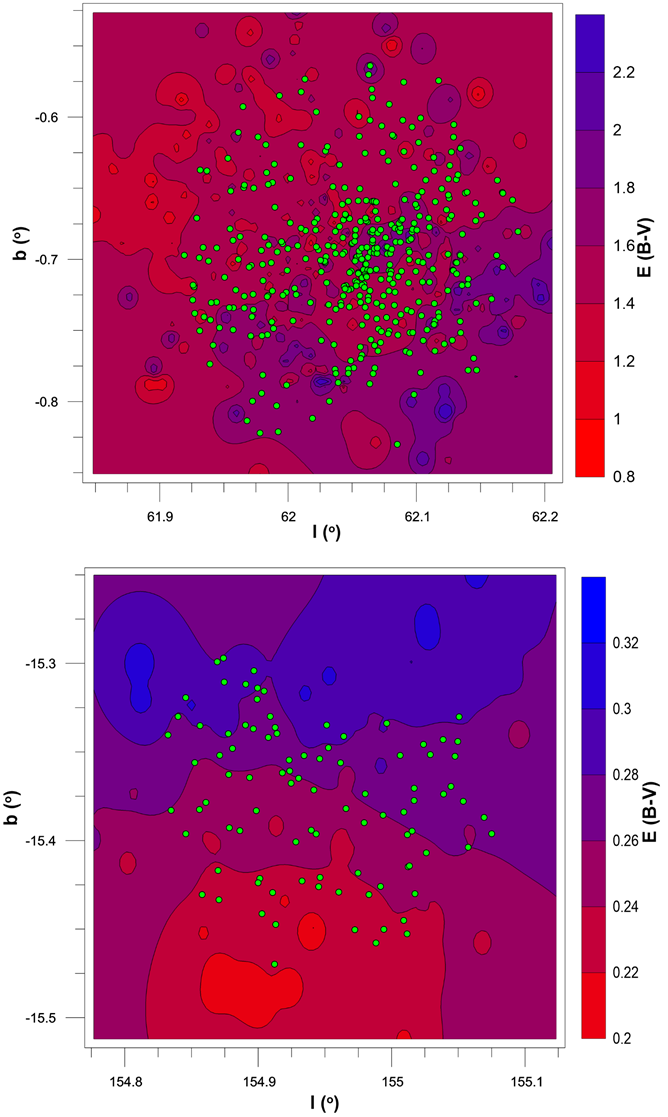}
\caption{The spatial distributions of member stars in Czernik~42 (top) and NGC~1342 (bottom) are presented in Galactic coordinates ($l, b$). Contour lines were generated based on color excess values derived from dust maps. The color scale in the right-hand panels indicates the magnitude of the color excess corresponding to the contour levels.}
\label{fig:dust-map} 
\end{figure}

\subsection{$UBV$ Based Color Excesses}
\label{sec: UBV-color_excess}
The color excesses $E(B-V)$ and $E(U-B)$ for Czernik 41 and NGC 1342 were determined using the ($U-B \times B-V$) TCDs. Initially, member stars with probabilities $P \geq 0.5$ within the defined limiting radii were selected, leading to the construction of ($U-B \times B-V$) TCDs for both OCs. For Czernik 41, 27 stars within the apparent magnitude range of $14\leq V~{\rm (mag)}\leq 17$ were identified, and for NGC 1342, 56 stars within the range $12\leq V~{\rm (mag)}\leq 19$ were selected. The solar metallicity ZAMS of \citet{Sung13} was shifted iteratively until it best matched the stars marked on the ($U-B \times B-V$) TCDs. To compare the results, the color excesses were calculated using the equation $E(U-B)/E(B-V)=0.72+0.05\times E(B-V)$ \citep{Garcia88} over the range $0<E(B-V)~{\rm (mag)}<2$, with steps of 0.01 mag. A $\chi^2$ minimization was applied, with the color excesses corresponding to the minimum $\chi^2$ value being accepted as the best fit. The derived color excess values for both OCs are as follows: for Czernik 41, $E(B-V) = 1.500 \pm 0.035$ mag and $E(U-B) = 1.080 \pm 0.025$ mag, and for NGC 1342, $E(B-V) = 0.270 \pm 0.043$ mag and $E(U-B) = 0.194 \pm 0.031$ mag. The ZAMS curves, which best fit the members of main-sequence stars in the ($U-B \times B-V$) TCDs, are shown in Figure~\ref{fig:tcds} as red dashed lines. The uncertainties in the color excesses, corresponding to one standard deviation ($1\sigma$), are indicated by green solid lines.

%FIGURE 10
\begin{figure}
\centering
\includegraphics[width=\columnwidth]{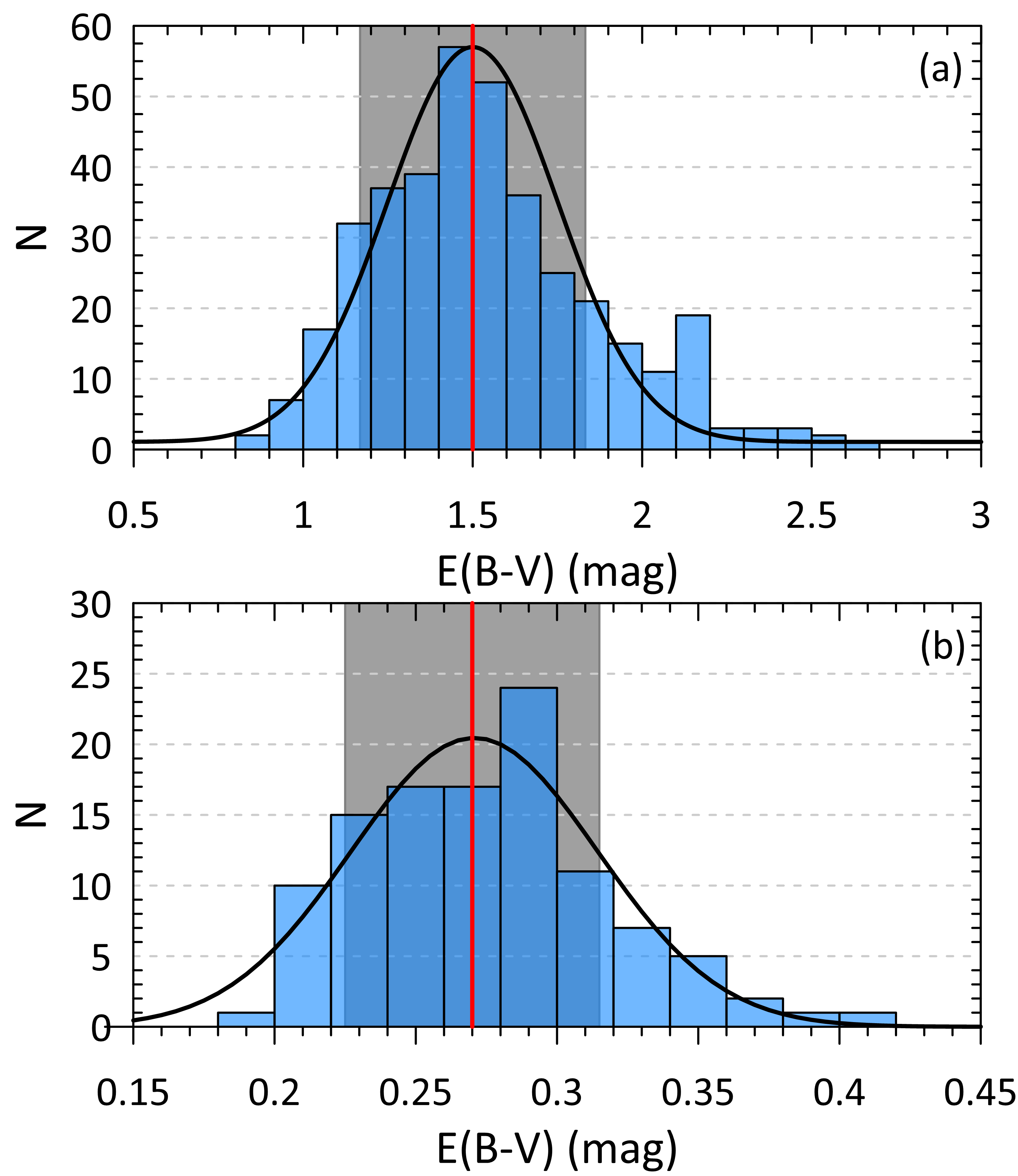}
\caption{Color excess histograms for the (a) Czernik 41 and (b) NGC 1342 OCs, estimated using the 3D-dust maps and distances between the Sun and individual stars calculated from trigonometric parallax data. The red line represents the mean color excess, while the gray-shaded area indicates the $\pm1\sigma$ standard deviation range.}
\label{fig:dust-map-histograms} 
\end{figure}

We used the \citet{Schlafly11} dust maps to check the reliability of the color excesses calculated for the clusters and also to assess the possibility of differential reddening within the OCs. Based on these maps and the equatorial coordinates of stars with high membership probabilities, we obtained $V$-band extinctions up to the Galactic edge. These $V$-band extinction values were then scaled to star-Sun distances using the relation given by \citet{Bahcall80}:
\begin{equation}
A_{\rm d}(V)=A_{\infty}(V)\times \left[1-\exp\left(\frac{-|d\times\sin b|}{H}\right)\right]
\label{equ: Eq1}
\end{equation} 
$b$  represents the Galactic latitude of the star while $d$  denotes its distance, determined using the trigonometric parallax measurements from the {\it Gaia} DR3 catalog via the relation $d{\rm (pc)}=1000/\varpi$ (mas). The parameter $H$ corresponds to the dust scale height \citep[$H=125^{+17}_{-7}$ pc,][]{Marshall06}. $A_{\infty}(V)$ represents the $V$-band extinction integrated along the line of sight to the Galactic boundary. The extinction value corresponding to the distance between the Sun and a star is given by $A_{\rm d}(V)$. The relation $E_{\rm d}(B-V)=3.1\times A_{\rm d}(V)$ was used to convert the calculated reduced $V$-band extinctions into reduced color excess. 

Figure~\ref{fig:dust-map} illustrates the spatial distribution of cluster member stars, color-coded according to their color excess values. The analysis reveals that the color excess values range from 0.8 to 2.4 mag for Czernik 41 and from 0.2 to 0.34 mag for NGC 1342. Figure~\ref{fig:dust-map-histograms} presents the histograms of the color excess distributions for the cluster members. Statistical analysis yields skewness values of 0.648 for Czernik 41 and 0.521 for NGC 1342, indicating a moderate positive skewness in the color excess distributions of both clusters. By fitting Gaussian curves to the distributions, the mean color excesses and corresponding standard deviations were determined to be (1.50, 0.33) mag for Czernik 41 and (0.27, 0.04) mag for NGC 1342. As can be seen in Figure~\ref{fig:dust-map}, Czernik 41 exhibits clear evidence of differential reddening, although the majority of the cluster members are concentrated around the median color excess value. In contrast, NGC 1342 shows a lower level of differential reddening, suggesting a more homogeneous distribution of interstellar extinction across the cluster.

%\newpage  %<----------- NOTE FOR LATER
\subsection{Photometric Metallicity}
\label{sec:metallicity}
The photometric metallicity calibration developed by \citet{Karaali11}, which is sensitive to ultraviolet (UV) excess for F-G spectral type main-sequence stars, was employed to determine the metallicities ([$\mathrm{Fe/H}$]) of the OCs. Through considering the calculated color excesses of the OCs, the intrinsic color indices $(U-B)_0$ and $(B-V)_0$ of the member stars were derived. F-G type main-sequence stars within the color range $0.3 \leq (B-V)_0~({\rm mag})\leq 0.6$, with membership probabilities $P \geq 0.5$, and located within the limit radii $(r_{\rm lim} \leq 11')$ of the OCs, were selected for the metallicity estimation \citep{Eker18, Eker20, Eker24, Eker25}. Since no {\it UBV} photometry was obtained for main-sequence stars in Czernik 41, there is no estimate of the metallicity from the two-color diagram for this cluster}, whereas for NGC 1342, the photometric metallicity was determined using nine main-sequence stars satisfying the selection criteria. A $(U-B)_0\times (B-V)_0$ TCD was constructed for these stars, and their data were compared with the Hyades main sequence to compute UV-excess values. The UV-excess, defined as $\delta = (U-B)_{0,\mathrm{H}} - (U-B)_{0,\mathrm{S}}$, represents the difference in $(U-B)_0$ indices, where the subscripts H and S refer to the Hyades and cluster stars, respectively. Following the approach described in \citet{Karaali11}, we normalized the UV-excess values to $(B-V)_0=0.6$ mag, obtaining $\delta_{0.6}$. The mean $\delta_{0.6}$ was determined by fitting a Gaussian function to the histogram of $\delta_{0.6}$ \citep{Karaali03a, Karaali03b}. The peak of each Gaussian fit provided the mean $\delta_{0.6}$ value, which was found to be $\delta_{0.6} =0.056 \pm 0.014$ mag for NGC 1342. The uncertainties reflect the standard deviation $\pm 1\sigma$ of the Gaussian fits. The photometric metallicity of NGC 1342 was calculated using the calibration equation of $[\mathrm{Fe/H}]=-14.316\times\delta_{0.6}^2 - 3.557\times\delta_{0.6} + 0.105$ presented by \citet{Karaali11}. The $(U-B)_0\times (B-V)_0$ TCD and the $\delta_{0.6}$ histograms for the selected stars in NGC 1342 are shown in Figure~\ref{fig:hyades}. The estimated photometric metallicity value for the cluster is $[\mathrm{Fe/H}]=-0.14\pm 0.07$ dex. Within the scope of the Large Sky Area Multi-Object Fiber Spectroscopic Telescope \citep[LAMOST;][]{Cui12, Deng12, Zhao12} spectroscopic sky survey, atmospheric parameters were derived for two stars in NGC 1342, each with a cluster membership probability of $P \geq 0.8$ \citep{Fu22}. The spectroscopic analyses yielded a mean [Fe/H] of $-0.15$ dex, which is consistent with the calculated cluster [Fe/H]. In this study, a photometric metallicity value based on UV-excess was determined to be $-0.14 \pm 0.07$ dex, in good agreement with the LAMOST results.

To incorporate the effects of metallicity into the age determination, the heavy-element abundance $(Z)$ of a given OC was derived using equations provided by Bovy\footnote{\url{ https://github.com/jobovy/isodist/blob/master/isodist/Isochrone.py}}, which are developed for {\sc parsec} isochrones \citep{Bressan12}. The relevant equations are:
\begin{equation}
\label{eq:02}
Z_{\rm x}={10^{{\rm [Fe/H]}+\log \left(Z_{\odot} / (1-0.248-2.78\times Z_{\odot}) \right)}}
\end{equation}      
and
\begin{equation}
\label{eq:03}
Z=\frac{(Z_{\rm x}-0.2485\times Z_{\rm x})}{(2.78\times Z_{\rm x}+1)}.
\end{equation}
Here, $Z_X$ is an intermediate processing function. $Z_\odot$ represents the solar heavy-element abundance, with a value of $0.0152$. $Z$ is a function of $Z_X$ and is considered in determining the heavy element abundance of the cluster. For NGC 1342, the value $Z$ was calculated as $0.012$, whereas for Czernik 41, $Z$ was assumed to be the value estimated from the method explained in Section~\ref{sec:mcmc}.
%---------------------------------------------------------------

% FIGURE 09
\begin{figure}
\centering
\includegraphics[scale=0.55, angle=0]{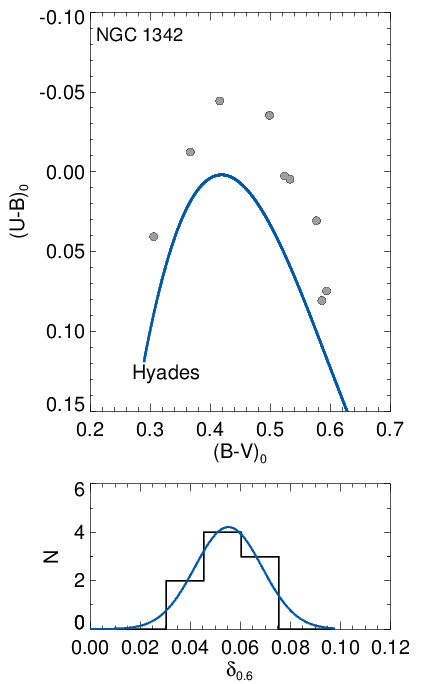}
\caption{TCD for nine F-G type main-sequence stars ($P \geq 0.5$) in NGC 1342 is shown in the top panel. The blue curve in this panel represents the Hyades main sequence. The bottom panel illustrates the histogram of normalized UV-excesses $(\delta_{0.6})$, with a Gaussian function fitted to the distribution, depicted by the blue curve.
\label{fig:hyades}} 
\end {figure}

%---------------------------------------------------------------

%\newpage
%-----------------------------------------------------------
\subsection{Astrophysical Parameters for Czernik 41 and NGC 1342}
To minimize parameter degeneracy during the determination of distance and age, the color excess and metallicity were independently derived using the {\it UBV} photometric system (see Sections~\ref{sec: UBV-color_excess} and ~\ref{sec:metallicity}) as the `classic method', where parameters are determined separately. These parameters were then held constant, allowing for the simultaneous determination of the distance modulus and age. In contrast, within the {\it Gaia} photometric system, all astrophysical parameters were derived simultaneously by utilizing a Markov-Chain Monte Carlo (MCMC) approach. All approaches were conducted using the stars with a membership probability $P \geq 0.5$ and located within the cluster's limiting radius ($r_{\rm lim} \leq 11'$). The numbers of stars so determined are 172 for Czernik 41 and 89 for NGC 1342 in the {\it UBV}-based analysis, while for the MCMC approach, they are 382 and 111, respectively.

%\newpage
\subsubsection{Parameter Estimation with Markov-Chain Monte Carlo}
\label{sec:mcmc}
We developed a new method based on the Monte Carlo approach to independently determine the distance ($d$), extinction ($A_{\rm G}$), age ($\tau$), and metallicity ($Z$) of an OC. This method attempts to reach a common solution for the cluster by treating each member star as an independent `walker' (or chain) over a specific isochrone grid. The grid used for this method was derived from version CMD 3.8\footnote{http://stev.oapd.inaf.it\/cgi-bin\/cmd} of the {\sc parsec} stellar isochrone library \citep{Bressan12}. The properties of the obtained grid are as follows: in the $\log \tau$ space, steps of 0.05 were employed within the range of $6 \leq \log \tau ~{\rm (yr)} \leq  10.13$, while in the metallicity space $Z$, an isochrone grid was created with steps of 0.0005 dex in the range of $0 \leq Z \leq 0.03$.

The input parameter space for the Monte Carlo sampling was set as follows:
\begin{itemize}
    \item Distance ($d$) ranging from 0 to 30 kpc,
    \item Extinction ($A_{\rm G}$) ranging from 0 to 10 magnitudes,
    \item Metallicity ($Z$) and age ($\tau$) spanning the full extent of the isochrone grid.
\end{itemize}

To interpolate the intermediate points of this constructed grid, the Delaunay function from the Python \texttt{Scipy} library \citep{Scipy} was used. This function enables a continuous space within the input grid by performing three-dimensional interpolation among the desired parameters of the input space. As a result, the obtained solution is free from the uncertainties introduced by the step size of the input grid and provides the continuous space required for the Monte Carlo method.

% Table 7
\begin{table}[t]
  \centering
  \caption{Fundamental parameters of the Czernik 41 and NGC 1342 derived using the MCMC for {\it Gaia} DR3 data.}
    \begin{tabular}{lcc}
    \hline
    & Czernik 41 & NGC 1342 \\
    \hline
    Parameter & $G \times (G_{\rm BP}-G_{\rm RP}$) & $G \times (G_{\rm BP}-G_{\rm RP}$) \\
    \hline
    $A_{\rm G}$ (mag)  & $3.70^{+0.54}_{-0.55}$        &  $0.54^{+0.22}_{-0.22}$ \\
    $d$         (pc)   & $2594^{+123}_{-158}$          &  $645^{+27}_{-40}$      \\
    $Z$                & $0.0176^{+0.0050}_{-0.0025}$  & $0.0126^{+0.0025}_{-0.0050}$ \\
    $\log t$    (yr)   &  $7.94^{+0.36}_{-0.48}$       & $8.99^{+0.16}_{-0.25}$  \\
    \hline
    \end{tabular}%
  \label{tab:07}%
\end{table}%

The relationship between the fundamental stellar parameters and the observed magnitudes in the Gaia photometric system can be expressed as follows:

\begin{equation}
    m_{\rm X} = M_{\rm X} (\tau, Z) + \mu + A_{\rm X}.
\end{equation}
%\moindent
where:
\begin{itemize}
    \item $m_{\rm X}$ represents the observed magnitude in band $X$ ($X \in \{G, G_{\text{BP}}, G_{\text{RP}}\}$),
    \item $M_{\rm X} (\tau, Z)$ is the absolute magnitude interpolated from the isochrone grid,
    \item $\mu = 5 \log d - 5$ is the distance modulus,
    \item $A_{\rm X} = k_{\rm X} A_{\rm G}$ is the extinction correction for band $X$, with extinction coefficients $k_{\rm X}$.
\end{itemize}

\noindent
Expanding for each photometric band:

\begin{align}
    G &= M_G (\tau, Z) + 5 \log d - 5 + A_{\rm G} \\
    G_{\text{BP}} &= M_{\text{BP}} (\tau, Z) + 5 \log d - 5 + k_{\text{BP}} A_{\rm G} \\
    G_{\text{RP}} &= M_{\text{RP}} (\tau, Z) + 5 \log d - 5 + k_{\text{RP}} A_{\rm G}
\end{align}
where $M_{\rm X} (\tau, Z)$ values are interpolated from the isochrone grid using the Delaunay triangulation method to ensure a smooth parameter space. The extinction coefficients $k_{\text{BP}}$ and $k_{\text{RP}}$ are 1.2955, 0.7586, respectively. These values are obtained by using \citet{Cardelli89}
and \citet{ODonnell94} extinction curve with $R_{\rm V}$=3.1.

The method aims to fit the observational data using a maximum likelihood approach by allowing each member star to act as a walker, traversing the input space while preserving the overall consistency. The likelihood function is:

\begin{equation}
    \mathcal{L} (\theta) = \prod_{i=1}^{N_{\text{stars}}} \prod_{X \in \{G, G_{\text{BP}}, G_{\text{RP}}\}} P \left( m_{X, i} | \theta \right)
\end{equation}
where $\theta = (\tau, Z, d, A_{\rm G})$ is the parameter set to be estimated. Assuming independent Gaussian errors, the probability distribution function is:

\begin{equation}
    P \left( m_{X, i} | \theta \right) = \frac{1}{\sqrt{2 \pi \sigma_{X, i}^2}} \exp \left( -\frac{(m_{X, i} - \hat{m}_{X, i})^2}{2 \sigma_{X, i}^2} \right)
\end{equation}
where $\hat{m}_{X, i}$ is the model-predicted magnitude:

\begin{equation}
    \hat{m}_{X, i} = M_X (\tau, Z) + 5 \log d - 5 + A_X
\end{equation}
and $\sigma_{\rm X, i}$ is the observational uncertainty in band $X$. The log-likelihood function, which is more convenient for optimization, is:

\begin{equation}
    \ln \mathcal{L} (\theta) = - \sum_{i=1}^{N_{\text{stars}}} \sum_{X} \left[ \frac{(m_{X, i} - \hat{m}_{X, i})^2}{2 \sigma_{X, i}^2} + \ln (\sqrt{2 \pi} \sigma_{X, i}) \right]
\end{equation}

A Monte Carlo sampling method is employed to explore the posterior distribution of the parameters. This method was implemented using the Python \texttt{emcee} library \citep{emcee}, a widely used affine-invariant MCMC sampler. Each cluster parameter is determined using as many walkers as the number of cluster members, with 5,000 iterations per walker. The final parameter estimates are obtained from the posterior distributions by taking the 16th, 50th, and 84th percentiles:

\begin{equation}
    \theta_{\text{best}} = \left( Q_{50} (\theta), \; Q_{16} (\theta), \; Q_{84} (\theta) \right)
\end{equation}
where $Q_p (\theta)$ represents the $p$-th percentile of the posterior distribution.

\begin{figure}
\centering
\includegraphics[width=0.5\textwidth]{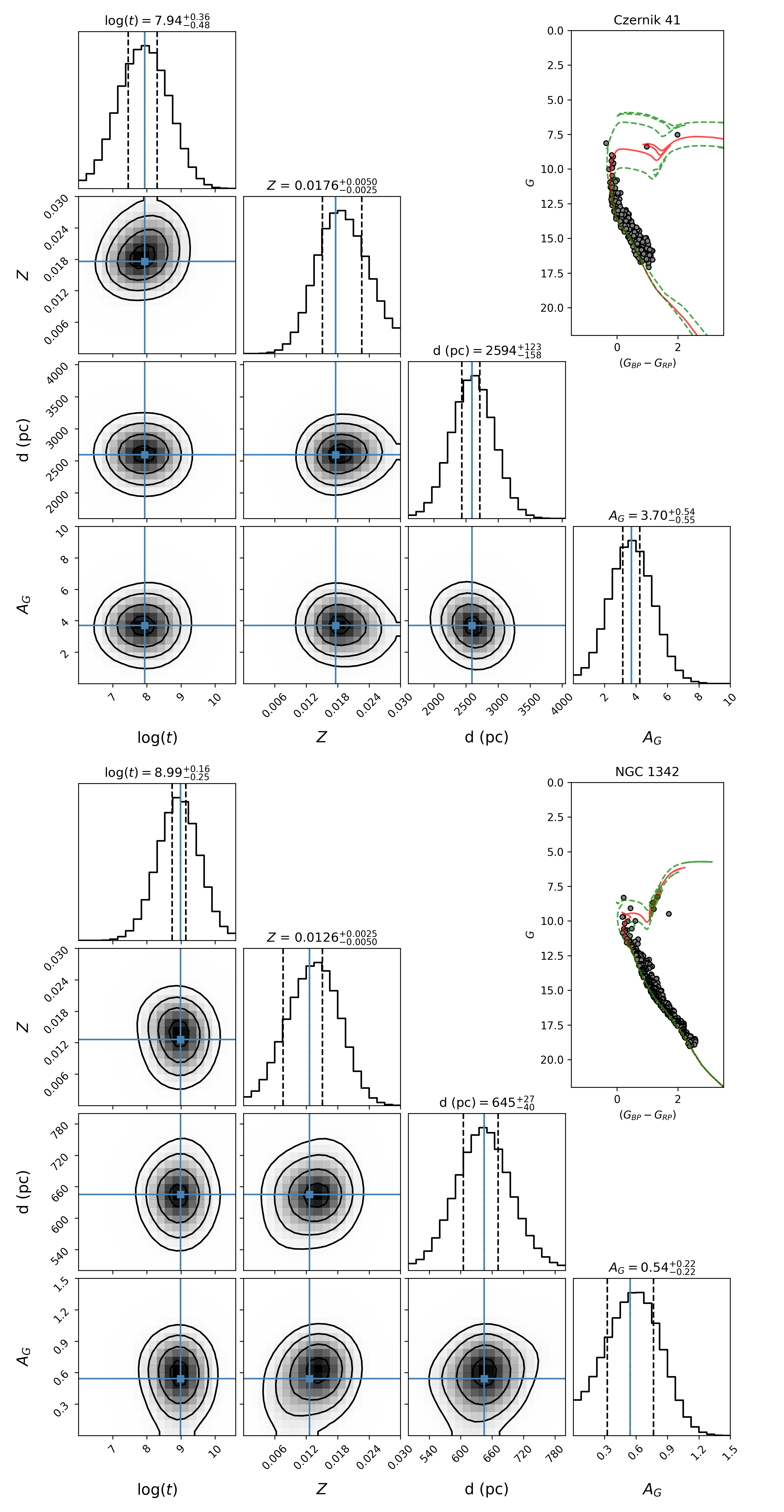}
\caption{Corner plot illustrating the two-dimensional joint and one-dimensional marginalized posterior distributions for the properties of Czernik 41 and NGC 1342. The contours in the two-dimensional joint posterior distributions represent confidence levels of 68\%, 90\%, and 95\%. In the one-dimensional posterior distributions, vertical lines indicate the median as well as the 16th and 84th percentile confidence intervals. The CMDs of the OCs are displayed on the right side of the corner plots. Black dots correspond to the most probable member stars. The red solid lines are the {\sc parsec} isochrones derived from the median values of the posterior distributions, and green dashed lines indicate calculated age errors of the OCs.}
\label{fig:figure_tenx} 
\end {figure}

\begin{figure*}
\centering
\includegraphics[scale=.60, angle=0]{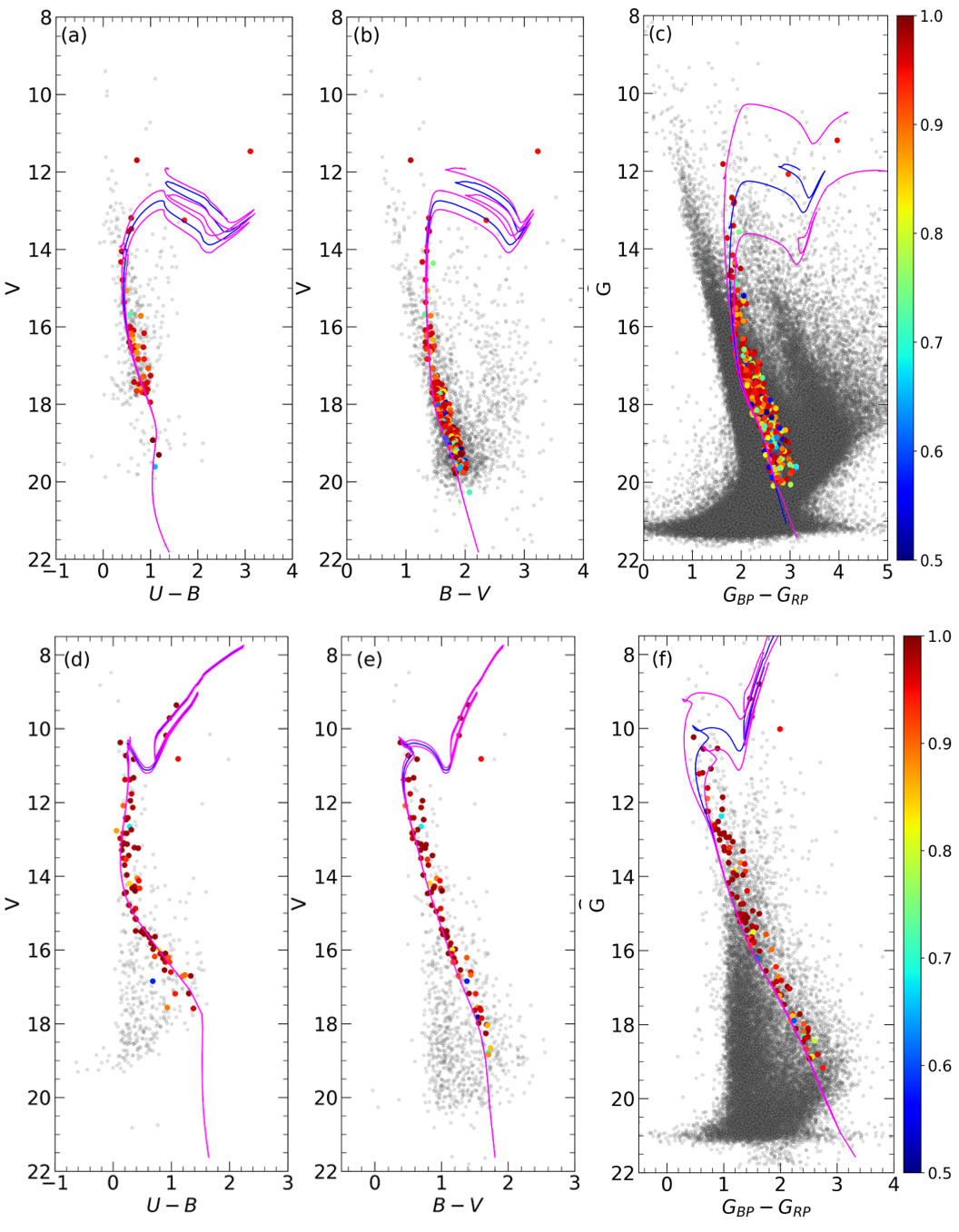}
\caption{$UBV$ and {\it Gaia} based CMDs for Czernik 41 (panels a, b, c) and NGC 1342 (panels d, e, f). The blue solid lines are the {\sc parsec} isochrones that best fit the observed data, providing estimates for the clusters' distance moduli and ages. Purple solid lines represent the uncertainties in the age estimates. The color scales and symbol meanings are consistent with those described in Figure \ref{fig:cmds}.
\label{fig:figure_ten} }
\end {figure*}

The method aims to ensure that the optimal solution for each star converges to the best solution for all stars, enabling the independent determination of the cluster's parameters.  As a result of the analysis, the ages, heavy-element abundances, distances, and extinction coefficients in the $G$ band ($A_{\rm G}$) for the two OCs were determined. The results are presented in Table~\ref{tab:07}. Figure~\ref{fig:figure_tenx} illustrates the posterior distributions for $A_{\rm G}$, $d$, $Z$ and $\log\ t$ derived for the two open clusters. The two-dimensional joint posterior distributions include contours that represent \(68\%\), \(90\%\), and \(95\%\) of the marginalized probability. In the one-dimensional marginalized distributions, vertical dashed lines indicate the 16th, 50th, and 84th percentiles, respectively. Using the parameters obtained from the MCMC analysis of {\it Gaia} data, CMDs were constructed, with the {\sc parsec} model overlaid for validation. As illustrated in Figures~\ref{fig:figure_ten}c and f, the estimated parameters show a strong agreement with the observed {\it Gaia}-based CMDs, demonstrating a high level of consistency between the data and the model.

% FIGURE 10

\citet{Bailer-Jones21} employed a Bayesian framework to infer the photogeometric distance from {\it Gaia} trigonometric parallaxes, incorporating interstellar extinction, color, and magnitude to construct posterior probability distributions. To associate individual cluster members with their corresponding entries in the \citet{Bailer-Jones21} catalog, we performed a cross-matching procedure based on the equatorial coordinates of each cluster member, adopting a cross-match radius of less than 1 arcsec. Our analysis confirms that this choice of radius effectively minimizes the risk of multiple matches, even in relatively crowded cluster environments. After identifying the sources in the catalog, we constructed histograms of the photogeometric distance provided by \citet{Bailer-Jones21}. A Gaussian function was then fitted to these distributions to determine the median distances for two OCs. According to this analysis, the photogeometric distance for Czernik 41 is found to be $d_{\rm BJ21}=2575\pm 322$ pc, while for NGC 1342, it is $d_{\rm BJ21}=646\pm 17$.

% FIGURE 11

\subsubsection{Distance Moduli and Age Estimation}
For the {\it UBV}-based analysis, the distance moduli ($\mu$), distance ($d$), and ages ($t$) of Czernik 41 and NGC 1342 were determined using the {\sc parsec} isochrones \citep{Bressan12}. CMDs in $V \times (U-B)$, and $V \times (B-V)$ were constructed using stars identified in the cluster regions. Stars with a membership probability $P \geq 0.5$ and located within the cluster's limiting radius as $r_{\rm lim} \leq 11'$ were highlighted on these CMDs. {\sc parsec} isochrones, generated for a range of ages consistent with the heavy-element abundances ($Z$) of each cluster, were adjusted to match the observed CMDs. For Czernik 41, the $Z$ value was adopted based on results obtained from an MCMC analysis (see Table~\ref{tab:07}), while for NGC 1342, we adopted the value estimated from the relation described in Equations~\ref{eq:02} and ~\ref{eq:03} in this study. Determining cluster ages and reddened distance modulus emphasized main-sequence stars, turn-off points, and red giants with $P\geq 0.5$. This ensured precise age estimation while identifying the {\sc parsec} isochrones that best represent the clusters, incorporating their metallicities and high-probability member stars. The shifts were applied in the $U-B$ and $B-V$ color indices using the color excess values $E(U-B)$ and $E(B-V)$, as well as in apparent magnitudes using the distance modulus equation $\mu_0 = V - M_{\rm V} - A_{\rm V} = 5 \times \log d - 5$ (where $A_{\rm V}$ is the extinction coefficient in $V$ band and $d$ is the distance to the Sun). The uncertainties in the cluster ages were determined by encompassing all high-probability member stars and defining lower and upper age bounds that align with the observed data. The analysis yielded reddened distance moduli for the Czernik 41 and NGC 1342 as $m_{\rm V} - M_{\rm V} = 16.594 \pm 0.128$ and $m_{\rm V} - M_{\rm V} = 9.880 \pm 0.136$ mag, respectively. These correspond to de-reddened distances of $d_{\rm iso} = 2485 \pm 151$~pc for Czernik 41 and $d_{\rm iso} = 645 \pm 42$~pc for NGC 1342. The uncertainties in the distance moduli and distances were calculated using the equations provided in \citet{Carraro17}. The results are presented in Table~\ref{tab:Final_table}. The {\sc parsec} isochrones, representing the best-fit parameters for reddened distance moduli and ages, along with the $V \times (U-B)$ and $V \times (B-V)$ CMDs of Czernik 41 and NGC 1342, are presented in Figure~\ref{fig:figure_ten}.

%---------------------------------------------------------------

\section{Kinematic and Orbital analysis}

Radial velocity ($V_{\rm R}$) measurements are essential parameters for the kinematic and orbital dynamics analysis of OCs. In this study, the mean radial velocities of Czernik 41 and NGC 1342 were determined using $V_{\rm R}$ data from the {\it Gaia} DR3 catalog. Stars within the clusters' limiting radii ($r_{\rm lim} \leq 11'$) and with membership probabilities $P \geq 0.5$ were selected for the analysis. The {\it Gaia} DR3 catalog contained $V_{\rm R}$ measurements for seven stars in Czernik 41 and 43 stars in NGC 1342. Weighted mean values were calculated to derive the clusters' $\langle V_{\rm R}\rangle$, using equations provided by \citet{Soubiran18}. The  $\langle V_{\rm R}\rangle$ of Czernik 41 was determined as $2.41 \pm 1.92$ km~s$^{-1}$, while for NGC 1342 it was calculated as $-10.48 \pm 0.20$ km~s$^{-1}$. The determination of the space velocity components and Galactic orbits of the clusters was performed using the {\it MWPotential2014} algorithm from the {\tt galpy} library \citep{Bovy15}. For Czernik~41 and NGC~1342, the input parameters for calculating the space velocity components and Galactic orbit parameters included equatorial coordinates, radial velocities derived in this study, mean proper motion (PM) components, distances from main-sequence fitting, and their associated uncertainties. In the calculations, the distance of the Sun from the Galactic center was assumed to be $R_{\rm gc} = 8$ kpc, with a rotational velocity of $V_{\rm rot} = 220$ km~s$^{-1}$ \citep{Bovy15, Bovy12}, and a vertical distance from the Galactic plane of $27 \pm 4$ pc \citep{Chen00}.

The orbital integrations of Czernik 41 and NGC 1342 were performed by tracing their current positions backwards in time for 2.5 Gyr with time steps of 1 Myr. From the analysis, key orbital parameters were determined, including the Galactocentric distance ($R_{\rm gc}$), apogalactic ($R_{\rm a}$) and perigalactic ($R_{\rm p}$) distances from the Galactic center, orbital eccentricities ($e$), maximum vertical distances from the Galactic plane ($z_{\rm max}$), space velocity components ($U$, $V$, $W$), and orbital periods ($T_{\rm P}$).

The traceback early orbital radius ($R_{\rm teo}$) method, recently introduced to the literature by \citet{Akbaba24}, has demonstrated that the birth regions of celestial objects within the Galaxy can be determined by tracing their past orbital paths under the assumption of a static Galactic potential. In contrast to the concept of the birth radius ($R_{\rm birth}$), which aims to determine the formation sites of astronomical objects within the Galaxy based on Galactic chemical evolution models involving numerous assumptions \citep[c.f.,][]{Minchev18, Yuxi24}, the traceback early orbital radius ($R_{\rm teo}$) provides information about the orbital position of an object’s birth region by considering its present-day astrometric data and age. In this study, the calculation of $R_{\rm teo}$ was performed alongside other kinematic and dynamical parameters. The dynamical orbital parameters computed for the two clusters are listed in Table ~\ref{tab:Final_table}. 

The space velocity components ($U$, $V$, $W$) for Czernik 41 were calculated as $ (71.90 \pm 4.41, -35.73 \pm 2.36, -7.03 \pm 1.01)$ km~s$^{-1}$. Similarly, for NGC 1342, the values were found to be $(8.24 \pm 0.19, -8.32 \pm 0.29, -0.44 \pm 0.24)$ km~s$^{-1}$. The components of space velocity are influenced both by the stars' positions within the Milky Way and by inherent observational deviations from the Sun. To mitigate these biases, corrections for differential rotation and adjustments to the local standard of rest (LSR) were applied to each space velocity component. In particular, the differential rotation effects for the Czernik 41 and NGC 1342 were corrected using the relations presented in \citet{Mihalas81}. This procedure yielded velocity corrections ($dU,~dV$) as (59.69, 2.80) km s$^{-1}$  for Czernik 41 and (7.17, -1.35) km s$^{-1}$ for NGC 1342. Since the $W$ space velocity component is unaffected by differential rotation no correction was necessary. For the LSR adjustment, the solar motion values from \citet{Coskunoglu11}, namely $(U, V, W)_{\odot}=(8.83\pm 0.24, 14.19\pm 0.34, 6.57\pm 0.21)$ km s$^{-1}$ were applied to adjust the space velocity components after differential velocity corrections were made. The total space velocities ($S_{\rm LSR}$) of both OCs were obtained using the relation $S_{\rm LSR}=\sqrt{U_{\rm LSR}^2+V_{\rm LSR}^2+W_{\rm LSR}^2}$, with results presented in Table \ref{tab:Final_table}. The $S_{\rm LSR}$ values indicate that both clusters are associated with the young thin-disc stellar population \citep{Leggett92}.

% FIGURE 12
\begin{figure*}
\centering
\includegraphics[width=\textwidth]{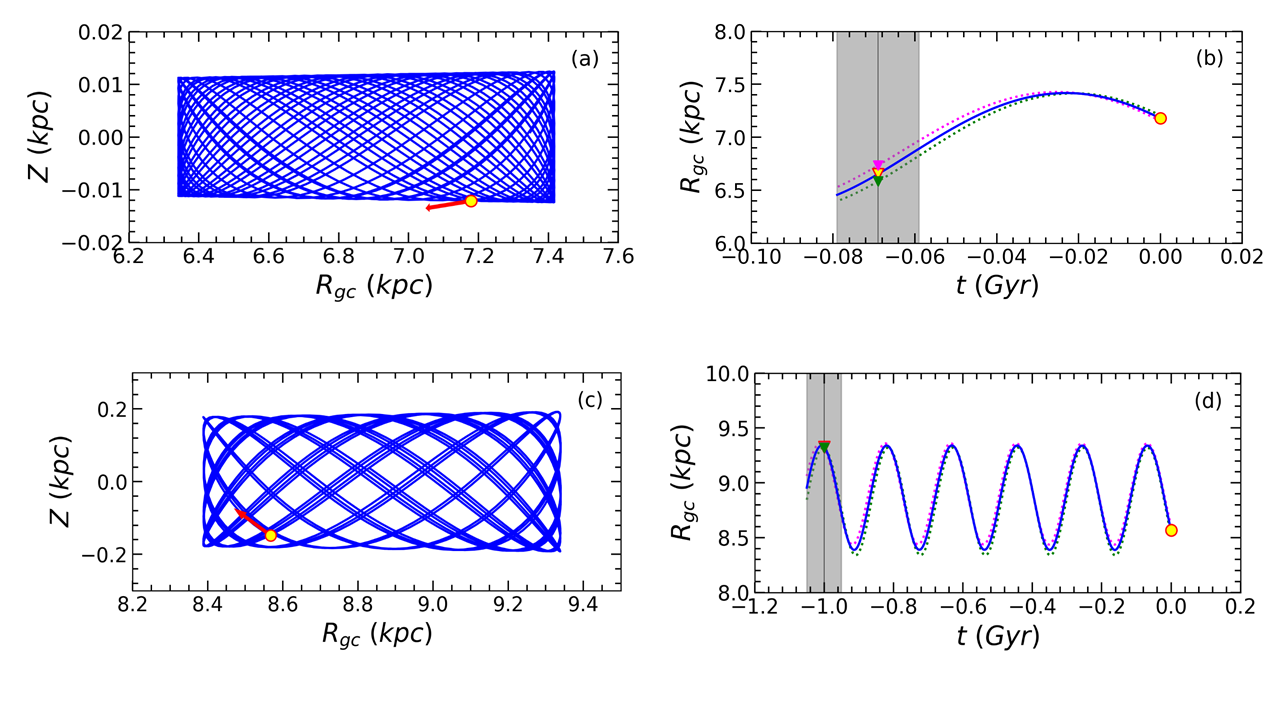}
\caption{The distances of Czernik 41 (a, b) and NGC 1342 (c, d) from the Galactic center, along with their orbital trajectories perpendicular to the Galactic plane, are presented. The yellow-filled circles and triangles represent the current and early orbital positions of the clusters, respectively, while the red arrows indicate the direction of motion vectors of the OCs. The red-shaded regions in the right panels illustrate the uncertainty areas associated with the ages of the OCs. The vertical solid lines correspond to the ages of both OCs, marking their respective early orbital positions.}
\label{fig:galactic_orbits}
\end {figure*}

The Galactic orbits of Czernik 41 and NGC 1342  are shown in Figure~\ref{fig:galactic_orbits}, concerning their vertical distance from the Galactic plane ($Z$) and the distance from the Galactic center ($R_{\rm gc}$). Additionally, the evolution of the clusters' distance from the Galactic center over time ($R_{\rm gc} \times t$) is presented \citep[e.g.][]{Tasdemir23, Yontan23c, Yucel25}. Panels a and c illustrate the motion of the clusters within the Galaxy as a function of their distance from the $R_{\rm gc}$ and the $Z$, respectively. Panels b and d depict the change in the clusters' distance from the Galactic center as a function of time \citep[e.g.][]{Yucel24, Evcil24, Elsanhoury25}. The current and early orbital positions of Czernik 41 and NGC 1342 are represented by yellow-filled circles and triangles, respectively, in Figure~\ref{fig:galactic_orbits}. The dashed pink and green lines, along with the corresponding triangles, illustrate the orbits and early orbital positions of the clusters, accounting for the upper and lower bounds of the input parameter errors. The upper panels of the figure demonstrate that Czernik 41 originated within the solar circle, following an orbit entirely traced within this region. In contrast, the lower panels of Figure~\ref{fig:galactic_orbits} show that NGC 1342 formed outside the solar circle and subsequently entered it, also fully tracing this path. The uncertainties in the ages of the OCs further influence their early orbital positions. The age uncertainties were determined to be 10 Myr for Czernik 41 and 50 Myr for NGC 1342. Taking these age uncertainties into account, the variation in the $R_{\rm teo}$ positions for both clusters is represented by the shaded regions in the right panels of Figure~\ref{fig:galactic_orbits}. Dynamical orbital analysis reveal that, when considering age uncertainties, the early orbital positions of clusters could vary within the ranges approximately $6.45 \leq R_{\rm gc}~{\rm (kpc)} \leq 6.90$ for Czernik 41 and $8.60 \leq R_{\rm gc}~{\rm (kpc)} \leq 8.95$ for NGC 1342. Results support the inference of birth regions for both clusters.  

The calculated Galactic orbit parameters of the two OCs were considered to determine their Galactic population types. The Galactic orbits of Czernik 41 and NGC 1342 exhibit eccentricities that do not exceed 0.08, with a maximum distances from the Galactic plane of $Z_{\rm max}=12\pm2$ pc and $Z_{\rm max}=192\pm8$ pc, respectively. These results also suggest that both OCs belong to the thin disk of the Galaxy \citep{Plevne15, Guctekin19}.

%---------------------------------------------------------------

\section{Investigating the Dynamical Behavior of the Clusters}
\subsection{Luminosity and Mass Functions}

\begin{figure}
\centering
\includegraphics[width=\columnwidth]{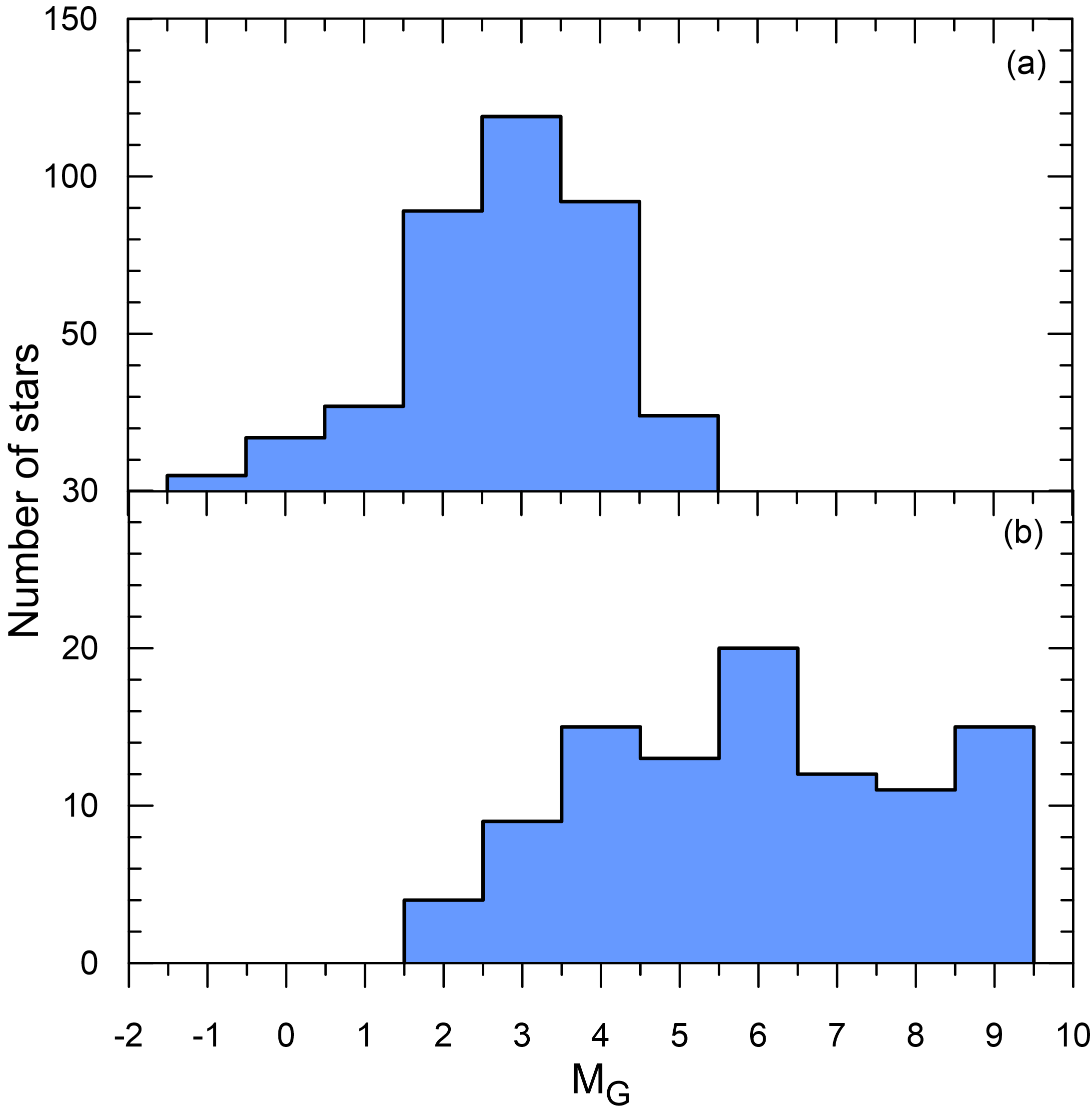}
\caption{The histograms of LFs for Czernik 41 (a) and NGC 1342 (b).}
\label{fig:luminosity_functions}
\end {figure}

In this study, the luminosity functions (LF) of both OCs were determined using main-sequence stars within the limiting radius defined for each cluster. The LFs of Czernik 41 and NGC 1342 were calculated using the cluster catalogs created in the {\it Gaia} system. To recap, the limiting radii of the OCs are $11'$, and the apparent magnitude range of the main-sequence stars with membership probabilities $P \geq 0.5$ within this region is $12\leq G~{\rm (mag)}\leq 20$ for Czernik 41 and $11\leq G~{\rm (mag)}\leq 19$ for NGC 1342. The absolute magnitudes $M_{\rm G}$ of these stars were calculated using the equation $M_{\rm G} = G - 5 \times \log d + 1.8626 \times E(G_{\rm BP}-G_{\rm RP})$ \citep{Cardelli89, ODonnell94}, where $d$ is the distance estimated above from the main-sequence fitting. As a result of these calculations, the $M_{\rm G}$ ranges for the stars in Czernik 41 and NGC 1342 were found to be $-1.5< M_{\rm G}~{\rm (mag)}\leq 5.5$ and $1.4< M_{\rm G}~{\rm (mag)}\leq 9.5$, respectively. The histograms of the LFs were then constructed by counting the number of stars corresponding to the unit $M_{\rm G}$ ranges (Figure~\ref{fig:luminosity_functions}). From Figure~\ref{fig:luminosity_functions}, it can be observed that in Czernik 41 (a) and NGC 1342 (b), the number of stars increases up to $M_{\rm G}$ of 3 and 6 mag, respectively, and then decreases after these magnitudes.

\begin{figure}
\centering
\includegraphics[width=\columnwidth]{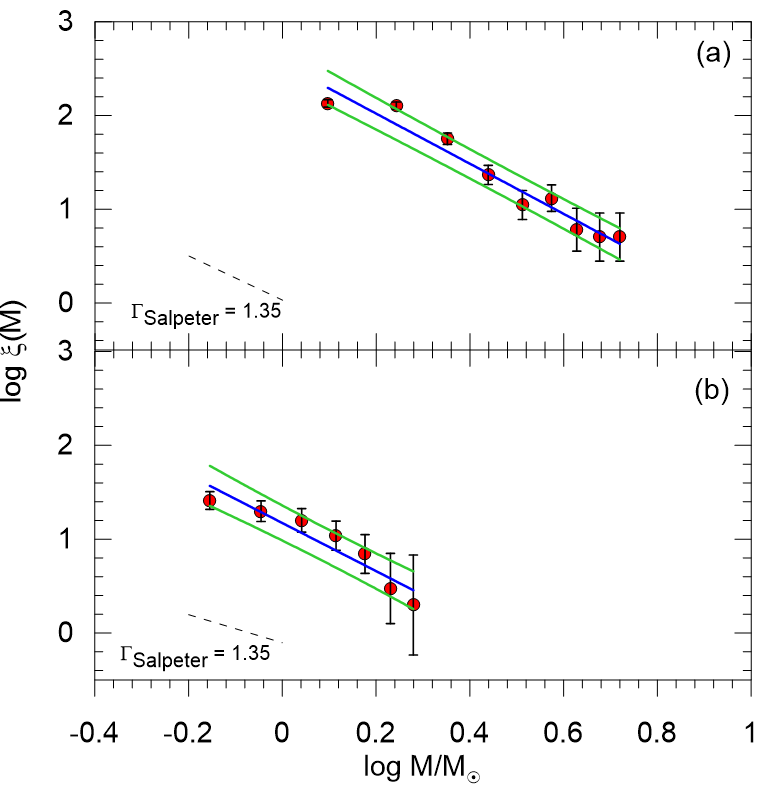}
\caption{For Czernik 41 (a) and NGC 1342 (b), the mass distributions (red points) are shown along with the blue linear line applied to the distribution. The green lines indicate the prediction levels ($\pm 1\sigma$) of the linear fit.}
\label{fig:mass_functions}
\end {figure}

Additionally, the present-day mass functions (PDMFs) of both OCs were derived. The stellar masses were determined from the theoretical {\sc parsec} models used in this study for the determination of the clusters' ages and distance moduli \citep{Bressan12}. The data files from the {\sc parsec} models include both the absolute magnitudes and mass values of the stars across various photometric systems. A polynomial function was applied to establish a relationship between the absolute magnitudes and stellar masses. These mass-luminosity relations were independently derived for each cluster and used to calculate the masses of the main-sequence stars identified in the LF analysis. The mass ranges for the main-sequence stars in Czernik 41 and NGC 1342 were determined to be $1.05 < M/M_{\odot} < 5.20$ and $0.40< M/M_{\odot} < 1.85$, respectively. The $G$-band apparent magnitudes corresponding to these mass intervals span from 12 to 20 mag for Czernik 41, and from 11 to 19 mag for NGC 1342. It is worth noting that photometry is expected to be complete even for the fainter stars in these magnitude ranges, supporting the robustness of the PDMF analysis. For the calculation of the PDMFs, the stellar mass distributions were first generated. The stars were grouped into mass bins of $0.1 M/M_{\odot}$ in both clusters. The logarithmic values of the number of stars in each mass bin were then calculated. These mass distributions are shown in Figure~\ref{fig:mass_functions}. The errors in the stellar mass values were computed using Poisson statistics ($1/\sqrt{N}$). For the determination of the clusters' PDMFs, the linear relation of $\log \left( \frac{dN}{dM} \right) = -(1 + \Gamma) \times \log M + \text{constant}$ from \citet{Salpeter55} was applied. $dN$ represents the number of stars per unit mass $dM$, $M$ is the central mass of the stars, and $\Gamma$ denotes the slope of the PDMF. The relationship used for the mass values is represented by a blue solid line in Figure~\ref{fig:mass_functions}. The slopes of the PDMFs for Czernik 41 and NGC 1342 were determined to be $\Gamma = 1.67 \pm 0.23$ and $\Gamma = 1.56 \pm 0.41$, respectively. These values are in agreement with the \citet{Salpeter55} value of $\Gamma = 1.35$.

% FIGURE 13

% FIGURE 14

\subsection{The Dynamical State of Mass Segregation}

Mass segregation, the process by which more massive stars tend to concentrate towards the center of a star cluster over time, is a fundamental aspect of the dynamical evolution of stellar systems. This phenomenon is driven by gravitational interactions among the stars within the cluster, and it is closely tied to relaxation processes occurring over time \citep{Sagar88, Raboud98, Fischer98}.  The relaxation time, often referred to as the half-mass relaxation time $(T_{\rm E})$, represents the characteristic time scale over which a cluster's stars undergo significant dynamical interactions. The interactions between cluster stars cause energy transfer from high-mass stars to low-mass stars, resulting in more massive stars gradually moving toward the cluster center while lower-mass stars move toward the cluster halo \citep[see, e.g.,][]{Heggie79, Hillenbrand98, Baumgardt03, Dib19, Bisht20, Pavlik22}. This time is crucial in determining the extent of mass segregation in a cluster. This energy exchange process leads to kinetic energy equipartition, where the velocity distribution of stars conforms to a Maxwellian distribution. The $T_{\rm E}$ is crucial in determining the extent of mass segregation in a cluster, as it dictates how quickly the stars redistribute within the cluster. $T_{\rm E}$ depends on several factors, including the number of stars ($N$), the half-mass radius (in parsecs) of the cluster ($R_{\rm h}$), and the mean mass of the stars ($\langle m \rangle$). It is given by \citet{Spitzer71} and typically expressed as:
\begin{equation}
T_{\rm E} = \frac{8.9 \times 10^{5} N^{1/2} R_{\rm h}^{3/2}}{\langle m\rangle^{1/2}\log(0.4N)}
\end{equation} 

The connection between relaxation time and mass segregation is further complicated by the cluster's age and the presence of external influences, such as tidal interactions or stellar feedback \citep{Kruijssen12}. In younger clusters, primordial mass segregation may dominate, as the stars are still settling into their equilibrium distribution. As clusters age, the relaxation process becomes more significant, and the stars gradually redistribute in a way that leads to further mass segregation, particularly in denser regions. Studies of mass segregation in OCs indicate that the dynamical state of a cluster plays a pivotal role in determining its observed mass segregation \citep{Allison09, Haroon25}. For example, younger clusters, or those with short relaxation times, may exhibit strong central concentrations of massive stars, while older clusters with longer relaxation times may show a more uniform distribution due to the effects of dynamical relaxation. This suggests that the degree of mass segregation in a cluster is not only influenced by its initial conditions but also by its evolutionary state.

In our analysis of both OCs examined in this study, we determined their relaxation times and compared them with the observed degree of mass segregation. Our sample selection included stars with membership probabilities of $P\geq 0.5$ and those located within the limiting radius of each cluster. Based on these criteria, we identified 382 stars in Czernik 41 and 111 stars in NGC 1342, with mass ranges of $1.05 < M/M_{\odot} \leq 9.90$ for Czernik 41 and $0.40 < M/M_{\odot} \leq 2.25$ for NGC 1342. The total mass of Czernik 41 was estimated as $M = 771 \, M_{\odot}$, while for NGC 1342, it was found to be $M = 112 \, M_{\odot}$. This corresponds to a mean stellar mass of $\langle m \rangle = 2.02 M/M_{\odot}$ for Czernik 41 and $\langle m \rangle = 1.01 M/M_{\odot}$ for NGC 1342. The half-mass radii were determined as $R_{\rm h} = 3.54$ pc for Czernik 41 and $R_{\rm h} = 0.85$ pc for NGC 1342. The estimated dynamical relaxation times are $T_{\rm E} = 37.31$ Myr for Czernik 41 and $T_{\rm E} = 4.44$ Myr for NGC 1342. Since the relaxation times are significantly shorter than the present-day ages of both OCs, as derived in this study (see Table~\ref{tab:Final_table}), we conclude that both OCs have reached a state of dynamical relaxation. However, the relaxation time values presented in this study should be considered as lower limits. They are based solely on stars detected within the photometric completeness range. The presence of unresolved low-mass stars and binary systems may increase the total mass and number of stars, thereby extending the estimated relaxation time. Therefore, for Czernik 41, it is possible that the relaxation time could be longer than the cluster age, implying that the cluster may not yet be fully relaxed dynamically.

\begin{figure}
\centering
\includegraphics[width=\columnwidth]{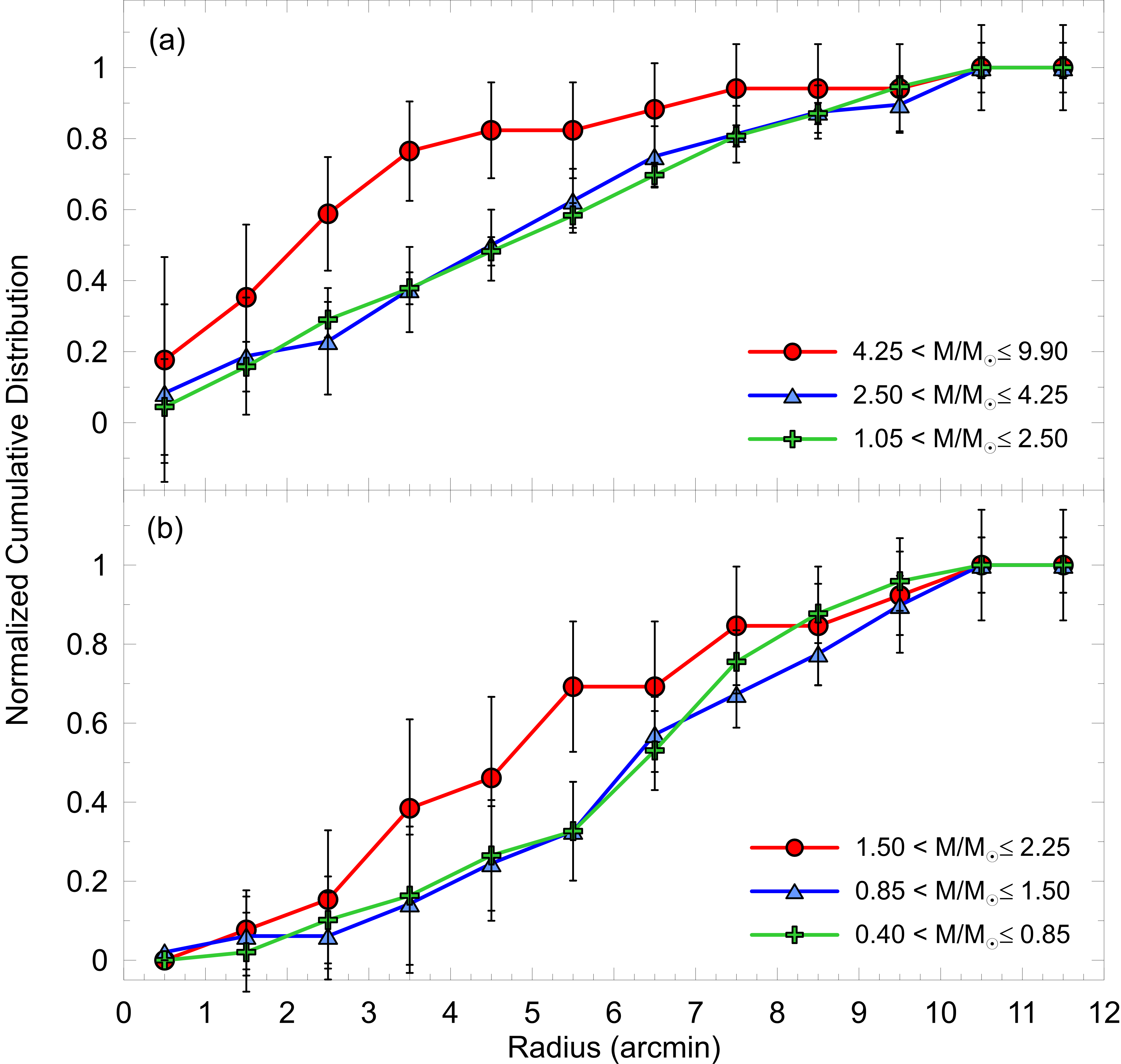}
\caption{\label{fig:radial_distributions}
The cumulative radial distribution of stars across various mass ranges for Czernik 41 (a) and NGC 1342 (b).}
\end {figure}

To understand the impact of the mass segregation effect in both OCs, we divided the masses of selected stars into three intervals containing almost an equal number of stars. The mass intervals are $1.10 < M/M_{\odot}\leq 2.50$ (low-mass), $2.50 < M/M_{\odot}\leq 4.25$ (intermediate-mass), and $4.25 < M/M_{\odot}\leq 9.90$ (high-mass) for Czernik 41 and $1.50 < M/M_{\odot}\leq 2.25$, $0.85 < M/M_{\odot}\leq 1.50$, and $0.40 < M/M_{\odot}\leq 0.85$ for NGC 1342. The normalized cumulative radial distributions of cluster membership stars in these different mass intervals are shown in Figure~\ref{fig:radial_distributions}. As shown in this figure, high-mass stars are concentrated in a region near the center of the OCs, whereas intermediate- and low-mass stars exhibit a similar distribution beyond the central region in both OCs. 

To quantify this segregation, we applied the two-sided Kolmogorov–Smirnov (K–S) test to compare the radial distributions of high-mass stars with those of the other mass groups. The resulting confidence levels of 95\% for Czernik 41 and 92\% for NGC 1342 indicate that the null hypothesis (stating that the compared populations are drawn from the same parent distribution) can be rejected at these significance levels. These results provide statistical evidence that in these clusters the high-mass stars are distributed differently from the lower-mass stars, consistent with the presence of mass segregation in both OCs.

%The Kolmogorov-Smirnov (K-S) test was used to analyze the mass distributions in the clusters and the confidence level for the mass separation effect was 95\% for Czernik 41 and 92\% for NGC 1342.

% FIGURE 15

% Table 8
\begin{table*}
  \centering
  \renewcommand{\arraystretch}{1}
  \caption{Fundamental parameters of Czernik 41 and NGC 1342.}
  \medskip
  {\small
        \begin{tabular}{lrr}
\hline
Parameter & Czernik 41 & NGC 1342 \\
\hline
($\alpha,~\delta)_{\rm J2000}$ (Sexagesimal)& 19:51:01.0, $+$25:17:24    & 03:31:34.6, $+37$:22:48  \\
($l, b)_{\rm J2000}$ (Decimal)              & 062.0238, $-$00.6899       & 154.9402, $-15$.3463     \\
$f_{0}$ (stars arcmin$^{-2}$)               & 7.622 $\pm$ 0.750          & 1.733 $\pm$ 0.240        \\
$f_{\rm bg}$ (stars arcmin$^{-2}$)          & 4.481 $\pm$ 0.171          & 1.766 $\pm$ 0.090        \\
$r_{\rm c}$ (arcmin)                        & 1.728 $\pm$ 0.266          & 2.155 $\pm$ 0.568        \\
$r_{\rm lim}$ (arcmin)                      & 11                         & 11                       \\
$r$ (pc)                                    & 7.95                       & 2.06                     \\
$C$                                         & 0.804 $\pm$ 0.062          & 0.708 $\pm$ 0.102                     \\
Cluster members ($P\geq0.5$)                & 382                        & 111                      \\
$\mu_{\alpha}\cos \delta$ (mas yr$^{-1}$)   & $-2.963 \pm 0.068$         & 0.419 $\pm$ 0.041        \\
$\mu_{\delta}$ (mas yr$^{-1}$)              & $-6.163 \pm 0.101$         & $-1.662 \pm 0.031$       \\
$\varpi$ (mas)                              & 0.381 $\pm$ 0.048          & 1.523  $\pm$ 0.051       \\
$d_{\varpi}$ (pc)                           & 2625 $\pm$ 331             & 657 $\pm$ 22             \\
$E(B-V)$ (mag)                              & 1.500 $\pm$ 0.035          & 0.270 $\pm$ 0.043        \\
$E(U-B)$ (mag)                              & 1.080 $\pm$ 0.025          & 0.194 $\pm$ 0.031        \\
$A_{\rm V}$ (mag)                           & 4.650 $\pm$ 0.109          & 0.837 $\pm$ 0.133        \\
$E({G_{\rm BP}-G_{\rm RP}})$ (mag)          & 1.986 $\pm$ 0.295          & 0.290 $\pm$ 0.118        \\
$A_{\rm G}$ (mag)                           & 3.700 $\pm$ 0.550          & 0.540 $\pm$ 0.220        \\
$[{\rm Fe/H}]$ (dex)                        & 0.07 $\pm$ 0.09            & -0.14 $\pm$ 0.07         \\
Z                                           & 0.0176 $\pm$ 0.0038        & 0.0112 $\pm$ 0.0018      \\
$m_V - M_V$ (mag)                           & 16.594 $\pm$ 0.128         & 9.880 $\pm$ 0.136        \\
$d_{\rm iso}$ (pc)                          & 2485 $\pm$ 151             & 645 $\pm$ 42             \\
Age (Myr)                                   & 69 $\pm$ 10                & 1000 $\pm$ 50            \\
$(X, Y, Z)_{\odot}$ (pc)                    & (1166, 2194, $-30$)        & ($-563$, 264, $-171$)    \\
$R_{\rm gc}$ (pc)                           & 7178 $\pm$ 27              & 8568 $\pm$ 37            \\
PDMF slope                                  &  1.67 $\pm$ 0.23           & 1.56 $\pm$ 0.41          \\
$V_{\rm R}$ (km s$^{-1}$)                   & 2.41 $\pm$ 1.92            & $-10.48 \pm 0.20$        \\
$U_{\rm LSR}$ (km s$^{-1}$)                 & 20.71 $\pm$ 4.42           & 9.57 $\pm$ 0.31          \\
$V_{\rm LSR}$ (km s$^{-1}$)                 & $-25.15 \pm 2.38$          & 6.41 $\pm$ 0.45          \\
$W_{\rm LSR}$ (km s$^{-1}$)                 & $-0.54 \pm 1.03$           & 6.05 $\pm$ 0.32          \\
$S_{_{\rm LSR}}$ (km s$^{-1}$)              & 32.58 $\pm$ 5.12           & 13.01 $\pm$ 0.63         \\
$R_{\rm a}$ (pc)                            & 7417 $\pm$ 6               & 9341 $\pm$ 19            \\
$R_{\rm p}$ (pc)                            & 6341 $\pm$ 34              & 8390 $\pm$ 50            \\
$Z_{\rm max}$ (pc)                          & 12 $\pm$ 2                 & 192 $\pm$ 8              \\
$e$                                         & 0.078 $\pm$ 0.002          & 0.054 $\pm$ 0.002        \\
$P_{\rm orb}$ (Myr)                         & 190 $\pm$ 1                & 250 $\pm$ 1              \\
$R_{\rm teo}$ (kpc)                         & 6.65 $\pm$ 0.07             & 9.32 $\pm$ 0.01          \\
\hline
        \end{tabular}%
    } %ends the small font selection
    \label{tab:Final_table}%
\end{table*}%

%----------------------------------------------------------------------------------------------------------

\section{Summary and Conclusion}
This study presents a comprehensive analysis of the photometric, astrometric, and spectroscopic data of the Czernik 41 and NGC 1342 OCs using CCD {\it UBV} observations and the {\it Gaia} DR3 catalog \citep{Gaia23}. Astrophysical, kinematical, and dynamical Galactic orbital parameters were derived exclusively from stars with a membership probability of $P \geq 0.5$. A key objective of this study is to investigate parameter degeneracy by applying both classical and MCMC methods to photometric and astrometric data.  The consistency between the results obtained with both classical and MCMC methods demonstrated the reliability of the derived astrophysical parameters. The results of the study are summarized as follows:

1) The RDPs were fitted with \citet{King62} models to derive structural parameters. The limiting radius for each cluster was visually set at $11'$, beyond which field stars dominate. Low central concentration, as indicated by the concentration parameters $C=0.804\pm 0.062$ for Czernik 41 and $C=0.708\pm 0.102$ for NGC 1342, may be characteristic of either dynamically young clusters that have not yet undergone significant mass segregation, or dynamically evolved clusters that have experienced mass loss and core expansion

2) Membership selection was refined using both astrometric ({\sc upmask}) and photometric criteria. For Czernik 41 and NGC 1342, 382, and 111 probable members ($P\geq 0.5$) were identified from {\it Gaia} DR3, while {\it UBV} photometry identified 172 and 89 members, respectively. The consistency between both selected criteria, with stars aligning along ZAMS-constrained regions, ensures a reliable basis for astrophysical parameter determination.

3) Proper motion distributions and vectorial motions revealed a denser stellar population in Czernik 41 compared to NGC 1342. The mean proper motion components ($\langle \mu_{\alpha}\cos\delta, \mu_{\delta} \rangle$) were ($-2.963\pm 0.068, -6.163\pm 0.101$) mas yr$^{-1}$ for Czernik 41 and ($0.419\pm 0.041, -1.662\pm 0.031$) mas yr$^{-1}$ for NGC 1342, consistent with the VPD of membership-selected stars. Mean trigonometric parallaxes ($\langle\varpi\rangle$) were $0.381 \pm 0.048$ and $1.523 \pm 0.051$ mas for Czernik 41 and NGC 1342, corresponding to distances ($d_{\rm \varpi}$) to be $2625\pm 331$ and $657\pm 22$ pc, respectively.

4) Reddening and metallicity of both OCs were estimated using different methods. For Czernik 41, reddening was derived from the TCD as $E(B-V) = 1.500 \pm 0.035$ mag. Due to the absence of suitable F-G main-sequence stars, metallicity was determined via an MCMC-based technique, yielding [Fe/H] $ = 0.07 \pm 0.09$ dex. The color excess and metallicity were obtained from TCDs, with $E(B-V)=0.270 \pm 0.043$ mag and [Fe/H] $ = -0.14 \pm 0.07$ dex for NGC 1342. These results show the methodological differences, particularly the use of MCMC for Czernik 41 due to the lack of metallicity-sensitive stars in {\it UBV} data.

5) Distances and ages of both OCs were determined using {\it UBV}-based {\sc parsec} isochrones and MCMC analysis with {\it Gaia} DR3. $V \times (U-B)$ and $V \times (B-V)$ CMDs were constructed for stars with $P \geq 0.5$, yielding distance moduli ($\mu_V$) of $16.594 \pm 0.128$ mag for Czernik 41 and $9.880 \pm 0.136$ mag for NGC 1342, corresponding to distances ($d_{\rm iso}$) of $2485 \pm 151$ pc and $645 \pm 42$ pc. The ages ($t$) were determined as 69 $\pm$ 10 Myr for Czernik 41 and 1000 $\pm$ 50 Myr for NGC 1342. MCMC provided an independent estimate of these parameters, accounting for parameter degeneracies (see Table~\ref{tab:Final_table}). The posterior distributions of $A_{\rm G}$, $d$, $Z$, and $\log\ t$ were consistent with the {\it UBV}-based results, confirming the reliability of both methods.

6) The radial velocity measurements of OC member stars from the {\it Gaia} DR3 catalog were used to determine the mean radial velocities of the OCs. The mean radial velocities of Czernik 41 and NGC 1342 were calculated to be $2.41 \pm 1.92$ (from 7 stars) and $-10.48 \pm 0.20$ km s$^{-1}$ (from 43 stars), respectively.

7) The Galactic orbital analysis of both OCs revealed distinct orbital paths. Czernik 41, originating within the solar circle, remains within this region, while NGC 1342, formed outside the solar circle, has since entered it. Also, the estimated $Z_{\rm max}$ of Czernik 41 and NGC 1342 are smaller than 200 pc, indicating that both OCs are members of the thin disk. 

8) The PDMF slopes were $\Gamma = 1.67 \pm 0.23$ for Czernik 41 and $\Gamma = 1.56 \pm 0.41$ for NGC 1342, falling within 1.4$\sigma$ of the \citet{Salpeter55} value, indicating a typical stellar mass distribution for both OCs.

9) The relaxation times for Czernik 41 and NGC 1342 were calculated as 37.31 Myr and 4.44 Myr, respectively, suggesting that both clusters have likely reached dynamical relaxation. However, for Czernik 41, this value should be considered a lower limit due to incompleteness in low-mass and binary star detections. Thus, its actual relaxation time may exceed its age, implying that it might not yet be fully dynamically relaxed. Mass segregation analysis showed high-mass stars concentrated at the center, while lower-mass stars were more evenly distributed. 

\software{Astrometry.net \citep{Lang10}, Galpy \citep{Bovy15}, IRAF \citep{Tody86, Tody93},  PyRAF \citep{Science12}, SExtractor \citep{Bertin96}, {\sc UPMASK} \citep{Krone-Martins14} .}

%---------------------------------------------------------------
\acknowledgments
 We would like to express our sincere gratitude to the anonymous referee for his/her valuable comments and suggestions, which were helpful in improving the quality of the manuscript. This study has been supported in part by the Scientific and Technological Research Council (T\"UB\.ITAK) 122F109. Funding was provided by the Scientific Research Projects Coordination Unit of Istanbul University as project numbers FYL-2022-39160 and 40044. We thank T\"UB\.ITAK for partial support towards using the T100 telescope via project 18CT100-1396. We also thank the on-duty observers and members of the technical staff at the T\"UB\.ITAK National Observatory for their support before and during the observations. This study is a part of the MSc thesis of Bur\c cin Tan\i k \"Ozt\"urk.  We also made use of VizieR and Simbad databases at CDS, Strasbourg, France. We made use of data from the European Space Agency (ESA) mission \emph{Gaia}\footnote{https://www.cosmos.esa.int/gaia}, processed by the \emph{Gaia} Data Processing and Analysis Consortium (DPAC)\footnote{https://www.cosmos.esa.int/web/gaia/dpac/consortium}. Funding for DPAC has been provided by national institutions, in particular the institutions participating in the \emph{Gaia} Multilateral Agreement. We are grateful for the analysis system IRAF, which was distributed by the National Optical Astronomy Observatory (NOAO). NOAO was operated by the Association of Universities for Research in Astronomy (AURA) under a cooperative agreement with the National Science Foundation. PyRAF is a product of the Space Telescope Science Institute, which is operated by AURA for NASA. We thank the University of Queensland for the collaboration software.

%%%%%%%%%%%%%%%%%%%% REFERENCES %%%%%%%%%%%%%%%%%%

\bibliography{open_clusters.bib}

\begin{thebibliography}{}
\expandafter\ifx\csname natexlab\endcsname\relax\def\natexlab#1{#1}\fi
\providecommand{\url}[1]{\href{#1}{#1}}
\providecommand{\dodoi}[1]{doi:~\href{http://doi.org/#1}{\nolinkurl{#1}}}
\providecommand{\doeprint}[1]{\href{http://ascl.net/#1}{\nolinkurl{http://ascl.net/#1}}}
\providecommand{\doarXiv}[1]{\href{https://arxiv.org/abs/#1}{\nolinkurl{https://arxiv.org/abs/#1}}}

\bibitem[{{Ak} {et~al.}(2016){Ak}, {Bostanc{\i}}, {Yontan}, {Bilir}, {G{\"u}ver}, {Ak}, {{\"U}rg{\"u}p}, \& {Paunzen}}]{Ak16}
{Ak}, T., {Bostanc{\i}}, Z.~F., {Yontan}, T., {et~al.} 2016, \apss, 361, 126, \dodoi{10.1007/s10509-016-2707-2}

\bibitem[{{Ak} {et~al.}(2024){Ak}, {Canbay}, \& {Yontan}}]{Ak24}
{Ak}, T., {Canbay}, R., \& {Yontan}, T. 2024, Physics and Astronomy Reports, 2, 58, \dodoi{10.26650/PAR.2024.00006}

\bibitem[{{Akbaba} {et~al.}(2024){Akbaba}, {Ak}, {Bilir}, {Plevne}, {{\~A}-nal Ta{\c{s}}}, \& {Seabroke}}]{Akbaba24}
{Akbaba}, F., {Ak}, T., {Bilir}, S., {et~al.} 2024, Astronomische Nachrichten, 345, e20240052, \dodoi{10.1002/asna.20240052}

\bibitem[{Akbulut {et~al.}(2021)Akbulut, Ak, Yontan, Bilir, Ak, Banks, Ulgen, \& Paunzen}]{Akbulut21}
Akbulut, B., Ak, S., Yontan, T., {et~al.} 2021, Astrophysics and Space Science, 366, 68, \dodoi{10.1007/s10509-021-03975-x}

\bibitem[{{Allison} {et~al.}(2009){Allison}, {Goodwin}, {Parker}, {de Grijs}, {Portegies Zwart}, \& {Kouwenhoven}}]{Allison09}
{Allison}, R.~J., {Goodwin}, S.~P., {Parker}, R.~J., {et~al.} 2009, \apjl, 700, L99, \dodoi{10.1088/0004-637X/700/2/L99}

\bibitem[{{Bahcall} \& {Soneira}(1980)}]{Bahcall80}
{Bahcall}, J.~N., \& {Soneira}, R.~M. 1980, \apjs, 44, 73, \dodoi{10.1086/190685}

\bibitem[{{Bailer-Jones} {et~al.}(2021){Bailer-Jones}, {Rybizki}, {Fouesneau}, {Demleitner}, \& {Andrae}}]{Bailer-Jones21}
{Bailer-Jones}, C.~A.~L., {Rybizki}, J., {Fouesneau}, M., {Demleitner}, M., \& {Andrae}, R. 2021, \aj, 161, 147, \dodoi{10.3847/1538-3881/abd806}

\bibitem[{{Banks} {et~al.}(2020){Banks}, {Yontan}, {Bilir}, \& {Canbay}}]{Banks20}
{Banks}, T., {Yontan}, T., {Bilir}, S., \& {Canbay}, R. 2020, Journal of Astrophysics and Astronomy, 41, 6, \dodoi{10.1007/s12036-020-9621-2}

\bibitem[{{Bastian} \& {Lardo}(2018)}]{Bastian18}
{Bastian}, N., \& {Lardo}, C. 2018, \araa, 56, 83, \dodoi{10.1146/annurev-astro-081817-051839}

\bibitem[{{Baumgardt} \& {Makino}(2003)}]{Baumgardt03}
{Baumgardt}, H., \& {Makino}, J. 2003, \mnras, 340, 227, \dodoi{10.1046/j.1365-8711.2003.06286.x}

\bibitem[{{Bertin} \& {Arnouts}(1996)}]{Bertin96}
{Bertin}, E., \& {Arnouts}, S. 1996, \aaps, 117, 393, \dodoi{10.1051/aas:1996164}

\bibitem[{{Bilir} {et~al.}(2010){Bilir}, {G{\"u}ver}, {Khamitov}, {Ak}, {Ak}, {Co{\c{s}}kuno{\u{g}}lu}, {Paunzen}, \& {Yaz}}]{Bilir10}
{Bilir}, S., {G{\"u}ver}, T., {Khamitov}, I., {et~al.} 2010, \apss, 326, 139, \dodoi{10.1007/s10509-009-0233-1}

\bibitem[{{Bilir} {et~al.}(2016){Bilir}, {Bostanc{\i}}, {Yontan}, {G{\"u}ver}, {Bak{\i}{\c{s}}}, {Ak}, {Ak}, {Paunzen}, \& {Eker}}]{Bilir16}
{Bilir}, S., {Bostanc{\i}}, Z.~F., {Yontan}, T., {et~al.} 2016, Advances in Space Research, 58, 1900, \dodoi{10.1016/j.asr.2016.06.039}

\bibitem[{{Bisht} {et~al.}(2020){Bisht}, {Zhu}, {Yadav}, {Durgapal}, \& {Rangwal}}]{Bisht20}
{Bisht}, D., {Zhu}, Q., {Yadav}, R.~K.~S., {Durgapal}, A., \& {Rangwal}, G. 2020, \mnras, 494, 607, \dodoi{10.1093/mnras/staa656}

\bibitem[{{Bostanc{\i}} {et~al.}(2015){Bostanc{\i}}, {Ak}, {Yontan}, {Bilir}, {G{\"u}ver}, {Ak}, {{\c{C}}ak{\i}rl{\i}}, {{\"O}zdarcan}, {Paunzen}, {De Cat}, {Fu}, {Zhang}, {Hou}, {Li}, {Wang}, {Zhang}, {Shi}, \& {Wu}}]{Bostanci15}
{Bostanc{\i}}, Z.~F., {Ak}, T., {Yontan}, T., {et~al.} 2015, \mnras, 453, 1095, \dodoi{10.1093/mnras/stv1665}

\bibitem[{{Bostanc{\i}} {et~al.}(2018){Bostanc{\i}}, {Yontan}, {Bilir}, {Ak}, {G{\"u}ver}, {Ak}, {Paunzen}, {Ba{\c{s}}aran}, {Vurgun}, {Akti}, {{\c{C}}elebi}, \& {{\"U}rg{\"u}p}}]{Bostanci18}
{Bostanc{\i}}, Z.~F., {Yontan}, T., {Bilir}, S., {et~al.} 2018, \apss, 363, 143, \dodoi{10.1007/s10509-018-3364-4}

\bibitem[{Bovy(2015)}]{Bovy15}
Bovy, J. 2015, The Astrophysical Journal Supplement Series, 216, 29, \dodoi{10.1088/0067-0049/216/2/29}

\bibitem[{Bovy \& Tremaine(2012)}]{Bovy12}
Bovy, J., \& Tremaine, S. 2012, The Astrophysical Journal, 756, 89, \dodoi{10.1088/0004-637X/756/1/89}

\bibitem[{Bressan {et~al.}(2012)Bressan, Marigo, Girardi, Salasnich, Dal~Cero, Rubele, \& Nanni}]{Bressan12}
Bressan, A., Marigo, P., Girardi, L., {et~al.} 2012, Monthly Notices of the Royal Astronomical Society, 427, 127, \dodoi{10.1111/j.1365-2966.2012.21948.x}

\bibitem[{{Cameron}(1985)}]{Cameron85}
{Cameron}, L.~M. 1985, \aap, 147, 39

\bibitem[{{Cantat-Gaudin} \& {Anders}(2020)}]{Cantat-Gaudin_Anders20}
{Cantat-Gaudin}, T., \& {Anders}, F. 2020, \aap, 633, A99, \dodoi{10.1051/0004-6361/201936691}

\bibitem[{Cantat-Gaudin {et~al.}(2018)Cantat-Gaudin, Jordi, Vallenari, Bragaglia, Balaguer-N{\'u}{\~n}ez, Soubiran, Bossini, Moitinho, Castro-Ginard, Krone-Martins, Casamiquela, Sordo, \& Carrera}]{Cantat-Gaudin18}
Cantat-Gaudin, T., Jordi, C., Vallenari, A., {et~al.} 2018, Astronomy and Astrophysics, 618, A93, \dodoi{10.1051/0004-6361/201833476}

\bibitem[{Cantat-Gaudin {et~al.}(2020)Cantat-Gaudin, Anders, Castro-Ginard, Jordi, Romero-G{\'o}mez, Soubiran, Casamiquela, Tarricq, Moitinho, Vallenari, Bragaglia, Krone-Martins, \& Kounkel}]{Cantat-Gaudin20}
Cantat-Gaudin, T., Anders, F., Castro-Ginard, A., {et~al.} 2020, Astronomy and Astrophysics, 640, A1, \dodoi{10.1051/0004-6361/202038192}

\bibitem[{{Cardelli} {et~al.}(1989){Cardelli}, {Clayton}, \& {Mathis}}]{Cardelli89}
{Cardelli}, J.~A., {Clayton}, G.~C., \& {Mathis}, J.~S. 1989, \apj, 345, 245, \dodoi{10.1086/167900}

\bibitem[{{Carraro} {et~al.}(2017){Carraro}, {Sales Silva}, {Moni Bidin}, \& {Vazquez}}]{Carraro17}
{Carraro}, G., {Sales Silva}, J.~V., {Moni Bidin}, C., \& {Vazquez}, R.~A. 2017, The Astronomical Journal, 153, 99, \dodoi{10.3847/1538-3881/153/3/99}

\bibitem[{{Castro-Ginard} {et~al.}(2021){Castro-Ginard}, {McMillan}, {Luri}, {Jordi}, {Romero-G{\'o}mez}, {Cantat-Gaudin}, {Casamiquela}, {Tarricq}, {Soubiran}, \& {Anders}}]{Castro-Ginard21}
{Castro-Ginard}, A., {McMillan}, P.~J., {Luri}, X., {et~al.} 2021, \aap, 652, A162, \dodoi{10.1051/0004-6361/202039751}

\bibitem[{{{\c{C}}akmak} {et~al.}(2024){{\c{C}}akmak}, {Yontan}, {Bilir}, {Banks}, {Michel}, {Soydugan}, {Ko{\c{c}}}, \& {Er{\c{c}}ay}}]{Cakmak24}
{{\c{C}}akmak}, H., {Yontan}, T., {Bilir}, S., {et~al.} 2024, Astronomische Nachrichten, 345, e20240054, \dodoi{10.1002/asna.20240054}

\bibitem[{{{\c{C}}{\i}nar} {et~al.}(2024){{\c{C}}{\i}nar}, {Tasdemir}, {Koc}, \& {Iyer}}]{Cinar24}
{{\c{C}}{\i}nar}, D.~C., {Tasdemir}, S., {Koc}, S., \& {Iyer}, S. 2024, Physics and Astronomy Reports, 2, 1, \dodoi{10.26650/PAR.2024.00002}

\bibitem[{{Chen} {et~al.}(2000){Chen}, {Nissen}, {Zhao}, {Zhang}, \& {Benoni}}]{Chen00}
{Chen}, Y.~Q., {Nissen}, P.~E., {Zhao}, G., {Zhang}, H.~W., \& {Benoni}, T. 2000, \aaps, 141, 491, \dodoi{10.1051/aas:2000124}

\bibitem[{{Conrad} {et~al.}(2017){Conrad}, {Scholz}, {Kharchenko}, {Piskunov}, {R{\"o}ser}, {Schilbach}, {de Jong}, {Schnurr}, {Steinmetz}, {Grebel}, {Zwitter}, {Bienaym{\'e}}, {Bland-Hawthorn}, {Gibson}, {Gilmore}, {Kordopatis}, {Kunder}, {Navarro}, {Parker}, {Reid}, {Seabroke}, {Siviero}, {Watson}, \& {Wyse}}]{Conrad17}
{Conrad}, C., {Scholz}, R.~D., {Kharchenko}, N.~V., {et~al.} 2017, \aap, 600, A106, \dodoi{10.1051/0004-6361/201630012}

\bibitem[{Co{\c{s}}kuno{\v{g}}lu {et~al.}(2011)Co{\c{s}}kuno{\v{g}}lu, Ak, Bilir, Karaali, Yaz, Gilmore, Seabroke, Bienaym{\'e}, Bland-Hawthorn, Campbell, Freeman, Gibson, Grebel, Munari, Navarro, Parker, Siebert, Siviero, Steinmetz, Watson, Wyse, \& Zwitter}]{Coskunoglu11}
Co{\c{s}}kuno{\v{g}}lu, B., Ak, S., Bilir, S., {et~al.} 2011, Monthly Notices of the Royal Astronomical Society, 412, 1237, \dodoi{10.1111/j.1365-2966.2010.17983.x}

\bibitem[{{Cui} {et~al.}(2012){Cui}, {Zhao}, {Chu}, {Li}, {Li}, {Zhang}, {Su}, {Yao}, {Wang}, {Xing}, {Li}, {Zhu}, {Wang}, {Gu}, {Luo}, {Xu}, {Zhang}, {Liu}, {Zhang}, {Yang}, {Cao}, {Chen}, {Chen}, {Chen}, {Chen}, {Chu}, {Feng}, {Gong}, {Hou}, {Hu}, {Hu}, {Hu}, {Jia}, {Jiang}, {Jiang}, {Jiang}, {Jin}, {Li}, {Li}, {Li}, {Liu}, {Liu}, {Lu}, {Mao}, {Men}, {Qi}, {Qi}, {Shi}, {Tang}, {Tao}, {Wang}, {Wang}, {Wang}, {Wang}, {Wang}, {Wang}, {Wang}, {Wang}, {Wang}, {Wang}, {Wang}, {Wang}, {Xu}, {Xu}, {Yang}, {Yu}, {Yuan}, {Yuan}, {Zhai}, {Zhang}, {Zhang}, {Zhang}, {Zhao}, {Zhou}, {Zhou}, {Zhu}, \& {Zou}}]{Cui12}
{Cui}, X.-Q., {Zhao}, Y.-H., {Chu}, Y.-Q., {et~al.} 2012, Research in Astronomy and Astrophysics, 12, 1197, \dodoi{10.1088/1674-4527/12/9/003}

\bibitem[{{Czernik}(1966)}]{Czernik66}
{Czernik}, M. 1966, \actaa, 16, 93

\bibitem[{{Deng} {et~al.}(2012){Deng}, {Newberg}, {Liu}, {Carlin}, {Beers}, {Chen}, {Chen}, {Christlieb}, {Grillmair}, {Guhathakurta}, {Han}, {Hou}, {Lee}, {L{\'e}pine}, {Li}, {Liu}, {Pan}, {Sellwood}, {Wang}, {Wang}, {Yang}, {Yanny}, {Zhang}, {Zhang}, {Zheng}, \& {Zhu}}]{Deng12}
{Deng}, L.-C., {Newberg}, H.~J., {Liu}, C., {et~al.} 2012, Research in Astronomy and Astrophysics, 12, 735, \dodoi{10.1088/1674-4527/12/7/003}

\bibitem[{Dias {et~al.}(2021)Dias, Monteir, Moitinho, L{\'e}pine, Carraro, Paunzen, Alessi, \& Villela}]{Dias21}
Dias, W.~S., Monteir, H., Moitinho, A., {et~al.} 2021, Monthly Notices of the Royal Astronomical Society, 504, 356, \dodoi{10.1093/mnras/stab770}

\bibitem[{Dias {et~al.}(2014)Dias, Monteiro, Caetano, L{\'e}pine, Assafin, \& Oliveira}]{Dias14}
Dias, W.~S., Monteiro, H., Caetano, T.~C., {et~al.} 2014, Astronomy and Astrophysics, 564, A79, \dodoi{10.1051/0004-6361/201323226}

\bibitem[{{Dib} \& {Henning}(2019)}]{Dib19}
{Dib}, S., \& {Henning}, T. 2019, \aap, 629, A135, \dodoi{10.1051/0004-6361/201834080}

\bibitem[{{Eker} \& {Bak{\i}\c{s}}(2025)}]{Eker25}
{Eker}, Z., \& {Bak{\i}\c{s}}, V. 2025, Physics and Astronomy Reports, 3, 43, \dodoi{10.26650/PAR.2025.00005}

\bibitem[{{Eker} {et~al.}(2024){Eker}, {Soydugan}, \& {Bilir}}]{Eker24}
{Eker}, Z., {Soydugan}, F., \& {Bilir}, S. 2024, Physics and Astronomy Reports, 2, 41, \dodoi{10.26650/PAR.2024.00001}

\bibitem[{Eker {et~al.}(2020)Eker, Soydugan, Bilir, G., Alpsoy, \& K{\"o}se}]{Eker20}
Eker, Z., Soydugan, F., Bilir, S., {et~al.} 2020, Monthly Notices of the Royal Astronomical Society, 496, 3887, \dodoi{10.1093/mnras/staa1659}

\bibitem[{Eker {et~al.}(2018)Eker, Bak{\i}{\c{s}}, Bilir, Soydugan, Steer, Soydugan, Bak{\i}{\c{s}}, Ali{\c{c}}avu{\c{s}}, Aslan, \& Alpsoy}]{Eker18}
Eker, Z., Bak{\i}{\c{s}}, V., Bilir, S., {et~al.} 2018, Monthly Notices of the Royal Astronomical Society, 479, 5491, \dodoi{10.1093/mnras/sty1834}

\bibitem[{{Elsanhoury} {et~al.}(2025){Elsanhoury}, {Haroon}, {Elkholy}, \& {{\c{C}}{\i}nar}}]{Elsanhoury25}
{Elsanhoury}, W.~H., {Haroon}, A.~A., {Elkholy}, E.~A., \& {{\c{C}}{\i}nar}, D.~C. 2025, Journal of Astrophysics and Astronomy, 46, 21, \dodoi{10.1007/s12036-025-10044-0}

\bibitem[{{Elson} {et~al.}(1987){Elson}, {Hut}, \& {Inagaki}}]{Elson87}
{Elson}, R., {Hut}, P., \& {Inagaki}, S. 1987, \araa, 25, 565, \dodoi{10.1146/annurev.aa.25.090187.003025}

\bibitem[{{Evcil} {et~al.}(2024){Evcil}, {Adalal{\i}}, {Alan}, {Canbay}, \& {Bilir}}]{Evcil24}
{Evcil}, S., {Adalal{\i}}, S., {Alan}, N., {Canbay}, R., \& {Bilir}, S. 2024, Astronomische Nachrichten, 345, e20240038, \dodoi{10.1002/asna.20240038}

\bibitem[{{Fischer} {et~al.}(1998){Fischer}, {Pryor}, {Murray}, {Mateo}, \& {Richtler}}]{Fischer98}
{Fischer}, P., {Pryor}, C., {Murray}, S., {Mateo}, M., \& {Richtler}, T. 1998, \aj, 115, 592, \dodoi{10.1086/300212}

\bibitem[{{Foreman-Mackey} {et~al.}(2013){Foreman-Mackey}, {Hogg}, {Lang}, \& {Goodman}}]{emcee}
{Foreman-Mackey}, D., {Hogg}, D.~W., {Lang}, D., \& {Goodman}, J. 2013, \pasp, 125, 306, \dodoi{10.1086/670067}

\bibitem[{{Frinchaboy} {et~al.}(2013){Frinchaboy}, {Thompson}, {Jackson}, {O'Connell}, {Meyer}, {Zasowski}, {Majewski}, {Chojnowksi}, {Johnson}, {Allende Prieto}, {Beers}, {Bizyaev}, {Brewington}, {Cunha}, {Ebelke}, {Garc{\'\i}a P{\'e}rez}, {Hearty}, {Holtzman}, {Kinemuchi}, {Malanushenko}, {Malanushenko}, {Marchante}, {M{\'e}sz{\'a}ros}, {Muna}, {Nidever}, {Oravetz}, {Pan}, {Schiavon}, {Schneider}, {Shetrone}, {Simmons}, {Snedden}, {Smith}, \& {Wilson}}]{Frinchaboy13}
{Frinchaboy}, P.~M., {Thompson}, B., {Jackson}, K.~M., {et~al.} 2013, \apjl, 777, L1, \dodoi{10.1088/2041-8205/777/1/L1}

\bibitem[{{Fu} {et~al.}(2022){Fu}, {Bragaglia}, {Liu}, {Zhang}, {Xu}, {Wang}, {Zhang}, {Zhong}, {Chang}, {Li}, {Chen}, {Chen}, {Wang}, {Gjergo}, {Wang}, {Yue}, \& {Zhang}}]{Fu22}
{Fu}, X., {Bragaglia}, A., {Liu}, C., {et~al.} 2022, \aap, 668, A4, \dodoi{10.1051/0004-6361/202243590}

\bibitem[{{Gaia Collaboration} {et~al.}(2016){Gaia Collaboration}, {Prusti}, {de Bruijne}, {Brown}, {Vallenari}, {Babusiaux}, {Bailer-Jones}, {Bastian}, {Biermann}, {Evans}, {Eyer}, {Jansen}, {Jordi}, {Klioner}, {Lammers}, {Lindegren}, {Luri}, {Mignard}, {Milligan}, {Panem}, {Poinsignon}, {Pourbaix}, {Randich}, {Sarri}, {Sartoretti}, {Siddiqui}, {Soubiran}, {Valette}, {van Leeuwen}, {Walton}, {Aerts}, {Arenou}, {Cropper}, {Drimmel}, {H{\o}g}, {Katz}, {Lattanzi}, {O'Mullane}, {Grebel}, {Holland}, {Huc}, {Passot}, {Bramante}, {Cacciari}, {Casta{\~n}eda}, {Chaoul}, {Cheek}, {De Angeli}, {Fabricius}, {Guerra}, {Hern{\'a}ndez}, {Jean-Antoine-Piccolo}, {Masana}, {Messineo}, {Mowlavi}, {Nienartowicz}, {Ord{\'o}{\~n}ez-Blanco}, {Panuzzo}, {Portell}, {Richards}, {Riello}, {Seabroke}, {Tanga}, {Th{\'e}venin}, {Torra}, {Els}, {Gracia-Abril}, {Comoretto}, {Garcia-Reinaldos}, {Lock}, {Mercier}, {Altmann}, {Andrae}, {Astraatmadja}, {Bellas-Velidis}, {Benson}, {Berthier}, {Blomme}, {Busso}, {Carry}, {Cellino}, {Clementini},
  {Cowell}, {Creevey}, {Cuypers}, {Davidson}, {De Ridder}, {de Torres}, {Delchambre}, {Dell'Oro}, {Ducourant}, {Fr{\'e}mat}, {Garc{\'\i}a-Torres}, {Gosset}, {Halbwachs}, {Hambly}, {Harrison}, {Hauser}, {Hestroffer}, {Hodgkin}, {Huckle}, {Hutton}, {Jasniewicz}, {Jordan}, {Kontizas}, {Korn}, {Lanzafame}, {Manteiga}, {Moitinho}, {Muinonen}, {Osinde}, {Pancino}, {Pauwels}, {Petit}, {Recio-Blanco}, {Robin}, {Sarro}, {Siopis}, {Smith}, {Smith}, {Sozzetti}, {Thuillot}, {van Reeven}, {Viala}, {Abbas}, {Abreu Aramburu}, {Accart}, {Aguado}, {Allan}, {Allasia}, {Altavilla}, {{\'A}lvarez}, {Alves}, {Anderson}, {Andrei}, {Anglada Varela}, {Antiche}, {Antoja}, {Ant{\'o}n}, {Arcay}, {Atzei}, {Ayache}, {Bach}, {Baker}, {Balaguer-N{\'u}{\~n}ez}, {Barache}, {Barata}, {Barbier}, {Barblan}, {Baroni}, {Barrado y Navascu{\'e}s}, {Barros}, {Barstow}, {Becciani}, {Bellazzini}, {Bellei}, {Bello Garc{\'\i}a}, {Belokurov}, {Bendjoya}, {Berihuete}, {Bianchi}, {Bienaym{\'e}}, {Billebaud}, {Blagorodnova}, {Blanco-Cuaresma}, {Boch},
  {Bombrun}, {Borrachero}, {Bouquillon}, {Bourda}, {Bouy}, {Bragaglia}, {Breddels}, {Brouillet}, {Br{\"u}semeister}, {Bucciarelli}, {Budnik}, {Burgess}, {Burgon}, {Burlacu}, {Busonero}, {Buzzi}, {Caffau}, {Cambras}, {Campbell}, {Cancelliere}, {Cantat-Gaudin}, {Carlucci}, {Carrasco}, {Castellani}, {Charlot}, {Charnas}, {Charvet}, {Chassat}, {Chiavassa}, {Clotet}, {Cocozza}, {Collins}, {Collins}, {Costigan}, {Crifo}, {Cross}, {Crosta}, {Crowley}, {Dafonte}, {Damerdji}, {Dapergolas}, {David}, {David}, {De Cat}, {de Felice}, {de Laverny}, {De Luise}, {De March}, {de Martino}, {de Souza}, {Debosscher}, {del Pozo}, {Delbo}, {Delgado}, {Delgado}, {di Marco}, {Di Matteo}, {Diakite}, {Distefano}, {Dolding}, {Dos Anjos}, {Drazinos}, {Dur{\'a}n}, {Dzigan}, {Ecale}, {Edvardsson}, {Enke}, {Erdmann}, {Escolar}, {Espina}, {Evans}, {Eynard Bontemps}, {Fabre}, {Fabrizio}, {Faigler}, {Falc{\~a}o}, {Farr{\`a}s Casas}, {Faye}, {Federici}, {Fedorets}, {Fern{\'a}ndez-Hern{\'a}ndez}, {Fernique}, {Fienga}, {Figueras}, {Filippi},
  {Findeisen}, {Fonti}, {Fouesneau}, {Fraile}, {Fraser}, {Fuchs}, {Furnell}, {Gai}, {Galleti}, {Galluccio}, {Garabato}, {Garc{\'\i}a-Sedano}, {Gar{\'e}}, {Garofalo}, {Garralda}, {Gavras}, {Gerssen}, {Geyer}, {Gilmore}, {Girona}, {Giuffrida}, {Gomes}, {Gonz{\'a}lez-Marcos}, {Gonz{\'a}lez-N{\'u}{\~n}ez}, {Gonz{\'a}lez-Vidal}, {Granvik}, {Guerrier}, {Guillout}, {Guiraud}, {G{\'u}rpide}, {Guti{\'e}rrez-S{\'a}nchez}, {Guy}, {Haigron}, {Hatzidimitriou}, {Haywood}, {Heiter}, {Helmi}, {Hobbs}, {Hofmann}, {Holl}, {Holland}, {Hunt}, {Hypki}, {Icardi}, {Irwin}, {Jevardat de Fombelle}, {Jofr{\'e}}, {Jonker}, {Jorissen}, {Julbe}, {Karampelas}, {Kochoska}, {Kohley}, {Kolenberg}, {Kontizas}, {Koposov}, {Kordopatis}, {Koubsky}, {Kowalczyk}, {Krone-Martins}, {Kudryashova}, {Kull}, {Bachchan}, {Lacoste-Seris}, {Lanza}, {Lavigne}, {Le Poncin-Lafitte}, {Lebreton}, {Lebzelter}, {Leccia}, {Leclerc}, {Lecoeur-Taibi}, {Lemaitre}, {Lenhardt}, {Leroux}, {Liao}, {Licata}, {Lindstr{\o}m}, {Lister}, {Livanou}, {Lobel}, {L{\"o}ffler},
  {L{\'o}pez}, {Lopez-Lozano}, {Lorenz}, {Loureiro}, {MacDonald}, {Magalh{\~a}es Fernandes}, {Managau}, {Mann}, {Mantelet}, {Marchal}, {Marchant}, {Marconi}, {Marie}, {Marinoni}, {Marrese}, {Marschalk{\'o}}, {Marshall}, {Mart{\'\i}n-Fleitas}, {Martino}, {Mary}, {Matijevi{\v{c}}}, {Mazeh}, {McMillan}, {Messina}, {Mestre}, {Michalik}, {Millar}, {Miranda}, {Molina}, {Molinaro}, {Molinaro}, {Moln{\'a}r}, {Moniez}, {Montegriffo}, {Monteiro}, {Mor}, {Mora}, {Morbidelli}, {Morel}, {Morgenthaler}, {Morley}, {Morris}, {Mulone}, {Muraveva}, {Musella}, {Narbonne}, {Nelemans}, {Nicastro}, {Noval}, {Ord{\'e}novic}, {Ordieres-Mer{\'e}}, {Osborne}, {Pagani}, {Pagano}, {Pailler}, {Palacin}, {Palaversa}, {Parsons}, {Paulsen}, {Pecoraro}, {Pedrosa}, {Pentik{\"a}inen}, {Pereira}, {Pichon}, {Piersimoni}, {Pineau}, {Plachy}, {Plum}, {Poujoulet}, {Pr{\v{s}}a}, {Pulone}, {Ragaini}, {Rago}, {Rambaux}, {Ramos-Lerate}, {Ranalli}, {Rauw}, {Read}, {Regibo}, {Renk}, {Reyl{\'e}}, {Ribeiro}, {Rimoldini}, {Ripepi}, {Riva}, {Rixon},
  {Roelens}, {Romero-G{\'o}mez}, {Rowell}, {Royer}, {Rudolph}, {Ruiz-Dern}, {Sadowski}, {Sagrist{\`a} Sell{\'e}s}, {Sahlmann}, {Salgado}, {Salguero}, {Sarasso}, {Savietto}, {Schnorhk}, {Schultheis}, {Sciacca}, {Segol}, {Segovia}, {Segransan}, {Serpell}, {Shih}, {Smareglia}, {Smart}, {Smith}, {Solano}, {Solitro}, {Sordo}, {Soria Nieto}, {Souchay}, {Spagna}, {Spoto}, {Stampa}, {Steele}, {Steidelm{\"u}ller}, {Stephenson}, {Stoev}, {Suess}, {S{\"u}veges}, {Surdej}, {Szabados}, {Szegedi-Elek}, {Tapiador}, {Taris}, {Tauran}, {Taylor}, {Teixeira}, {Terrett}, {Tingley}, {Trager}, {Turon}, {Ulla}, {Utrilla}, {Valentini}, {van Elteren}, {Van Hemelryck}, {van Leeuwen}, {Varadi}, {Vecchiato}, {Veljanoski}, {Via}, {Vicente}, {Vogt}, {Voss}, {Votruba}, {Voutsinas}, {Walmsley}, {Weiler}, {Weingrill}, {Werner}, {Wevers}, {Whitehead}, {Wyrzykowski}, {Yoldas}, {{\v{Z}}erjal}, {Zucker}, {Zurbach}, {Zwitter}, {Alecu}, {Allen}, {Allende Prieto}, {Amorim}, {Anglada-Escud{\'e}}, {Arsenijevic}, {Azaz}, {Balm}, {Beck}, {Bernstein},
  {Bigot}, {Bijaoui}, {Blasco}, {Bonfigli}, {Bono}, {Boudreault}, {Bressan}, {Brown}, {Brunet}, {Bunclark}, {Buonanno}, {Butkevich}, {Carret}, {Carrion}, {Chemin}, {Ch{\'e}reau}, {Corcione}, {Darmigny}, {de Boer}, {de Teodoro}, {de Zeeuw}, {Delle Luche}, {Domingues}, {Dubath}, {Fodor}, {Fr{\'e}zouls}, {Fries}, {Fustes}, {Fyfe}, {Gallardo}, {Gallegos}, {Gardiol}, {Gebran}, {Gomboc}, {G{\'o}mez}, {Grux}, {Gueguen}, {Heyrovsky}, {Hoar}, {Iannicola}, {Isasi Parache}, {Janotto}, {Joliet}, {Jonckheere}, {Keil}, {Kim}, {Klagyivik}, {Klar}, {Knude}, {Kochukhov}, {Kolka}, {Kos}, {Kutka}, {Lainey}, {LeBouquin}, {Liu}, {Loreggia}, {Makarov}, {Marseille}, {Martayan}, {Martinez-Rubi}, {Massart}, {Meynadier}, {Mignot}, {Munari}, {Nguyen}, {Nordlander}, {Ocvirk}, {O'Flaherty}, {Olias Sanz}, {Ortiz}, {Osorio}, {Oszkiewicz}, {Ouzounis}, {Palmer}, {Park}, {Pasquato}, {Peltzer}, {Peralta}, {P{\'e}turaud}, {Pieniluoma}, {Pigozzi}, {Poels}, {Prat}, {Prod'homme}, {Raison}, {Rebordao}, {Risquez}, {Rocca-Volmerange}, {Rosen},
  {Ruiz-Fuertes}, {Russo}, {Sembay}, {Serraller Vizcaino}, {Short}, {Siebert}, {Silva}, {Sinachopoulos}, {Slezak}, {Soffel}, {Sosnowska}, {Strai{\v{z}}ys}, {ter Linden}, {Terrell}, {Theil}, {Tiede}, {Troisi}, {Tsalmantza}, {Tur}, {Vaccari}, {Vachier}, {Valles}, {Van Hamme}, {Veltz}, {Virtanen}, {Wallut}, {Wichmann}, {Wilkinson}, {Ziaeepour}, \& {Zschocke}}]{Gaia16}
{Gaia Collaboration}, {Prusti}, T., {de Bruijne}, J.~H.~J., {et~al.} 2016, \aap, 595, A1, \dodoi{10.1051/0004-6361/201629272}

\bibitem[{{Gaia Collaboration} {et~al.}(2018){Gaia Collaboration}, {Brown}, {Vallenari}, {Prusti}, {de Bruijne}, {Babusiaux}, {Bailer-Jones}, {Biermann}, {Evans}, {Eyer}, {Jansen}, {Jordi}, {Klioner}, {Lammers}, {Lindegren}, {Luri}, {Mignard}, {Panem}, {Pourbaix}, {Randich}, {Sartoretti}, {Siddiqui}, {Soubiran}, {van Leeuwen}, {Walton}, {Arenou}, {Bastian}, {Cropper}, {Drimmel}, {Katz}, {Lattanzi}, {Bakker}, {Cacciari}, {Casta{\~n}eda}, {Chaoul}, {Cheek}, {De Angeli}, {Fabricius}, {Guerra}, {Holl}, {Masana}, {Messineo}, {Mowlavi}, {Nienartowicz}, {Panuzzo}, {Portell}, {Riello}, {Seabroke}, {Tanga}, {Th{\'e}venin}, {Gracia-Abril}, {Comoretto}, {Garcia-Reinaldos}, {Teyssier}, {Altmann}, {Andrae}, {Audard}, {Bellas-Velidis}, {Benson}, {Berthier}, {Blomme}, {Burgess}, {Busso}, {Carry}, {Cellino}, {Clementini}, {Clotet}, {Creevey}, {Davidson}, {De Ridder}, {Delchambre}, {Dell'Oro}, {Ducourant}, {Fern{\'a}ndez-Hern{\'a}ndez}, {Fouesneau}, {Fr{\'e}mat}, {Galluccio}, {Garc{\'\i}a-Torres},
  {Gonz{\'a}lez-N{\'u}{\~n}ez}, {Gonz{\'a}lez-Vidal}, {Gosset}, {Guy}, {Halbwachs}, {Hambly}, {Harrison}, {Hern{\'a}ndez}, {Hestroffer}, {Hodgkin}, {Hutton}, {Jasniewicz}, {Jean-Antoine-Piccolo}, {Jordan}, {Korn}, {Krone-Martins}, {Lanzafame}, {Lebzelter}, {L{\"o}ffler}, {Manteiga}, {Marrese}, {Mart{\'\i}n-Fleitas}, {Moitinho}, {Mora}, {Muinonen}, {Osinde}, {Pancino}, {Pauwels}, {Petit}, {Recio-Blanco}, {Richards}, {Rimoldini}, {Robin}, {Sarro}, {Siopis}, {Smith}, {Sozzetti}, {S{\"u}veges}, {Torra}, {van Reeven}, {Abbas}, {Abreu Aramburu}, {Accart}, {Aerts}, {Altavilla}, {{\'A}lvarez}, {Alvarez}, {Alves}, {Anderson}, {Andrei}, {Anglada Varela}, {Antiche}, {Antoja}, {Arcay}, {Astraatmadja}, {Bach}, {Baker}, {Balaguer-N{\'u}{\~n}ez}, {Balm}, {Barache}, {Barata}, {Barbato}, {Barblan}, {Barklem}, {Barrado}, {Barros}, {Barstow}, {Bartholom{\'e} Mu{\~n}oz}, {Bassilana}, {Becciani}, {Bellazzini}, {Berihuete}, {Bertone}, {Bianchi}, {Bienaym{\'e}}, {Blanco-Cuaresma}, {Boch}, {Boeche}, {Bombrun}, {Borrachero},
  {Bossini}, {Bouquillon}, {Bourda}, {Bragaglia}, {Bramante}, {Breddels}, {Bressan}, {Brouillet}, {Br{\"u}semeister}, {Brugaletta}, {Bucciarelli}, {Burlacu}, {Busonero}, {Butkevich}, {Buzzi}, {Caffau}, {Cancelliere}, {Cannizzaro}, {Cantat-Gaudin}, {Carballo}, {Carlucci}, {Carrasco}, {Casamiquela}, {Castellani}, {Castro-Ginard}, {Charlot}, {Chemin}, {Chiavassa}, {Cocozza}, {Costigan}, {Cowell}, {Crifo}, {Crosta}, {Crowley}, {Cuypers}, {Dafonte}, {Damerdji}, {Dapergolas}, {David}, {David}, {de Laverny}, {De Luise}, {De March}, {de Martino}, {de Souza}, {de Torres}, {Debosscher}, {del Pozo}, {Delbo}, {Delgado}, {Delgado}, {Di Matteo}, {Diakite}, {Diener}, {Distefano}, {Dolding}, {Drazinos}, {Dur{\'a}n}, {Edvardsson}, {Enke}, {Eriksson}, {Esquej}, {Eynard Bontemps}, {Fabre}, {Fabrizio}, {Faigler}, {Falc{\~a}o}, {Farr{\`a}s Casas}, {Federici}, {Fedorets}, {Fernique}, {Figueras}, {Filippi}, {Findeisen}, {Fonti}, {Fraile}, {Fraser}, {Fr{\'e}zouls}, {Gai}, {Galleti}, {Garabato}, {Garc{\'\i}a-Sedano}, {Garofalo},
  {Garralda}, {Gavel}, {Gavras}, {Gerssen}, {Geyer}, {Giacobbe}, {Gilmore}, {Girona}, {Giuffrida}, {Glass}, {Gomes}, {Granvik}, {Gueguen}, {Guerrier}, {Guiraud}, {Guti{\'e}rrez-S{\'a}nchez}, {Haigron}, {Hatzidimitriou}, {Hauser}, {Haywood}, {Heiter}, {Helmi}, {Heu}, {Hilger}, {Hobbs}, {Hofmann}, {Holland}, {Huckle}, {Hypki}, {Icardi}, {Jan{\ss}en}, {Jevardat de Fombelle}, {Jonker}, {Juh{\'a}sz}, {Julbe}, {Karampelas}, {Kewley}, {Klar}, {Kochoska}, {Kohley}, {Kolenberg}, {Kontizas}, {Kontizas}, {Koposov}, {Kordopatis}, {Kostrzewa-Rutkowska}, {Koubsky}, {Lambert}, {Lanza}, {Lasne}, {Lavigne}, {Le Fustec}, {Le Poncin-Lafitte}, {Lebreton}, {Leccia}, {Leclerc}, {Lecoeur-Taibi}, {Lenhardt}, {Leroux}, {Liao}, {Licata}, {Lindstr{\o}m}, {Lister}, {Livanou}, {Lobel}, {L{\'o}pez}, {Managau}, {Mann}, {Mantelet}, {Marchal}, {Marchant}, {Marconi}, {Marinoni}, {Marschalk{\'o}}, {Marshall}, {Martino}, {Marton}, {Mary}, {Massari}, {Matijevi{\v{c}}}, {Mazeh}, {McMillan}, {Messina}, {Michalik}, {Millar}, {Molina}, {Molinaro},
  {Moln{\'a}r}, {Montegriffo}, {Mor}, {Morbidelli}, {Morel}, {Morris}, {Mulone}, {Muraveva}, {Musella}, {Nelemans}, {Nicastro}, {Noval}, {O'Mullane}, {Ord{\'e}novic}, {Ord{\'o}{\~n}ez-Blanco}, {Osborne}, {Pagani}, {Pagano}, {Pailler}, {Palacin}, {Palaversa}, {Panahi}, {Pawlak}, {Piersimoni}, {Pineau}, {Plachy}, {Plum}, {Poggio}, {Poujoulet}, {Pr{\v{s}}a}, {Pulone}, {Racero}, {Ragaini}, {Rambaux}, {Ramos-Lerate}, {Regibo}, {Reyl{\'e}}, {Riclet}, {Ripepi}, {Riva}, {Rivard}, {Rixon}, {Roegiers}, {Roelens}, {Romero-G{\'o}mez}, {Rowell}, {Royer}, {Ruiz-Dern}, {Sadowski}, {Sagrist{\`a} Sell{\'e}s}, {Sahlmann}, {Salgado}, {Salguero}, {Sanna}, {Santana-Ros}, {Sarasso}, {Savietto}, {Schultheis}, {Sciacca}, {Segol}, {Segovia}, {S{\'e}gransan}, {Shih}, {Siltala}, {Silva}, {Smart}, {Smith}, {Solano}, {Solitro}, {Sordo}, {Soria Nieto}, {Souchay}, {Spagna}, {Spoto}, {Stampa}, {Steele}, {Steidelm{\"u}ller}, {Stephenson}, {Stoev}, {Suess}, {Surdej}, {Szabados}, {Szegedi-Elek}, {Tapiador}, {Taris}, {Tauran}, {Taylor},
  {Teixeira}, {Terrett}, {Teyssandier}, {Thuillot}, {Titarenko}, {Torra Clotet}, {Turon}, {Ulla}, {Utrilla}, {Uzzi}, {Vaillant}, {Valentini}, {Valette}, {van Elteren}, {Van Hemelryck}, {van Leeuwen}, {Vaschetto}, {Vecchiato}, {Veljanoski}, {Viala}, {Vicente}, {Vogt}, {von Essen}, {Voss}, {Votruba}, {Voutsinas}, {Walmsley}, {Weiler}, {Wertz}, {Wevers}, {Wyrzykowski}, {Yoldas}, {{\v{Z}}erjal}, {Ziaeepour}, {Zorec}, {Zschocke}, {Zucker}, {Zurbach}, \& {Zwitter}}]{Gaia18}
{Gaia Collaboration}, {Brown}, A.~G.~A., {Vallenari}, A., {et~al.} 2018, Astronomy and Astrophysics, 616, A1, \dodoi{10.1051/0004-6361/201833051}

\bibitem[{{Gaia Collaboration} {et~al.}(2023){Gaia Collaboration}, {Vallenari}, {Brown}, {Prusti}, {de Bruijne}, {Arenou}, {Babusiaux}, {Biermann}, {Creevey}, {Ducourant}, {Evans}, {Eyer}, {Guerra}, {Hutton}, {Jordi}, {Klioner}, {Lammers}, {Lindegren}, {Luri}, {Mignard}, {Panem}, {Pourbaix}, {Randich}, {Sartoretti}, {Soubiran}, {Tanga}, {Walton}, {Bailer-Jones}, {Bastian}, {Drimmel}, {Jansen}, {Katz}, {Lattanzi}, {van Leeuwen}, {Bakker}, {Cacciari}, {Casta{\~n}eda}, {De Angeli}, {Fabricius}, {Fouesneau}, {Fr{\'e}mat}, {Galluccio}, {Guerrier}, {Heiter}, {Masana}, {Messineo}, {Mowlavi}, {Nicolas}, {Nienartowicz}, {Pailler}, {Panuzzo}, {Riclet}, {Roux}, {Seabroke}, {Sordo}, {Th{\'e}venin}, {Gracia-Abril}, {Portell}, {Teyssier}, {Altmann}, {Andrae}, {Audard}, {Bellas-Velidis}, {Benson}, {Berthier}, {Blomme}, {Burgess}, {Busonero}, {Busso}, {C{\'a}novas}, {Carry}, {Cellino}, {Cheek}, {Clementini}, {Damerdji}, {Davidson}, {de Teodoro}, {Nu{\~n}ez Campos}, {Delchambre}, {Dell'Oro}, {Esquej},
  {Fern{\'a}ndez-Hern{\'a}ndez}, {Fraile}, {Garabato}, {Garc{\'\i}a-Lario}, {Gosset}, {Haigron}, {Halbwachs}, {Hambly}, {Harrison}, {Hern{\'a}ndez}, {Hestroffer}, {Hodgkin}, {Holl}, {Jan{\ss}en}, {Jevardat de Fombelle}, {Jordan}, {Krone-Martins}, {Lanzafame}, {L{\"o}ffler}, {Marchal}, {Marrese}, {Moitinho}, {Muinonen}, {Osborne}, {Pancino}, {Pauwels}, {Recio-Blanco}, {Reyl{\'e}}, {Riello}, {Rimoldini}, {Roegiers}, {Rybizki}, {Sarro}, {Siopis}, {Smith}, {Sozzetti}, {Utrilla}, {van Leeuwen}, {Abbas}, {{\'A}brah{\'a}m}, {Abreu Aramburu}, {Aerts}, {Aguado}, {Ajaj}, {Aldea-Montero}, {Altavilla}, {{\'A}lvarez}, {Alves}, {Anders}, {Anderson}, {Anglada Varela}, {Antoja}, {Baines}, {Baker}, {Balaguer-N{\'u}{\~n}ez}, {Balbinot}, {Balog}, {Barache}, {Barbato}, {Barros}, {Barstow}, {Bartolom{\'e}}, {Bassilana}, {Bauchet}, {Becciani}, {Bellazzini}, {Berihuete}, {Bernet}, {Bertone}, {Bianchi}, {Binnenfeld}, {Blanco-Cuaresma}, {Blazere}, {Boch}, {Bombrun}, {Bossini}, {Bouquillon}, {Bragaglia}, {Bramante}, {Breedt},
  {Bressan}, {Brouillet}, {Brugaletta}, {Bucciarelli}, {Burlacu}, {Butkevich}, {Buzzi}, {Caffau}, {Cancelliere}, {Cantat-Gaudin}, {Carballo}, {Carlucci}, {Carnerero}, {Carrasco}, {Casamiquela}, {Castellani}, {Castro-Ginard}, {Chaoul}, {Charlot}, {Chemin}, {Chiaramida}, {Chiavassa}, {Chornay}, {Comoretto}, {Contursi}, {Cooper}, {Cornez}, {Cowell}, {Crifo}, {Cropper}, {Crosta}, {Crowley}, {Dafonte}, {Dapergolas}, {David}, {David}, {de Laverny}, {De Luise}, \& {De March}}]{Gaia23}
{Gaia Collaboration}, {Vallenari}, A., {Brown}, A.~G.~A., {et~al.} 2023, \aap, 674, A1, \dodoi{10.1051/0004-6361/202243940}

\bibitem[{{Garcia} {et~al.}(1988){Garcia}, {Claria}, \& {Levato}}]{Garcia88}
{Garcia}, B., {Claria}, J.~J., \& {Levato}, H. 1988, \apss, 143, 317, \dodoi{10.1007/BF00637143}

\bibitem[{{Gokmen} {et~al.}(2023){Gokmen}, {Eker}, {Yontan}, {Bilir}, {Ak}, {Ak}, {Banks}, \& {Sarajedini}}]{Gokmen23}
{Gokmen}, S., {Eker}, Z., {Yontan}, T., {et~al.} 2023, \aj, 166, 263, \dodoi{10.3847/1538-3881/ad08b0}

\bibitem[{{Hao} {et~al.}(2022){Hao}, {Xu}, {Wu}, {Lin}, {Bian}, {Li}, \& {Liu}}]{Hao22}
{Hao}, C.~J., {Xu}, Y., {Wu}, Z.~Y., {et~al.} 2022, \aap, 668, A13, \dodoi{10.1051/0004-6361/202244570}

\bibitem[{{Hao} {et~al.}(2021){Hao}, {Xu}, {Hou}, {Bian}, {Li}, {Wu}, {He}, {Li}, \& {Liu}}]{Hao21}
{Hao}, C.~J., {Xu}, Y., {Hou}, L.~G., {et~al.} 2021, \aap, 652, A102, \dodoi{10.1051/0004-6361/202140608}

\bibitem[{{Haroon} {et~al.}(2025){Haroon}, {Elsanhoury}, {Elkholy}, {Saad}, \& {{\c{C}}{\i}nar}}]{Haroon25}
{Haroon}, A.~A., {Elsanhoury}, W.~H., {Elkholy}, E.~A., {Saad}, A.~S., \& {{\c{C}}{\i}nar}, D.~C. 2025, \physscr, 100, 055006, \dodoi{10.1088/1402-4896/adbf71}

\bibitem[{Heggie(1979)}]{Heggie79}
Heggie, D.~C. 1979, Monthly Notices of the Royal Astronomical Society, 188, 525, \dodoi{10.1093/mnras/188.3.525}

\bibitem[{{Herschel}(1802)}]{Herchel02}
{Herschel}, W. 1802, Phil. Trans. R. Soc., 92, 477, \dodoi{https://doi.org/10.1098/rstl.1802.0021}

\bibitem[{{Hidayat} {et~al.}(2019){Hidayat}, {Arifyanto}, {Aprilia}, \& {Hakim}}]{Hidayat19}
{Hidayat}, W., {Arifyanto}, M.~I., {Aprilia}, \& {Hakim}, M.~I. 2019, in Journal of Physics Conference Series, Vol. 1231, Journal of Physics Conference Series (IOP), 012032, \dodoi{10.1088/1742-6596/1231/1/012032}

\bibitem[{{Hill} \& {Zaritsky}(2006)}]{Hill06}
{Hill}, A., \& {Zaritsky}, D. 2006, \aj, 131, 414, \dodoi{10.1086/498647}

\bibitem[{{Hillenbrand} \& {Hartmann}(1998)}]{Hillenbrand98}
{Hillenbrand}, L.~A., \& {Hartmann}, L.~W. 1998, \apj, 492, 540, \dodoi{10.1086/305076}

\bibitem[{{Hoag} {et~al.}(1961){Hoag}, {Johnson}, {Iriarte}, {Mitchell}, {Hallam}, \& {Sharpless}}]{Hoag61}
{Hoag}, A.~A., {Johnson}, H.~L., {Iriarte}, B., {et~al.} 1961, Publ. Us. Nav. Obs. XVII part VII, 17, 347

\bibitem[{{Janes} {et~al.}(2013){Janes}, {Barnes}, {Meibom}, \& {Hoq}}]{Janes13}
{Janes}, K., {Barnes}, S.~A., {Meibom}, S., \& {Hoq}, S. 2013, \aj, 145, 7, \dodoi{10.1088/0004-6256/145/1/7}

\bibitem[{{Joshi} {et~al.}(2024){Joshi}, {Deepak}, \& {Malhotra}}]{Joshi24}
{Joshi}, Y.~C., {Deepak}, \& {Malhotra}, S. 2024, Frontiers in Astronomy and Space Sciences, 11, 1348321, \dodoi{10.3389/fspas.2024.1348321}

\bibitem[{Karaali {et~al.}(2003{\natexlab{a}})Karaali, Ak, Bilir, Karata{\c{s}}, \& Gilmore}]{Karaali03b}
Karaali, S., Ak, S.~G., Bilir, S., Karata{\c{s}}, Y., \& Gilmore, G. 2003{\natexlab{a}}, Monthly Notices of the Royal Astronomical Society, 343, 1013, \dodoi{10.1046/j.1365-8711.2003.06743.x}

\bibitem[{Karaali {et~al.}(2011)Karaali, Bilir, Ak, Yaz, \& Co{\c{s}}kuno{\u{g}}lu}]{Karaali11}
Karaali, S., Bilir, S., Ak, S., Yaz, E., \& Co{\c{s}}kuno{\u{g}}lu, B. 2011, Publications of the Astronomical Society of Australia, 28, 95, \dodoi{10.1071/AS10026}

\bibitem[{Karaali {et~al.}(2003{\natexlab{b}})Karaali, Bilir, Karata{\c{s}}, \& Ak}]{Karaali03a}
Karaali, S., Bilir, S., Karata{\c{s}}, Y., \& Ak, S.~G. 2003{\natexlab{b}}, Publications of the Astronomical Society of Australia, 20, 165, \dodoi{10.1071/AS02028}

\bibitem[{Kharchenko {et~al.}(2005)Kharchenko, Piskunov, R{\"o}ser, Schilbach, \& Scholz}]{Kharchenko05}
Kharchenko, N.~V., Piskunov, A.~E., R{\"o}ser, S., Schilbach, E., \& Scholz, R.~D. 2005, Astronomy and Astrophysics, 438, 1163, \dodoi{10.1051/0004-6361:20042523}

\bibitem[{{Kharchenko} {et~al.}(2012){Kharchenko}, {Piskunov}, {Schilbach}, {R{\"o}ser}, \& {Scholz}}]{Kharchenko12}
{Kharchenko}, N.~V., {Piskunov}, A.~E., {Schilbach}, E., {R{\"o}ser}, S., \& {Scholz}, R.~D. 2012, \aap, 543, A156, \dodoi{10.1051/0004-6361/201118708}

\bibitem[{Kharchenko {et~al.}(2013)Kharchenko, Piskunov, Schilbach, R{\"o}ser, \& Scholz}]{Kharchenko13}
Kharchenko, N.~V., Piskunov, A.~E., Schilbach, E., R{\"o}ser, S., \& Scholz, R.~D. 2013, Astronomy and Astrophysics, 558, A53, \dodoi{10.1051/0004-6361/201322302}

\bibitem[{King(1962)}]{King62}
King, I. 1962, Astronomical Journal, 67, 471, \dodoi{10.1086/108756}

\bibitem[{{Ko{\c{c}}} {et~al.}(2022){Ko{\c{c}}}, {Yontan}, {Bilir}, {Canbay}, {Ak}, {Banks}, {Ak}, \& {Paunzen}}]{Koc22}
{Ko{\c{c}}}, S., {Yontan}, T., {Bilir}, S., {et~al.} 2022, The Astronomical Journal, 163, 191, \dodoi{10.3847/1538-3881/ac58a0}

\bibitem[{Krone-Martins \& Moitinho(2014)}]{Krone-Martins14}
Krone-Martins, A., \& Moitinho, A. 2014, Astronomy and Astrophysics, 561, A57, \dodoi{10.1051/0004-6361/201321143}

\bibitem[{{Kruijssen} {et~al.}(2012){Kruijssen}, {Maschberger}, {Moeckel}, {Clarke}, {Bastian}, \& {Bonnell}}]{Kruijssen12}
{Kruijssen}, J.~M.~D., {Maschberger}, T., {Moeckel}, N., {et~al.} 2012, \mnras, 419, 841, \dodoi{10.1111/j.1365-2966.2011.19748.x}

\bibitem[{{Lada} \& {Lada}(2003)}]{Lada03}
{Lada}, C.~J., \& {Lada}, E.~A. 2003, \araa, 41, 57, \dodoi{10.1146/annurev.astro.41.011802.094844}

\bibitem[{Landolt(2009)}]{Landolt09}
Landolt, A.~U. 2009, The Astronomical Journal, 137, 4186, \dodoi{10.1088/0004-6256/137/5/4186}

\bibitem[{{Lang} {et~al.}(2010){Lang}, {Hogg}, {Mierle}, {Blanton}, \& {Roweis}}]{Lang10}
{Lang}, D., {Hogg}, D.~W., {Mierle}, K., {Blanton}, M., \& {Roweis}, S. 2010, \aj, 139, 1782, \dodoi{10.1088/0004-6256/139/5/1782}

\bibitem[{Leggett(1992)}]{Leggett92}
Leggett, S.~K. 1992, Astrophysical Journal Supplement, 82, 351, \dodoi{10.1086/191720}

\bibitem[{Liu \& Pang(2019)}]{Liu19}
Liu, L., \& Pang, X. 2019, The Astrophysical Journal Supplement Series, 245, 32, \dodoi{10.3847/1538-4365/ab530a}

\bibitem[{{Loktin} \& {Beshenov}(2003)}]{Loktin03}
{Loktin}, A.~V., \& {Beshenov}, G.~V. 2003, Astronomy Reports, 47, 6, \dodoi{10.1134/1.1538491}

\bibitem[{Loktin \& Popova(2017)}]{Loktin17}
Loktin, A.~V., \& Popova, M.~E. 2017, Astrophysical Bulletin, 72, 257, \dodoi{10.1134/S1990341317030154}

\bibitem[{{Lu} {et~al.}(2024){Lu}, {Minchev}, {Buck}, {Khoperskov}, {Steinmetz}, {Libeskind}, {Cescutti}, {Freeman}, \& {Ratcliffe}}]{Yuxi24}
{Lu}, Y.~L., {Minchev}, I., {Buck}, T., {et~al.} 2024, \mnras, 535, 392, \dodoi{10.1093/mnras/stae2364}

\bibitem[{{Maciejewski} \& {Niedzielski}(2007)}]{Maciejewski07}
{Maciejewski}, G., \& {Niedzielski}, A. 2007, \aap, 467, 1065, \dodoi{10.1051/0004-6361:20066588}

\bibitem[{{Marshall} {et~al.}(2006){Marshall}, {Robin}, {Reyl{\'e}}, {Schultheis}, \& {Picaud}}]{Marshall06}
{Marshall}, D.~J., {Robin}, A.~C., {Reyl{\'e}}, C., {Schultheis}, M., \& {Picaud}, S. 2006, \aap, 453, 635, \dodoi{10.1051/0004-6361:20053842}

\bibitem[{{Maurya} \& {Joshi}(2020)}]{Maurya20}
{Maurya}, J., \& {Joshi}, Y.~C. 2020, \mnras, 494, 4713, \dodoi{10.1093/mnras/staa893}

\bibitem[{{McKee} \& {Ostriker}(2007)}]{McKee07}
{McKee}, C.~F., \& {Ostriker}, E.~C. 2007, \araa, 45, 565, \dodoi{10.1146/annurev.astro.45.051806.110602}

\bibitem[{{Medina} {et~al.}(2021){Medina}, {Lemasle}, \& {Grebel}}]{Medina21}
{Medina}, G.~E., {Lemasle}, B., \& {Grebel}, E.~K. 2021, \mnras, 505, 1342, \dodoi{10.1093/mnras/stab1267}

\bibitem[{{Mermilliod} {et~al.}(2008){Mermilliod}, {Mayor}, \& {Udry}}]{Mermilliod08}
{Mermilliod}, J.~C., {Mayor}, M., \& {Udry}, S. 2008, \aap, 485, 303, \dodoi{10.1051/0004-6361:200809664}

\bibitem[{{Mihalas} \& {Binney}(1981)}]{Mihalas81}
{Mihalas}, D., \& {Binney}, J. 1981, {Galactic astronomy. Structure and kinematics} (W.H. Freeman \& Co.)

\bibitem[{{Minchev} {et~al.}(2018){Minchev}, {Anders}, {Recio-Blanco}, {Chiappini}, {de Laverny}, {Queiroz}, {Steinmetz}, {Adibekyan}, {Carrillo}, {Cescutti}, {Guiglion}, {Hayden}, {de Jong}, {Kordopatis}, {Majewski}, {Martig}, \& {Santiago}}]{Minchev18}
{Minchev}, I., {Anders}, F., {Recio-Blanco}, A., {et~al.} 2018, \mnras, 481, 1645, \dodoi{10.1093/mnras/sty2033}

\bibitem[{{Miocchi} {et~al.}(2013){Miocchi}, {Lanzoni}, {Ferraro}, {Dalessandro}, {Vesperini}, {Pasquato}, {Beccari}, {Pallanca}, \& {Sanna}}]{Miocchi13}
{Miocchi}, P., {Lanzoni}, B., {Ferraro}, F.~R., {et~al.} 2013, \apj, 774, 151, \dodoi{10.1088/0004-637X/774/2/151}

\bibitem[{{Mu{\~n}oz} {et~al.}(2010){Mu{\~n}oz}, {Geha}, \& {Willman}}]{Munoz10}
{Mu{\~n}oz}, R.~R., {Geha}, M., \& {Willman}, B. 2010, \aj, 140, 138, \dodoi{10.1088/0004-6256/140/1/138}

\bibitem[{{Netopil} {et~al.}(2022){Netopil}, {Oralhan}, {{\c{C}}akmak}, {Michel}, \& {Karata{\c{s}}}}]{Netopil22}
{Netopil}, M., {Oralhan}, {\.I}.~A., {{\c{C}}akmak}, H., {Michel}, R., \& {Karata{\c{s}}}, Y. 2022, \mnras, 509, 421, \dodoi{10.1093/mnras/stab2961}

\bibitem[{{O'Donnell}(1994)}]{ODonnell94}
{O'Donnell}, J.~E. 1994, \apj, 422, 158, \dodoi{10.1086/173713}

\bibitem[{{Pavl{\'\i}k} \& {Vesperini}(2022)}]{Pavlik22}
{Pavl{\'\i}k}, V., \& {Vesperini}, E. 2022, \mnras, 515, 1830, \dodoi{10.1093/mnras/stac1776}

\bibitem[{{Pe{\~n}a} {et~al.}(1994){Pe{\~n}a}, {Peniche}, {Bravo}, \& {Yam}}]{Pena94}
{Pe{\~n}a}, J.~H., {Peniche}, R., {Bravo}, H., \& {Yam}, O. 1994, \rmxaa, 28, 7

\bibitem[{{Piskunov} {et~al.}(2006){Piskunov}, {Kharchenko}, {R{\"o}ser}, {Schilbach}, \& {Scholz}}]{Piskunov06}
{Piskunov}, A.~E., {Kharchenko}, N.~V., {R{\"o}ser}, S., {Schilbach}, E., \& {Scholz}, R.~D. 2006, \aap, 445, 545, \dodoi{10.1051/0004-6361:20053764}

\bibitem[{{Plevne} {et~al.}(2015){Plevne}, {Ak}, {Karaali}, {Bilir}, {Ak}, \& {Bostanci}}]{Plevne15}
{Plevne}, O., {Ak}, T., {Karaali}, S., {et~al.} 2015, \pasa, 32, e043, \dodoi{10.1017/pasa.2015.44}

\bibitem[{{Portegies Zwart} {et~al.}(2010){Portegies Zwart}, {McMillan}, \& {Gieles}}]{Portegies-Zwart10}
{Portegies Zwart}, S.~F., {McMillan}, S. L.~W., \& {Gieles}, M. 2010, \araa, 48, 431, \dodoi{10.1146/annurev-astro-081309-130834}

\bibitem[{{Raboud} \& {Mermilliod}(1998)}]{Raboud98}
{Raboud}, D., \& {Mermilliod}, J.~C. 1998, \aap, 333, 897, \dodoi{10.48550/arXiv.astro-ph/9802284}

\bibitem[{{Rangwal} {et~al.}(2024){Rangwal}, {Arya}, {Subramaniam}, {Singh}, \& {Liu}}]{Rangwal24}
{Rangwal}, G., {Arya}, A., {Subramaniam}, A., {Singh}, K.~P., \& {Liu}, X. 2024, arXiv e-prints, arXiv:2410.15305, \dodoi{10.48550/arXiv.2410.15305}

\bibitem[{{Reddy} {et~al.}(2015){Reddy}, {Giridhar}, \& {Lambert}}]{Reddy15}
{Reddy}, A. B.~S., {Giridhar}, S., \& {Lambert}, D.~L. 2015, \mnras, 450, 4301, \dodoi{10.1093/mnras/stv908}

\bibitem[{{Reddy} {et~al.}(2020){Reddy}, {Giridhar}, \& {Lambert}}]{Reddy20}
---. 2020, Journal of Astrophysics and Astronomy, 41, 38, \dodoi{10.1007/s12036-020-09658-3}

\bibitem[{{Robin} {et~al.}(2003){Robin}, {Reyl{\'e}}, {Derri{\`e}re}, \& {Picaud}}]{Robin03}
{Robin}, A.~C., {Reyl{\'e}}, C., {Derri{\`e}re}, S., \& {Picaud}, S. 2003, \aap, 409, 523, \dodoi{10.1051/0004-6361:20031117}

\bibitem[{{Robin} {et~al.}(2012){Robin}, {Luri}, {Reyl{\'e}}, {Isasi}, {Grux}, {Blanco-Cuaresma}, {Arenou}, {Babusiaux}, {Belcheva}, {Drimmel}, {Jordi}, {Krone-Martins}, {Masana}, {Mauduit}, {Mignard}, {Mowlavi}, {Rocca-Volmerange}, {Sartoretti}, {Slezak}, \& {Sozzetti}}]{Robin12}
{Robin}, A.~C., {Luri}, X., {Reyl{\'e}}, C., {et~al.} 2012, \aap, 543, A100, \dodoi{10.1051/0004-6361/201118646}

\bibitem[{{Roeser} {et~al.}(2010){Roeser}, {Demleitner}, \& {Schilbach}}]{Roeser10}
{Roeser}, S., {Demleitner}, M., \& {Schilbach}, E. 2010, \aj, 139, 2440, \dodoi{10.1088/0004-6256/139/6/2440}

\bibitem[{{Ruprecht}(1966)}]{Ruprecht66}
{Ruprecht}, J. 1966, Bulletin of the Astronomical Institutes of Czechoslovakia, 17, 33

\bibitem[{{Sagar} {et~al.}(1988){Sagar}, {Miakutin}, {Piskunov}, \& {Dluzhnevskaia}}]{Sagar88}
{Sagar}, R., {Miakutin}, V.~I., {Piskunov}, A.~E., \& {Dluzhnevskaia}, O.~B. 1988, \mnras, 234, 831, \dodoi{10.1093/mnras/234.4.831}

\bibitem[{{Salpeter}(1955)}]{Salpeter55}
{Salpeter}, E.~E. 1955, \apj, 121, 161, \dodoi{10.1086/145971}

\bibitem[{{Sarajedini} {et~al.}(1995){Sarajedini}, {Milone}, \& {Lowe}}]{Sarajedini95}
{Sarajedini}, A., {Milone}, A.~A.~E., \& {Lowe}, C.~R. 1995, in American Astronomical Society Meeting Abstracts, Vol. 186, American Astronomical Society Meeting Abstracts \#186, 49.03

\bibitem[{{Schlafly} \& {Finkbeiner}(2011)}]{Schlafly11}
{Schlafly}, E.~F., \& {Finkbeiner}, D.~P. 2011, \apj, 737, 103, \dodoi{10.1088/0004-637X/737/2/103}

\bibitem[{{Science Software Branch at STScI}(2012)}]{Science12}
{Science Software Branch at STScI}. 2012, {PyRAF: Python alternative for IRAF}, Astrophysics Source Code Library, record ascl:1207.011

\bibitem[{{Sestito} {et~al.}(2008){Sestito}, {Bragaglia}, {Randich}, {Pallavicini}, {Andrievsky}, \& {Korotin}}]{Sestito08}
{Sestito}, P., {Bragaglia}, A., {Randich}, S., {et~al.} 2008, \aap, 488, 943, \dodoi{10.1051/0004-6361:200809650}

\bibitem[{{Skrutskie} {et~al.}(2006){Skrutskie}, {Cutri}, {Stiening}, {Weinberg}, {Schneider}, {Carpenter}, {Beichman}, {Capps}, {Chester}, {Elias}, {Huchra}, {Liebert}, {Lonsdale}, {Monet}, {Price}, {Seitzer}, {Jarrett}, {Kirkpatrick}, {Gizis}, {Howard}, {Evans}, {Fowler}, {Fullmer}, {Hurt}, {Light}, {Kopan}, {Marsh}, {McCallon}, {Tam}, {Van Dyk}, \& {Wheelock}}]{Skrutzie16}
{Skrutskie}, M.~F., {Cutri}, R.~M., {Stiening}, R., {et~al.} 2006, \aj, 131, 1163, \dodoi{10.1086/498708}

\bibitem[{Soubiran {et~al.}(2018)Soubiran, Cantat-Gaudin, Romero-G{\'o}mez, Casamiquela, Jordi, Vallenari, Antoja, Balaguer-N{\'u}{\~n}ez, Bossini, Bragaglia, Carrera, Castro-Ginard, Figueras, Heiter, Katz, Krone-Martins, Le~Campion, Moitinho, \& Sordo}]{Soubiran18}
Soubiran, C., Cantat-Gaudin, T., Romero-G{\'o}mez, M., {et~al.} 2018, Astronomy and Astrophysics, 619, A155, \dodoi{10.1051/0004-6361/201834020}

\bibitem[{{Spina} {et~al.}(2022){Spina}, {Magrini}, \& {Cunha}}]{Spina22}
{Spina}, L., {Magrini}, L., \& {Cunha}, K. 2022, Universe, 8, 87, \dodoi{10.3390/universe8020087}

\bibitem[{{Spitzer}(1987)}]{Spitzer87}
{Spitzer}, L. 1987, {Dynamical evolution of globular clusters}

\bibitem[{{Spitzer} \& {Hart}(1971)}]{Spitzer71}
{Spitzer}, Jr., L., \& {Hart}, M.~H. 1971, \apj, 164, 399, \dodoi{10.1086/150855}

\bibitem[{{Strobel}(1991{\natexlab{a}})}]{Strobel-Poland91}
{Strobel}, A. 1991{\natexlab{a}}, Astronomische Nachrichten, 312, 177, \dodoi{10.1002/asna.2113120306}

\bibitem[{{Strobel}(1991{\natexlab{b}})}]{Strobel91}
---. 1991{\natexlab{b}}, \aap, 247, 35

\bibitem[{Sung {et~al.}(2013)Sung, Lim, Bessell, Kim, Hur, Chun, \& Byeong-Gon}]{Sung13}
Sung, H., Lim, B., Bessell, M.~S., {et~al.} 2013, Journal of Korean Astronomical Society, 46, 103, \dodoi{10.5303/JKAS.2013.46.3.103}

\bibitem[{{Ta{\c{s}}demir} {et~al.}(2025){Ta{\c{s}}demir}, {{\c{C}}{\i}nar}, {Canbay}, {Ta{\c{s}}tan}, {Elsanhoury}, \& {Haroon}}]{Tasdemir25}
{Ta{\c{s}}demir}, S., {{\c{C}}{\i}nar}, D.~C., {Canbay}, R., {et~al.} 2025, Physics and Astronomy Reports, 3, 1, \dodoi{10.26650/PAR.2025.00003}

\bibitem[{{Ta{\c{s}}demir} \& {Yontan}(2023)}]{Tasdemir23}
{Ta{\c{s}}demir}, S., \& {Yontan}, T. 2023, Physics and Astronomy Reports, 1, 1, \dodoi{10.26650/PAR.2023.00001}

\bibitem[{Tarricq {et~al.}(2021)Tarricq, C., Casamiquela, Cantat-Gaudin, Chemin, Anders, Antoja, omero G{\'o}mez, Figueras, Jordi, Bragaglia, Balaguer-N{\'u}{\~n}ez, Carrera, Castro-Ginard, Moitinho, Ramos, \& Bossini}]{Tarricq21}
Tarricq, Y., C., S., Casamiquela, L., {et~al.} 2021, Astronomy and Astrophysics, 647, A19, \dodoi{10.1051/0004-6361/202039388}

\bibitem[{{Tody}(1986)}]{Tody86}
{Tody}, D. 1986, in Society of Photo-Optical Instrumentation Engineers (SPIE) Conference Series, Vol. 627, Instrumentation in astronomy VI, ed. D.~L. {Crawford}, 733, \dodoi{10.1117/12.968154}

\bibitem[{{Tody}(1993)}]{Tody93}
{Tody}, D. 1993, in Astronomical Society of the Pacific Conference Series, Vol.~52, Astronomical Data Analysis Software and Systems II, ed. R.~J. {Hanisch}, R.~J.~V. {Brissenden}, \& J.~{Barnes}, 173

\bibitem[{{Tun{\c{c}}el G{\"u}{\c{c}}tekin} {et~al.}(2019){Tun{\c{c}}el G{\"u}{\c{c}}tekin}, {Bilir}, {Karaali}, {Plevne}, \& {Ak}}]{Guctekin19}
{Tun{\c{c}}el G{\"u}{\c{c}}tekin}, S., {Bilir}, S., {Karaali}, S., {Plevne}, O., \& {Ak}, S. 2019, Advances in Space Research, 63, 1360, \dodoi{10.1016/j.asr.2018.10.041}

\bibitem[{{V{\'a}zquez} {et~al.}(2010){V{\'a}zquez}, {Moitinho}, {Carraro}, \& {Dias}}]{Vazquez10}
{V{\'a}zquez}, R.~A., {Moitinho}, A., {Carraro}, G., \& {Dias}, W.~S. 2010, \aap, 511, A38, \dodoi{10.1051/0004-6361/200811583}

\bibitem[{Virtanen {et~al.}(2020)Virtanen, Gommers, Oliphant, Haberland, Reddy, Cournapeau, Burovski, Peterson, Weckesser, Bright, {van der Walt}, Brett, Wilson, Millman, Mayorov, Nelson, Jones, Kern, Larson, Carey, Polat, Feng, Moore, {VanderPlas}, Laxalde, Perktold, Cimrman, Henriksen, Quintero, Harris, Archibald, Ribeiro, Pedregosa, {van Mulbregt}, \& {SciPy 1.0 Contributors}}]{Scipy}
Virtanen, P., Gommers, R., Oliphant, T.~E., {et~al.} 2020, Nature Methods, 17, 261, \dodoi{10.1038/s41592-019-0686-2}

\bibitem[{{Yontan}(2023a)}]{Yontan23a}
{Yontan}, T. 2023a, \aj, 165, 79, \dodoi{10.3847/1538-3881/aca6f0}

\bibitem[{{Yontan} {et~al.}(2023b){Yontan}, {Bilir}, {{\c{C}}akmak}, {Ra{\'u}l}, {Banks}, {Soydugan}, {Canbay}, \& {Ta{\c{s}}demir}}]{Yontan23b}
{Yontan}, T., {Bilir}, S., {{\c{C}}akmak}, H., {et~al.} 2023b, Advances in Space Research, 72, 1454, \dodoi{10.1016/j.asr.2023.04.015}

\bibitem[{{Yontan} \& {Canbay}(2023c)}]{Yontan23c}
{Yontan}, T., \& {Canbay}, R. 2023c, Physics and Astronomy Reports, 1, 65, \dodoi{10.26650/PAR.2023.00008}

\bibitem[{{Yontan} {et~al.}(2015){Yontan}, {Bilir}, {Bostanc{\i}}, {Ak}, {Karaali}, {G{\"u}ver}, {Ak}, {Duran}, \& {Paunzen}}]{Yontan15}
{Yontan}, T., {Bilir}, S., {Bostanc{\i}}, Z.~F., {et~al.} 2015, \apss, 355, 267, \dodoi{10.1007/s10509-014-2175-5}

\bibitem[{{Yontan} {et~al.}(2019){Yontan}, {Bilir}, {Bostanc{\i}}, {Ak}, {Ak}, {G{\"u}ver}, {Paunzen}, {{\"U}rg{\"u}p}, {{\c{C}}elebi}, {Akti}, \& {G{\"o}kmen}}]{Yontan19}
---. 2019, \apss, 364, 152, \dodoi{10.1007/s10509-019-3640-y}

\bibitem[{{Yontan} {et~al.}(2021){Yontan}, {Bilir}, {Ak}, {Akbulut}, {Canbay}, {Banks}, {Paunzen}, {Ak}, \& {Bostanc{\i}}}]{Yontan21}
{Yontan}, T., {Bilir}, S., {Ak}, T., {et~al.} 2021, Astronomische Nachrichten, 342, 538, \dodoi{10.1002/asna.202113837}

\bibitem[{{Yontan} {et~al.}(2022){Yontan}, {{\c{C}}akmak}, {Bilir}, {Banks}, {Ra{\'u}l}, {Canbay}, {Ko{\c{c}}}, {Ta{\c{s}}demir}, {Er{\c{c}}ay}, {Tan{\i}k Ozt{\"u}rk}, \& {Dursun}}]{Yontan22}
{Yontan}, T., {{\c{C}}akmak}, T., {Bilir}, S., {et~al.} 2022, \rmxaa, 58, 333, \dodoi{10.22201/ia.01851101p.2022.58.02.14}

\bibitem[{{Y{\"u}cel} {et~al.}(2025){Y{\"u}cel}, {Bak{\i}{\c{s}}}, {Canbay}, {Alan}, {Banks}, \& {Bilir}}]{Yucel25}
{Y{\"u}cel}, G., {Bak{\i}{\c{s}}}, V., {Canbay}, R., {et~al.} 2025, \aj, 169, 71, \dodoi{10.3847/1538-3881/ad9d46}

\bibitem[{{Y\"ucel} {et~al.}(2024){Y\"ucel}, {Canbay}, \& {Bak{\i}\c s}}]{Yucel24}
{Y\"ucel}, G., {Canbay}, R., \& {Bak{\i}\c s}, V. 2024, Physics and Astronomy Reports, 2, 18, \dodoi{10.26650/PAR.2024.00003}

\bibitem[{{Zacharias} {et~al.}(2004){Zacharias}, {Urban}, {Zacharias}, {Wycoff}, {Hall}, {Monet}, \& {Rafferty}}]{Zacharias04}
{Zacharias}, N., {Urban}, S.~E., {Zacharias}, M.~I., {et~al.} 2004, \aj, 127, 3043, \dodoi{10.1086/386353}

\bibitem[{{Zhao} {et~al.}(2012){Zhao}, {Zhao}, {Chu}, {Jing}, \& {Deng}}]{Zhao12}
{Zhao}, G., {Zhao}, Y.-H., {Chu}, Y.-Q., {Jing}, Y.-P., \& {Deng}, L.-C. 2012, Research in Astronomy and Astrophysics, 12, 723, \dodoi{10.1088/1674-4527/12/7/002}

\bibitem[{{Zhong} {et~al.}(2020){Zhong}, {Chen}, {Wu}, {Li}, {Bai}, \& {Hou}}]{Zhong20}
{Zhong}, J., {Chen}, L., {Wu}, D., {et~al.} 2020, \aap, 640, A127, \dodoi{10.1051/0004-6361/201937131}

\bibitem[{{Zhou} \& {Chen}(2021)}]{Zhou21}
{Zhou}, X., \& {Chen}, X. 2021, \mnras, 504, 4768, \dodoi{10.1093/mnras/stab1209}

\end{thebibliography}
\bibliographystyle{aasjournal}

\end{document}